\documentclass[preprint2]{aastex}
\shorttitle{CAESAR}
\shortauthors{Davison}
\usepackage{rotating}
\usepackage{pdflscape}
\usepackage{epsfig}
\usepackage{natbib}
\usepackage{subfigure}
\usepackage{amsmath}
\usepackage{longtable}
\usepackage{mathabx}

\begin{document}

\title{A 3D Search for Companions to 12 Nearby M-Dwarfs}
\author{Davison, Cassy L.$^1$, White, R.J. $^1$, Henry, T.J.$^2$, Riedel, A.R.$^{3,4}$, Jao, W-C.$^1$, Bailey, J.I., III$^5$, Quinn, S.N.$^1$, Cantrell, J.R.$^1$, Subasavage, J.~P.$^6$ and Winters, J.G.$^1$}
\affil{$^1$Department of Physics and Astronomy, Georgia State University, Atlanta, GA, 30303, USA}
\affil{$^2$RECONS Institute, Chambersburg, PA 17201, USA}
\affil{$^3$Department of Physics and Astronomy, Hunter College, New York, NY, 10065, USA}
\affil{$^4$American Museum of Natural History, New York, NY 10024}
\affil{$^5$University of Michigan, Department of Astronomy, Ann Arbor, Mi, 48109, USA}
\affil{$^6$United States Naval Observatory, Flagstaff, AZ, 86002, USA}
\begin{abstract}  
We present a carefully vetted equatorial ($\pm$ 30$^\circ$ Decl.) sample of all known single (within 4$\arcsec$) mid M-dwarfs (M2.5V-M8.0V) extending out to 10 pc; their proximity and low masses make them ideal targets for planet searches.  For this sample of 58 stars, we provide \textit{V$_J$}, \textit{R$_{KC}$}, \textit{I$_{KC}$} photometry, new low dispersion optical ($6000 - 9000$\,\AA) spectra from which uniform spectral types are determined, multi-epoch H$\alpha$ equivalent widths, and gravity sensitive $Na\,I$ indices.  For 12 of these 58 stars, strict limits are placed on the presence of stellar and sub-stellar companions, based on a pioneering program described here that utilizes precise infrared radial velocities and optical astrometric measurements in an effort to search for Jupiter-mass, brown dwarf and stellar-mass companions.  Our infrared radial velocity precision using CSHELL at NASA's IRTF is $\sim$90 m s$^{-1}$ over timescales from 13 days to 5 years.  With our spectroscopic results the mean companion masses that we rule out of existence are 1.5 M$_{JUP}$ or greater in 10 day orbital periods and 7 M$_{JUP}$ or greater in 100 day orbital periods.  We use these spectra to determine rotational velocities and absolute radial velocities of these twelve stars. Our mean astrometric precision using RECONS \footnote[6]{www.recons.org}(Research Consortium on Nearby Stars) data from 0.9-m telescope at Cerro Tololo Inter-American Observatory is $\sim$3 milli-arcseconds over baselines ranging from 9 to 13 years.  With our astrometric results the mean companion masses that we rule out of existence are greater than 11.5 M$_{JUP}$ with an orbital period of 4 years and greater than 7.5 M$_{JUP}$ with an orbital period of 8 years.  Although we do not detect companions around our sub-sample of 12 stars, we demonstrate that our two techniques probe a regime that is commonly missed in other companion searches of late type stars.  
\end{abstract}
\section{Introduction}
Given human-kind's search for life in the universe, there is great motivation to find Earth-size and Earth-mass planets in the habitable zones of stars.  Recent studies have determined that Earth size planets are common around M-dwarfs.  \citet{2013arXiv1303.3013M} estimate an occurrence rate of 1.5 planets per M-dwarf with periods less than 90 days and radii larger than 0.5R$_{EARTH}$, using the list of 4000 stars with temperatures below 4000 K assembled in \citet{2013ApJS..204...24B} that were observed with the \textit{Kepler} mission \citep{2010Sci...327..977B, 2010cosp...38.2513K}. This estimate is consistent with but slightly higher than previous studies, which measure occurance rates of approximately one planet per M-dwarf \citep{2011ApJ...742...38Y, 2012ApJ...753...90M, 2013ApJ...764..105S, 2013ApJ...767...95D}.  Given the apparent abundance of Earth-size planets orbiting M-dwarfs, which dominate the stellar population \citep[75$\%$;][]{2006AJ....132.2360H}, \citet{2013ApJ...767...95D} predict the nearest non-transiting planet in the habitable zone orbiting an M-dwarf is within 5 pc of th Sun, with 95$\%$ confidence.  However, how suitable these nearby planets within the classically defined habitable zone \citep[e.g.][]{1993Icar..101..108K} may be for life is still under debate \citep[e.g.][]{2007AsBio...7...30T, 2011ASPC..448..391B, 2014arXiv1407.8174G}.  Nevertheless, given their ubiquity and proximity, M-dwarfs are ideal stars to search for Earth-size and Earth-mass planets in stellar habitable zones.

M-dwarfs have been favorite targets of precision searches for low mass planets, because a planetary companion will induce a greater reflex motion on a low mass star than a Sun-like star, making it easier to detect.  However, not all M-dwarfs are equally suitable targets for the precision measurements needed to find Earth-mass companions.  Some M-dwarfs have close stellar or substellar companions that may inhibit the detection of Earth-mass planets. The dynamically disruptive effects of these companions could also preclude the existence of Earth-mass planets; the lack of short period giant planets in multiple planet systems corroborates this hypothesis \citep{2011ApJ...732L..24L, 2012MNRAS.421.2342S, 2013ApJS..204...24B}.  Some M-stars also exhibit high chromospheric activity and large rotational velocities, which can hinder the achievable RV precision \citep{2003ApJ...583..451M, 2008ApJ...684.1390R, 2009ApJ...704..975J, 2013arXiv1310.5820R}; this is especially problematic for mid to late (M3 and cooler) M-dwarfs.  Given the above considerations, we argue that the best stars to target for Earth-mass planet searches are likely the lowest mass stars (mid to late M-dwarfs) that do not have a disruptive companion, and are both inactive and slowly rotating.

Yet, the statistics characterizing companions around mid to late M-dwarfs are still incomplete. Preliminary surveys show that Jupiter-mass companions are rare around M-dwarfs.  Using radial velocity (RV) measurements and high contrast imaging, \citet{2014ApJ...781...28M} found that 6.5$\pm$3.0$\%$ of M-dwarfs (M0-M5.5) host a giant planet (1-13 M$_{JUP}$) with a semi-major axis smaller than 20 AU, but this sample only included 18 M-dwarfs of M4 or later in their survey of 111 M-dwarfs.  Another large M-dwarf survey of 102 stars \citep{Bonfils2013} only included M-dwarfs with V$<$14 and finds the giant planet ($m$sin$i$ = 100-1000 M$_{EARTH}$) frequency to be 2$\%$ for orbital periods between 10 and 100 days, and the super-Earth ($m$sin$i$ = 1-10 M$_{EARTH}$) frequency with orbital periods between 10 and 100 days to be significantly higher at 52$\%$.  Many of the M-dwarfs not surveyed are faint and chromospherically active, which limits the achievable RV precision and chances to find Earth-mass planets.  As a result, these two surveys, which are representative of other spectroscopic M-dwarf surveys, include very few mid to late M-dwarfs and can report only preliminary statistics for planet occurrence around such stars.

In order to obtain an unbiased assessment of the companion fraction of the nearest mid M-dwarfs, we construct a volume-limited sample out to 10 pc.   Based on new, uniform optical spectra, we present this sample of 58 stars in \S2.  We report spectral types, H$\alpha$ equivalent widths, and $Na\,I$ indices for these stars in \S3.  We list our \textit{V$_J$}, \textit{R$_{KC}$}, \textit{I$_{KC}$}\footnote[7]{Subscripts: J indicates Johnson and KC indicates Kron-Cousins.} (hereafter without subscripts) photometry in \S4.  In \S5, we focus on a sub-sample of 12 stars, for which we obtain  high dispersion infrared spectroscopic data described in \S6 and astrometric data in \S7.  Results for the remaining stars in our volume-limited sample will be presented in a subsequent paper.  We describe our Monte Carlo technique and rule out the existence of massive gas-giant companions, brown dwarfs and stellar companions in \S8, and we conclude with a brief summary in \S9.

\section{Sample Selection}
\subsection{An Equatorial Sample of Nearby Mid M-Dwarfs}
Beginning with a parent volume-limited sample of stars extending out to 10 pc \citep[unpublished list;][]{2006AJ....132.2360H}, we assembled a sample of mid M-dwarfs for detailed study outlined in Table~\ref{table:sample}.  These stars are within the declination range of $\pm$ 30$^\circ$and are thus accessible to the majority of observing facilities in both the northern and southern hemispheres.  
To be inclusive, we define mid M-stars using three independent measures, including optical spectral type, (V-K) color and absolute magnitude.  Starting with the complete 10 pc sample, we include stars meeting one of the following criteria: spectral types between M3.5V and M8.0V classified by the RECONS system (described in \S3), V-K $=$5.0$-$9.0 mag or M$_V$$=$12.0$-$19.0 mag.  There are 69 systems that meet at least one of these criteria\footnote[8]{We include GJ 628 with V-K=4.99 in our sample, as its V-K value may be greater than 5.0, given our photometric uncertainties.}. Of those systems, 13 are known binaries, five are known triples, one is a quintuple system (GJ 644ABCD-GJ 643) and one is a known multi-planet host \citep[GJ 876;][]{2010ApJ...719..890R}.  The 21 close-separation ($<$4$\arcsec$) mid M-dwarf binaries\footnote[9]{This list includes four mid M-dwarf binaries that are wide companions to more massive stars (LP 771-096BC, GJ 569BC, GJ 695 BC, GJ 867BD).} within the distance and declination range of this sample are listed in Table~\ref{table:binaries}.  Also listed are 2MASS coordinates, parallaxes, spectral types, absolute \textit{V} magnitudes, \textit{V}, \textit{R}, \textit{I} apparent magnitudes, near infrared photometry from 2MASS (\textit{J}, \textit{H}, \textit{K$_{s}$} apparent magnitudes) and the configuration of the system.

A sub-sample of mid M-dwarfs is constructed that excludes these close binaries ($<$4$\arcsec$) and the planetary host GJ 876.  This sub-sample includes 7 mid M-dwarfs that are wide companions to higher mass stars (GJ 105B, GJ 166C, GJ 283B, GJ 644C, GJ 752B, GJ 896B, GJ 1230B).  At the start of this program, 60 mid M-dwarfs met the sample requirements.  Subsequently, GJ 867B has been determined to be a single-lined spectroscopic binary with a period of 1.795 days \citep{2014AJ....147...26D}.  Likewise, LHS 1610 has been claimed to be a spectroscopic binary \citep{Bonfils2013}, but its orbital properties are unknown.  Table~\ref{table:sample} lists the astrometric, photometric, spectroscopic and physical properties of the remaining 58 stars, which we refer to as effectively single equatorial mid M-dwarfs. Mass estimates in Table~\ref{table:sample} are based on mass luminosity relations of \citet{1993AJ....106..773H} and \citet{1999ApJ...512..864H}; typical errors using these relations are close to 20\%.  Because the long term goal of this program is to conduct a more comprehensive 3D search for companion to these stars, we refer to this sample as CAESAR, which stands for a Companion Assessment of Equatorial Stars with Astrometry and Radial velocity.  

Of the 58 stars identified above as possible targets for precision planet searches among equatorial mid M-dwarfs, only 33 of the stars have been included in past spectroscopic searches for planetary companions \citep{2012MNRAS.424..591B, 2012A&A...538A.141R, 2012ApJS..203...10T, Bonfils2013, 2014ApJ...781...28M}.  In addition, four of the 58 stars do not have known projected rotational velocity ($v$sin$i$) values, and thus could be rapidly rotating (see Table~\ref{table:sample}).

\section{Optical Spectroscopic Measurements}
\subsection{Observations}
We obtained optical ($6000 - 9000$\,\AA) spectra of all 58 stars in the CAESAR sample between 2003$-$2006 and 2009$-$2011 using the Cerro Tololo Inter-American Observatory (CTIO) 1.5-m Richey-Chretien Spectrograph (RCSpec) with the Loral 1200 $\times$ 800 CCD camera, as part of the broader RECONS spectral-typing program \citep[e.g.][]{Henry2004, 2008AJ....136..840J}. The spectra were obtained with the \#32 grating in first order with a 15.1$^\circ$ tilt, which yields a spectral resolution of 8.6\,\AA; the spectra were acquired through the OG570 order blocking filter.  For consistency checks and to mitigate the effects of cosmic rays, two spectra of each target were taken consecutively.  In addition, the majority of stars have spectroscopic observations on multiple epochs.  To assist with spectral classification, at least one flux standard was observed each night, and an ensemble collection of spectral standards from \citet{2002AJ....123.2002H} were observed.  Some of the stars in the CAESAR sample are used as standards, namely GJ 283B, GJ 644C, GJ 752B, GJ 1065, GJ 1111, GJ 1154, GJ 1207, LHS 292, LHS 2090 and LHS 3799.

The data were reduced with standard IRAF techniques; bias subtraction and dome and/or sky flat fielding were performed on the data using calibration frames taken at the beginning of each night.  Fringing was effectively removed from the data using a combination of dome and sky flats.  One flux standard per night was used for absolute flux calibration.  Spectra were wavelength calibrated using consecutively recorded HeAr arc spectra.  Further details regarding reduction and extraction are given in \citet{Henry2004}.

\subsection{Results}

\subsubsection{Spectral Types}
\begin{figure}
\begin{center}
\includegraphics[scale=0.45,angle=0]{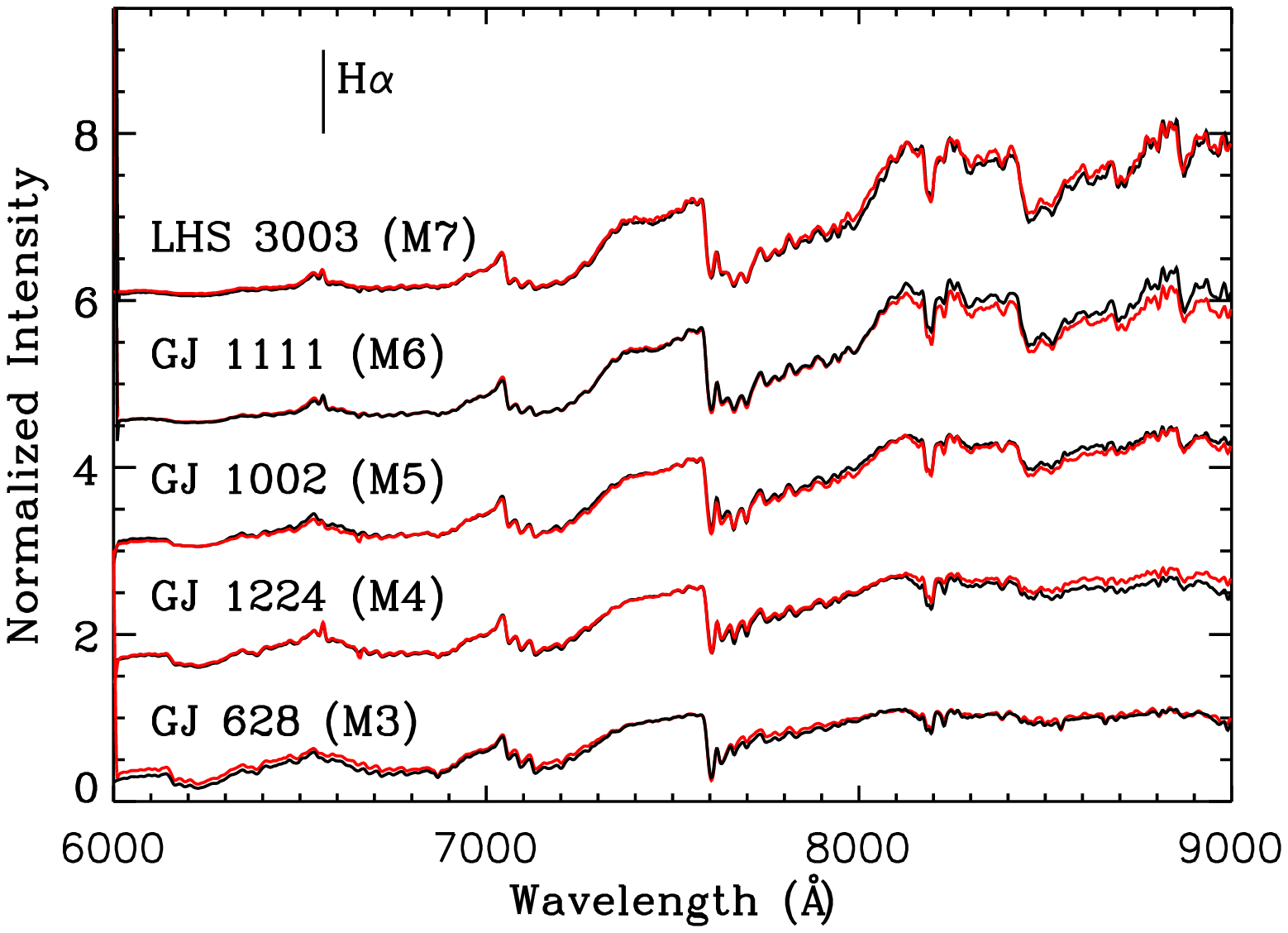}
\caption[fig1] {CTIO spectra of five CAESAR stars (black) normalized to 7500\,\AA.  The comparison spectra (red) are from the library of standards from \citet{2002AJ....123.2002H}.  Small discrepancies at the longest and shortest wavelengths are due to minor errors in the flux calibrations.  The vertical line represents the position of H$\alpha$ (6563 \,\AA).}
\label{fig:opticalspectrum}
\end{center}
\end{figure}

To assign a spectral type, the wavelength calibrated spectra are resampled via interpolation onto a fixed 1\,\AA~grid.  The H$\alpha$ and telluric features, based on the sky transmission map of \citet{Hinkle2003} are then given a spectral weight of zero to essentially remove these features from our analysis.  The spectra are normalized to a value of 1 at 7500\,\AA.  Then, the spectra are then compared to the library of observed standards (see Figure~\ref{fig:opticalspectrum}), and the adopted spectral type is that of the standard that yields the lowest standard deviation of the target spectrum divided by the standard spectrum over the spectral range ($6000 - 9000$\,\AA).  The determined spectral types range from M2.5 to M8.0, and are listed in Table~\ref{table:sample}.  The uncertainty in all cases is $\pm$0.5 spectral sub-classes, based on consistency between multiple epochs.  Additional details of the spectral type determinations are provided in \citet{2014AJ....147...85R}.

\subsubsection{H$\alpha$ Emission}
To assess the amount of chromospheric activity that these stars exhibit, we measure equivalent widths of H$\alpha$ 6563\,\AA.  To do this, we first subtract the linear continuum from the region between 6550\,\AA~and 6575\,\AA~and then assume the strongest maximum (or minimum) 
within this region is the H$\alpha$ line.  After determining its location, we integrate both the continuum and spectrum over 22\,\AA, with the central wavelength corresponding to the H$\alpha$ line to determine the equivalent width.  Upon visual inspection of the spectra, we could not reliably determine emission or absorption lines smaller than 0.5\,\AA.  Therefore, we conservatively use this number as the error value on the H$\alpha$ measurements, which are given in Table~\ref{table:halpha}; we adopt the standard convention of denoting emission with a negative sign.  If any star exhibits H$\alpha$ equivalent widths less than the the typical noise level of 0.5\,\AA~in at least one epoch, we categorized it as an emission-line star.  This is denoted with an `e' next to the star's spectral type in Table~\ref{table:sample}.

We classify 35/58 of the stars from the CAESAR sample as emission-line stars.  We also note that the fraction of emission-line stars increases with later spectral type.  37\% of the 35 stars with spectral types M4 or earlier are emission-line stars, whereas 100\% of the 9 stars with spectral type M6 or later are emission-line stars.  Our fraction of early M-dwarfs with emission is comparable to that of \citet{2012AJ....143...93R} who find 41\% of the 115 stars M3 to M4.5 to be active emission-line stars.  Likewise, \citet{2000AJ....120.1085G} reports an increase in emission-line stars with lower mass and finds 100\% of the M7 stars are emission-line stars.  He also notes that this trend breaks down past M7 and that the fraction of later type stars that show chromospheric activity is significantly less.

\subsubsection{Surface Gravity Indices}

To assess the surface gravity of these stars, as a possible tracer of their evolutionary states, the gravity sensitive $Na\,I$ doublet \citep[defined in][]{2004MNRAS.355..363L} is measured; this feature is well known to be weak in giants and strong in dwarfs \citep[e.g.][]{2007ApJ...657..511A, 2012AJ....143..114S}.  For the $Na\,I$ doublet index, we use the wavelength region between 8148\AA~and 8200\AA.  We conservatively estimate our index error to be 0.05; this value is the average difference value of our measured $Na\,I$ doublet indices for all of our stars with more than one epoch.  These values are given in Table~\ref{table:halpha}.   While these indices derived from the low-resolution spectra are a good indicator of the evolutionary state of the star, high-resolution spectra are needed to ascertain a more quantitative measure of the surface gravity.  Also, we do caution that these lines can be affected by metallicity and stellar activity; metal poor stars and chromospherically active stars will systematically have lower equivalent widths (EW) \citep[e.g.][]{1999AJ....117.1341H}.  

All of the stars in the ensemble sample exhibit average $Na\,I$ doublet indices in a typical range for main sequence stars \citep[$>$1.1;][]{2004MNRAS.355..363L,2007ApJ...657..511A}; we therefore classify all 58 CAESAR stars as on the main sequence.  This is denoted by adding a `V' to the spectral type listed in Table~\ref{table:sample}.

\section{Optical Photometry}
Prior to our measurements, four of the 58 stars from CAESAR sample did not have complete sets of \textit{V}, \textit{R} and \textit{I} photometry.  The remaining stars had photometric measurements presented in eight different publications.  Therefore, to establish a uniform, homogenous, set of photometric measurements for the ensemble sample of 58 stars, we obtained optical photometric observations using the 0.9-m telescope at CTIO.  For all of our photometry frames, we use the center 1024x1024 pixels on the Tektronix 2048x2048 CCD.  The CCD chip has a plate scale of 0$\farcs$401 pixel$^{-1}$, which gives a field of view (FOV) of 6.8 by 6.8 arcminutes \citep{2003AJ....125..332J}.  All frames were collected at an airmass less than 2 and with the target star having a signal-to-noise ratio (SNR) $>$ 100.  We use 10 or more standard stars from \citet{Landolt1992}, \citet{2007AJ....133.2502L}, and \citet{Graham1982} to create extinction curves each night, and transformation equations to obtain \textit{V}, \textit{R} and \textit{I} photometry for all our target stars and reference stars used for astrometric measurements described in \S7.2.  During the course of observations, we used two different \textit{V} filters for photometry.  \citet{Jao2011} demonstrate that both \textit{V} filters give effectively identical \textit{V} band photometry for standard stars; therefore it is suitable for us to combine photometry from the two filters.  For additional details on the photometric reduction and its associated errors, see \citet{Jao2005} and \citet{Winters2011}.  

The \textit{V}, \textit{R}, \textit{I} photometry and the number of nights on which observations were made are reported in Table~\ref{table:sample}.  Errors at \textit{V}, \textit{R} and \textit{I} are 0.02$-$0.03 mag.  A comparison of five stars (GJ 300, GJ 406, GJ 555, GJ 628 and GJ 729) with \citet{1990A&AS...83..357B} indicates that the two datasets are consistent to 0.04 mag.  Using our photometric measurements and the parallax data given in Table~\ref{table:sample}, the absolute magnitude errors range from 0.03 to 0.08 mag.


\section{A Companion Search of a 12 Star Subset}

We present results of a 3D companion search on a sub-sample of 12 CAESAR stars, including G 99-49, GJ 300, GJ 406, GJ 555, GJ 628, GJ 729, GJ 1002, GJ 1065, GJ 1224, GJ 1286, LHS 1723, and LHS 3799, which are the most data-rich in our sample.  The remaining stars in the ensemble sample will be presented in a subsequent paper.  These twelve stars have astrometry baselines, ranging from 9 to over 13 years, and have at least 5 infrared radial velocity measurements spanning from almost 2 weeks to 5 years.  The spectral classes of these stars range from M3.0 to M5.0.

\section{High Dispersion Infrared Spectroscopy}
\subsection{Observations}
All infrared spectroscopic observations were obtained using CSHELL \citep{1990SPIE.1235..131T, 1993SPIE.1946..313G} located on the 3-m telescope at NASA's Infrared Telescope Facility (IRTF).  CSHELL is a long-slit echelle spectrograph that uses a circular variable filter to isolate a single order onto a 256x256 InSb detector.  Spectra are centered at 2.298 microns (vacuum) and cover approximately a 50\,\AA~window.  Telluric methane absorption features from the Earth's atmosphere that are superimposed on the photospheric $^{12}$CO R branch lines at 2.3 microns are used as an absolute wavelength reference \citep[e.g.][]{Blake2010, Bailey2012}.  The design resolving power is 100,000 per pixel.  We use CSHELL in the high resolution mode (0.5$\arcsec$ slit=2.5 pixels), which yields a predicted spectral resolving power of R $\sim$ 40,000.  The measured resolving power determined from the model fits to telluric absorption features (described later) is $\sim$57,000.  This number is significantly higher than the predicted resolving power for CSHELL using the 0.5$\arcsec$ slit.  We note this discrepancy because CSHELL has an adjustable slit; it may be that the slit is smaller than the designed 0.5$\arcsec$.  \citet{Crockett2011} and \citet{2008ApJ...687L.103P} also report a higher than predicted spectral resolving power (R$\sim$46,000) for CSHELL in high resolution mode.  Also, we note a similar effect of determining a higher spectral resolving power for Keck is determined in \citet{Bailey2012}, and may be a feature of the analysis code.

Each night, two spectra of a given star were obtained in succession at two different positions along the slit, separated by 10$\arcsec$.  Hereafter, we refer to these two positions as nod A and nod B. Observations were obtained between 2008 November and 2014 January. 

Our exposure times ranged from 180 to 1200 seconds per nod position, and were set to yield SNRs of 125 per pixel (or optimally, a combined SNR of 175+) for most of our targets.  For the faintest three stars (\textit{K$_{s}$} $>$ 8.0) in our sample of 12 stars (GJ 1065, GJ 1224, GJ 1286), a SNR of 125 per pixel was not achieved as the maximum exposure time is set to 1200 seconds to limit cosmic ray events and the dark current.  

At the beginning of each night, we obtained a minimum of 30 flat and 30 dark images each with an integration time of ten seconds.  Also, on November 14, 2009, we collected an additional 30 flat and 30 dark images with an integration time of 20 seconds, which were used in creating a bad pixel mask described in \S6.2.  Most nights we also observed bright stars of spectral type early A, as these stars exhibit no intrinsic absorption or emission lines in this wavelength region and therefore can be used to identify telluric features.  These telluric standards are used to characterize the instrumental profile and wavelength solution.  When first collecting the data, we did not realize how sensitive our final RV measurements were to the initial solutions for the instrumental profile and wavelength solution obtained from the telluric standards.  After reducing part of the data, we determined that A star observations obtained nightly yield the best precision.  In a few cases, we only obtained a few A star observations per run leaving us with four nights that contain no A star observations.  The four nights the telluric standards were not observed are marked in Table~\ref{table:rv} by an asterisk next to the date.

We aimed to observe each target at least four nights within a single observing run ($\sim$ 1$-$3 weeks) in order to search for companions with orbital periods of less than a week.  Because of inclement weather we were not able to achieve this cadence for all targets.  On subsequent runs, we re-observed our targets at least once to search for companions with longer orbital periods, except for GJ 1065.  For the sample of 12 stars studied here, we obtained between 5$-$12 RV epochs for each star, spanning a temporal baseline between 13 days and 1884 days (see Table~\ref{table:rv}).

\subsection{Image Reduction and Spectral Extraction}
We subtracted each nodded pair of images from one another to remove sky emission, dark current and detector bias assuming that changes in the detector or spectrograph properties were negligible over the timescale when the nodded pair of images were obtained.  After completing the nod-subtraction, we corrected each image for flat fielding.  Corrections for flat fielding were performed by generating a nightly master flat field image from all flat field images obtained on a particular night.  The master flat field images were created by first subtracting the median dark image of the same exposure time from each of the flat field images.  Then, each flat field image was normalized to the central 15$\%$ of the array, which was the brightest section of the array and the least affected by deviant pixels.  After normalizing the image, all images were median combined.  

We then applied a bad pixel mask to our spectra to remove dead and hot pixels from the data.  To identify dead pixels, we located any pixel five times below the standard deviation of the median pixel value of the master flat field array.  To locate hot pixels, we subtracted two times the count value of the 10 second exposure master flat field image from the 20 second exposure master flat field image, and then normalized this number to the 20 second exposure master flat field. Because these pixels should increase linearly with time and therefore have the same values, we identified any pixel with values greater than three times the standard deviation of this median pixel value of the difference image to be a hot pixel.  All deviant pixels identified from the bad pixel mask are assigned interpolated values using the neighboring pixels.

We optimally extracted each spectrum following the procedures in \citet{1986PASP...98..609H} as implemented for nod-subtracted spectra in \citet{Bailey2012}.  The code used to analyze the data in this work is a modified version of that described in \citet{Bailey2012}, tuned to work for CSHELL data.  The advantage of optimally extracting spectra over the standard extraction is that the optimal extraction minimizes the noisy contribution of the profile wings and eliminates and/or mitigates noise features within the spectral profile caused by cosmic rays and deviant pixels not excluded with the bad pixel mask.  

To obtain an optimally extracted spectrum, we summed the pixels from the nod subtracted images over the cross-dispersion to give the standard spectrum.  We then fitted a second order polynomial to map the curvature of the order on the detector.  Next, we fitted the spectral profile of the standard spectrum with a Gaussian to model our spatial profiles at each pixel step (column) along the order parallel to the dispersion direction of the nod-subtracted spectral image.  From this, the variance of the profile was determined.  Then, we summed the pixels weighted by the variance image of the spectrum's spatial profile to create our two dimensional optimally extracted spectrum.  For low SNR data, we implemented a clipping routine that interpolates over a pixel that is more than 5$\sigma$ above or below the running average of the five pixels next to the pixel in question to remove any remaining deviant features.

To obtain an estimate of the SNR for each spectrum, we use a simplified version of the CCD equation from \citet{1981SPIE..290...28M} modified to account for the noise introduced by subtracting pairs of images.  When performing a nod pair subtraction we remove the sky background, dark current and bias simultaneously. Therefore, we cannot distinguish between these values and refer to them collectively as the uncertainty in background.  Executing a pair subtraction means that we have to deal with this uncertainty in the background twice and the read noise associated with each of those background estimates.  In most cases, the  background of the first image (Image A) should equal (within the uncertainty) the background of the second image (Image B).  Therefore, we simply double the noise contribution from the background and the read noise.  The equation to calculate the SNR per pixel is as follows:

\begin{equation}
SNR = \frac{S_{e}}{\sqrt{S_{e} + 2 \cdot n (B_{e}+ R_{e} ^2)}}\nonumber
\end{equation}
\noindent
where S$_{e}$ is the total number of counts per integrated column of a spectrum extracted from a nod-subtracted image in electrons, n is the number of pixels in the spatial direction that are integrated over during the extraction, B$_{e}$ is the integrated background counts of the corresponding sky image before image subtraction is performed, in electrons per pixel, and R$_{e}$$^{2}$ is the read noise set to be 30 electrons/pixel from \citet{1993SPIE.1946..313G}.

Following the above description, we determine the SNR for each integrated pixel of the spectrum and set the final SNR for the spectrum to be the mean of these values.  The SNRs are then added in quadrature for the nod A and nod B measurements to give a combined SNR value.

Because of occasional poor weather conditions leading to low SNR, not all observations are suitable for precision RV analysis.  We require the SNR of the individual spectrum in the nod pair to be greater than 50 and the reduced $\chi$$^{2}$ estimate of our modeling prescription (\S6.3) to be below 3.5.

\subsection{Method to Determine Spectral Properties}

We fit each observation to high resolution spectral models that are convolved to the resolution of CSHELL.  Each model spectrum is formed by combining a synthetic stellar spectrum and an empirical telluric spectrum.  The synthetic stellar spectra are created from NextGen models \citep{1999ApJ...512..377H}.  The telluric model spectra are extracted from observations of the Sun from an ultra high resolution KPNO/FTS telluric spectrum \citep{1991aass.book.....L}.  We adopt the stellar template closest in temperature to our star, using the assigned spectral types and the temperature scale of \citet{Kraus2007}.  We fix the surface gravity log(g) to 4.8 dex (cgs) for all of our stars, which is consistent with measurements assembled in \citet{2008ApJ...689.1127M} and \citet{2004ApJ...604..741H} for field M-dwarfs.

The model spectrum consists of 19 free parameters to fit.  The linear limb darkening coefficient is set to 0.6 for all stars, which is appropriate for cool stars at infrared wavelengths \citep{Claret2000}.  Three of the parameters make up a quadratic polynomial that characterizes the wavelength solution.  Nine of the parameters are Gaussians used to model the line spread function (LSF) of the spectrum; we assume that the LSF along the order is constant.  The remaining six parameters are the depth of the telluric features, the depth of the stellar features, the projected rotational velocity ($v$sin$i$), the RV, a normalization constant, and a linear normalization term.

We fit the empirical telluric spectrum to our rapidly rotating A star for each night we procured observations of A stars.  From this measurement, we estimate the wavelength solution and the instrumental profile.  The instrumental profile is solved for by interpolating the input spectrum onto a log-linear wavelength grid and convolving it with a Gaussian kernel set to the spectral resolution of the instrument determined by fitting the telluric spectrum. We tested both single and multiple Gaussian functions to obtain the instrumental profile of CSHELL.  We favor multiple Gaussians to fit the instrumental profile, as we had better agreement between the RV estimates from the AB nod pairs and smaller RV dispersions overall.  The Gaussian kernel is composed of one central Gaussian and four satellite Gaussians on each side following closely the technique described in \citet{Valenti1995}.  We set the positions of the centers of the Gaussians and the widths of the satellite Gaussians so that the curves barely overlap.  The amplitude of the central Gaussian is constrained by the normalization factor, while the width of the central Gaussian is allowed to vary.  The amplitudes of the satellite Gaussians are allowed to vary.  Optimization of these values is accomplished by minimizing the variance weighted reduced chi-squared as described in \citet{Bailey2012} to best reproduce the observed spectrum.  In the cases when no A stars were observed on a night, we use the mean values determined on nights close to the night when A stars were observed.

\begin{figure}
\begin{center}
\includegraphics[scale=.35,angle=90]{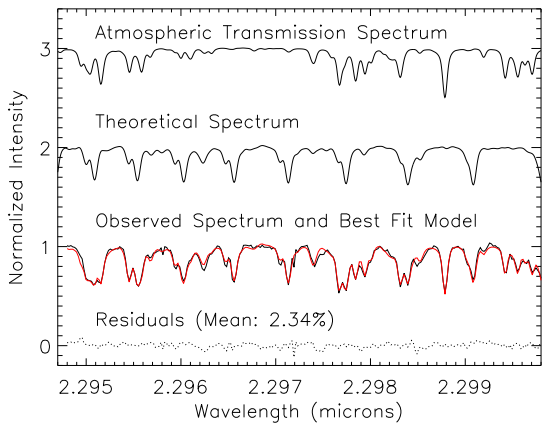}
\caption[fig2] {Spectral Modeling of GJ 300.  Spectra are modeled by combining a telluric spectrum (top spectrum) with a synthetic stellar spectrum (2$^{nd}$ spectrum); the telluric spectrum provides an absolute wavelength reference.  The CSHELL spectra of GJ 300 are shown (black) in comparison with the the best fit (red; 3$^{rd}$ spectrum).  The residuals of the fit are shown (bottom spectrum).}
\label{fig:spectrum}
\end{center}
\end{figure}

After fitting the telluric spectrum, the nine parameters used to characterize the LSF are kept constant for all remaining fits.  We use an iterative process where we fit the target spectrum to the combined synthetic stellar model and empirical telluric model.  On the first iteration, we fit the wavelength solution, the depth of the telluric spectrum, the RV, the normalization constant, and the linear normalization term.  With an improved guess on our second iteration, we allow the $v$sin$i$, the depth of the telluric model, the depth of spectral model and the two normalization constants to fluctuate.  The $v$sin$i$ is determined following the description provided in \citet{2005PASP..117..711G}. We adopt the average $v$sin$i$ value from this iteration for all epochs as the $v$sin$i$ value for the star.  Finally, we repeat the first iterative process allowing the wavelength solution, the depth of the telluric spectrum, the RV, the normalization constant, and the linear normalization term to vary in order to determine the absolute RV of the star.  Computationally, the optimization of the model spectrum is completed using AMOEBA, which is a routine used for minimization of multiple variables using the downhill simplex method of \citet{Nelder1965}.  We note that AMOEBA is very sensitive to initial guesses and is given user specified ranges to restrict the answers to physically reasonable solutions.  An example of an optimally extracted spectrum fit to our telluric and stellar models is shown in Figure~\ref{fig:spectrum}.

Rather than use the full 256 pixels along the order, the modeling analysis is restricted to pixels between 10 and 245, which corresponds to a small continuum area on the spectra.  These boundaries are set to prevent strong absorption features from moving in and out of the analysis region on different epochs, because of different barycentric corrections.  Partial features that are cut off by the edge of the chip can cause our RV value to change on the order of 100 m s$^{-1}$.  Using the restricted pixel range, our average precision improved by 27$\%$ for the 12 stars analyzed here.

\label{fig:GJ300_spectrum.ps}


\subsection{Spectroscopic Results}

The RV results of the spectroscopic modeling are given in Table~\ref{table:rv} with objects listed alphabetically.  Multiple RV measurements on a single night are averaged to provide a single epoch value.  All RV measurements are corrected to the Solar System Barycenter using a correction prescription accurate to $\sim$ 1 m s$^{-1}$ (G. Basri; priv. communication).

Under the assumption that the stars do not have companions, the observed dispersion is thought to be caused by a combination of theoretical photon noise error, intrinsic stellar error and instrumental error ($\sigma$$_{obs}$$^{2}$ $=$ $\sigma$$_{photon}$$^{2}$ $+$ $\sigma$$_{stellar}$$^{2}$  $+$ $\sigma$$_{instr}$$^{2}$).  The theoretical error for each spectrum is calculated based on the prescription by \citet{Butler1996}.  For our 12 stars, the average theoretical photon noise error is 57 m s$^{-1}$, with a standard deviation of 14 m s$^{-1}$.  The stellar error is assumed to be zero, as we are observing a field population of slowly rotating M-dwarfs.  This is supported by results of \citet{Bonfils2013} showing that the observed RV dispersions based on optical spectral for nine of the stars in this sub-sample are below 10 m s$^{-1}$.  We solve the equation above to determine the instrumental error ($\sigma$$_{instr}$$^{2}$ $=$ $\sigma$$_{obs}$$^{2}$ $-$ $\sigma$$_{photon}$$^{2}$) for each star, in which the observed dispersion is the standard deviation of the nightly radial velocity measurements and the theoretical photon error for each star is the average of the nightly theoretical photon errors.  The average instrumental error for this subset is 73 m s$^{-1}$, with a standard deviation of 42 m s$^{-1}$.  We adopt this number as the instrumental error for all targets in our sub-sample. In Table~\ref{table:rv}, we report the final error assigned to each measurement, which is calculated as the instrumental error added in quadrature with the theoretical photon noise error.

In Table~\ref{table:rvsum}, we summarize the infrared spectroscopic results, including the absolute RV, the number of epochs, the time span of observations, the standard deviation of the RV measurements and the $v$sin$i$ value and its uncertainty.  The absolute RV is the mean of the RV measurements from different nights.  We note that systematic uncertainties in the adopted synthetic template (log(g), T$_{eff}$), and the wavelength region used in the fit can cause RV shifts of $\sim$ 100 m s$^{-1}$.  Therefore, we set the uncertainty of the absolute RV measurements to be 100 m s$^{-1}$ for all 12 stars.  All of our stars have previous absolute radial velocity measurements and those measurements are less than 3 sigma from our measurements \citep{2002AJ....123.3356G, 2002ApJS..141..503N}. 

The $v$sin$i$ value is the average of the nightly best fit $v$sin$i$ measurements.  The error on the $v$sin$i$ value is calculated as the standard deviation of the best fit nightly $v$sin$i$ measurements.  We do caution that the spectral resolving power of CSHELL is not high enough to fully resolve the lines of the slowest rotators.   Line broadening becomes measurable for $v$sin$i$ values in excess of 3 km s$^{-1}$, therefore we set this value as our $v$sin$i$ detection limit.  This detection threshold is in line with those reported by \citet{2012AJ....143...93R} of 3 km s$^{-1}$ and \citet{2010AJ....139..504B} of 2.5 km s$^{-1}$ for similar resolution spectra (R=45,000-48,000).  We detect rotational broadening above our detection threshold for two stars, G 99-49 and GJ 729, out of the twelve.  The previous $v$sin$i$ value for G 99-49 of 7.4$\pm$0.8 km s$^{-1}$ by \citet{1998A&A...331..581D} is within 2 sigma of our measurement of 5.8$\pm$0.3 km s$^{-1}$.  Likewise, the previous $v$sin$i$ value for GJ 729 of 4.0$\pm$0.3 km s$^{-1}$ by \citet{2010AJ....139..504B} is within 0.7 sigma of our measurement of 3.8$\pm$0.6 km s$^{-1}$.

\begin{figure}%
\begin{center}%

\begin{subfigure}%
               
		\includegraphics[scale=.30,angle=0]{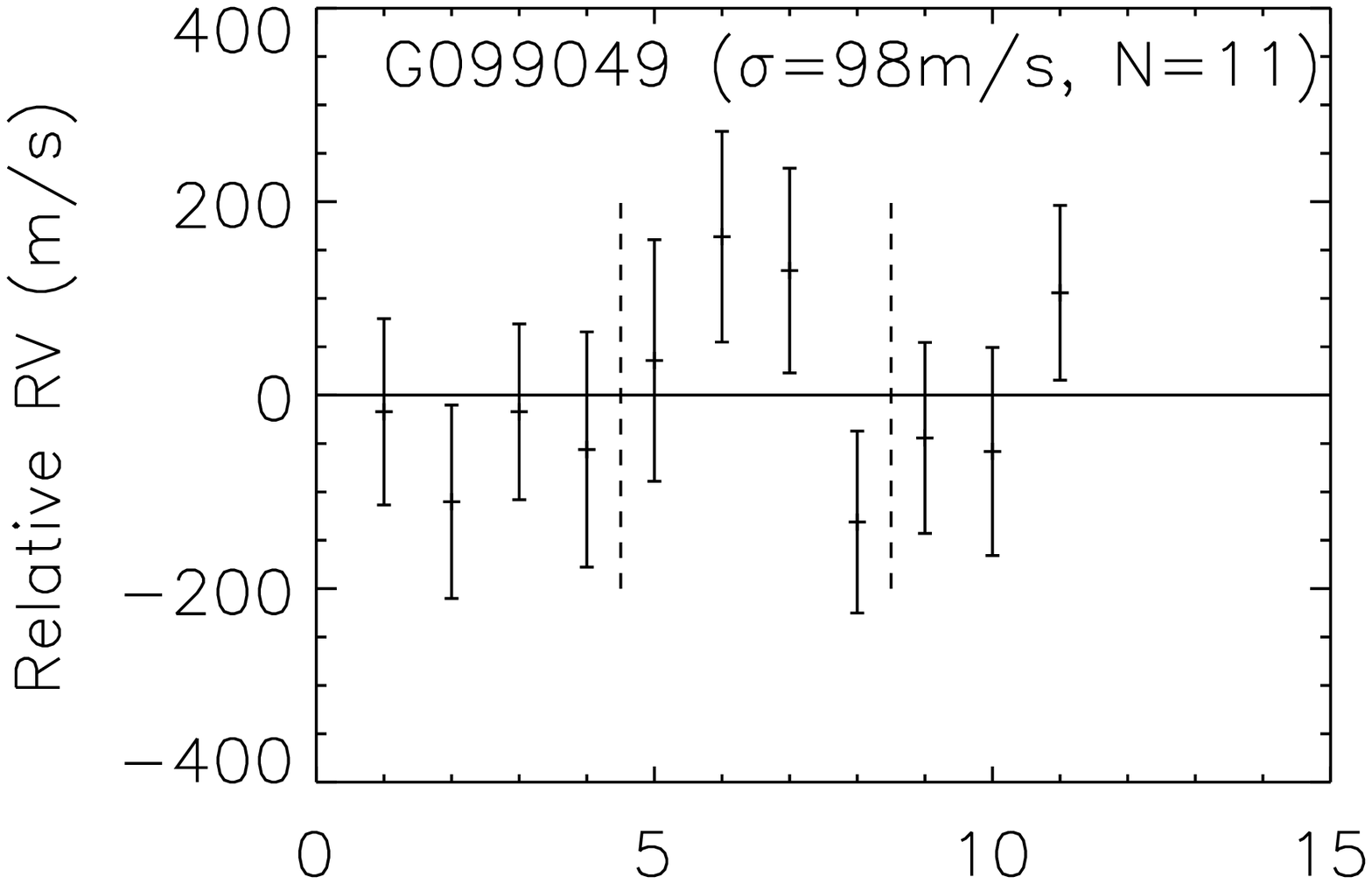} 
		\includegraphics[scale=.30,angle=0]{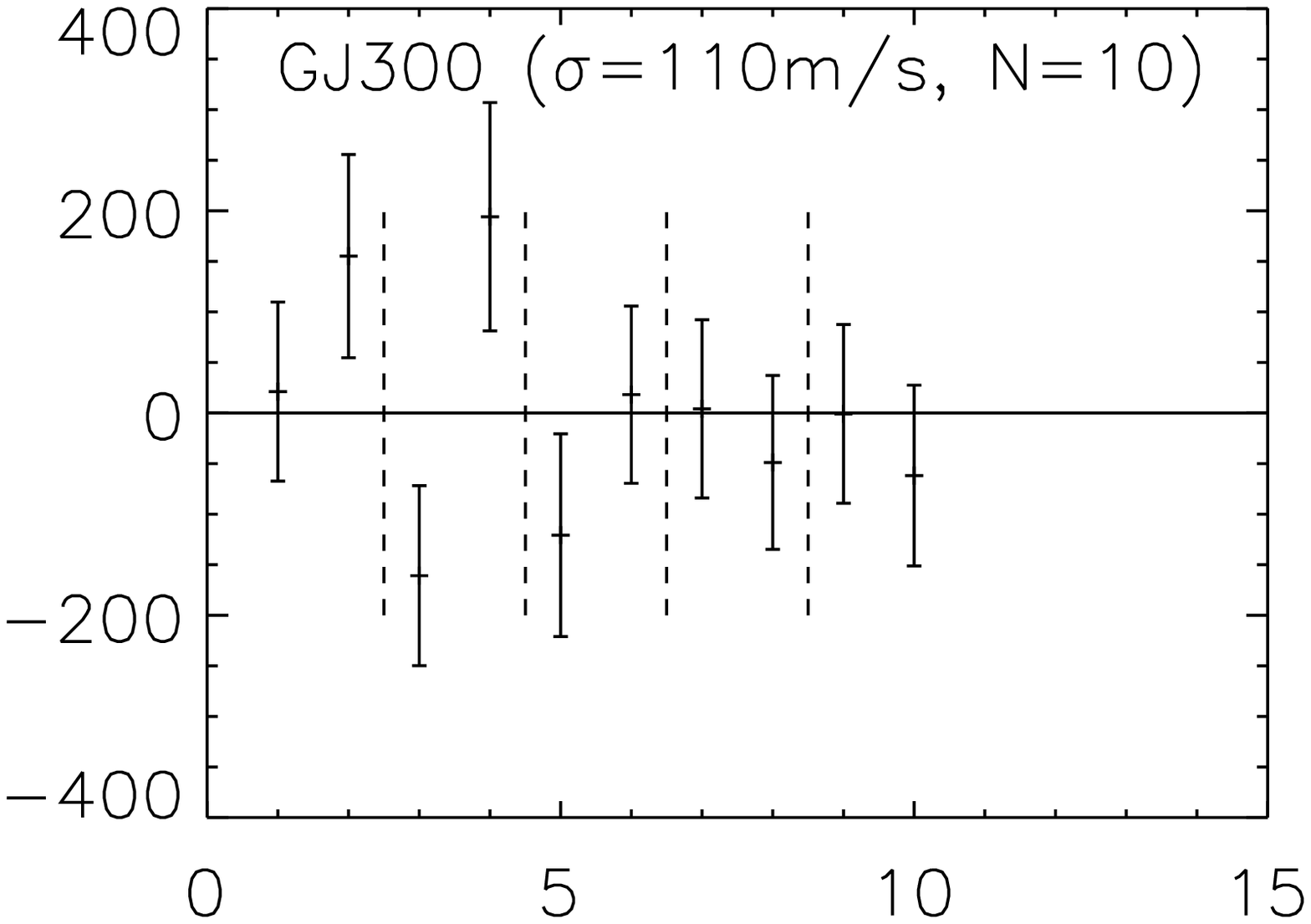}
                \includegraphics[scale=.30,angle=0]{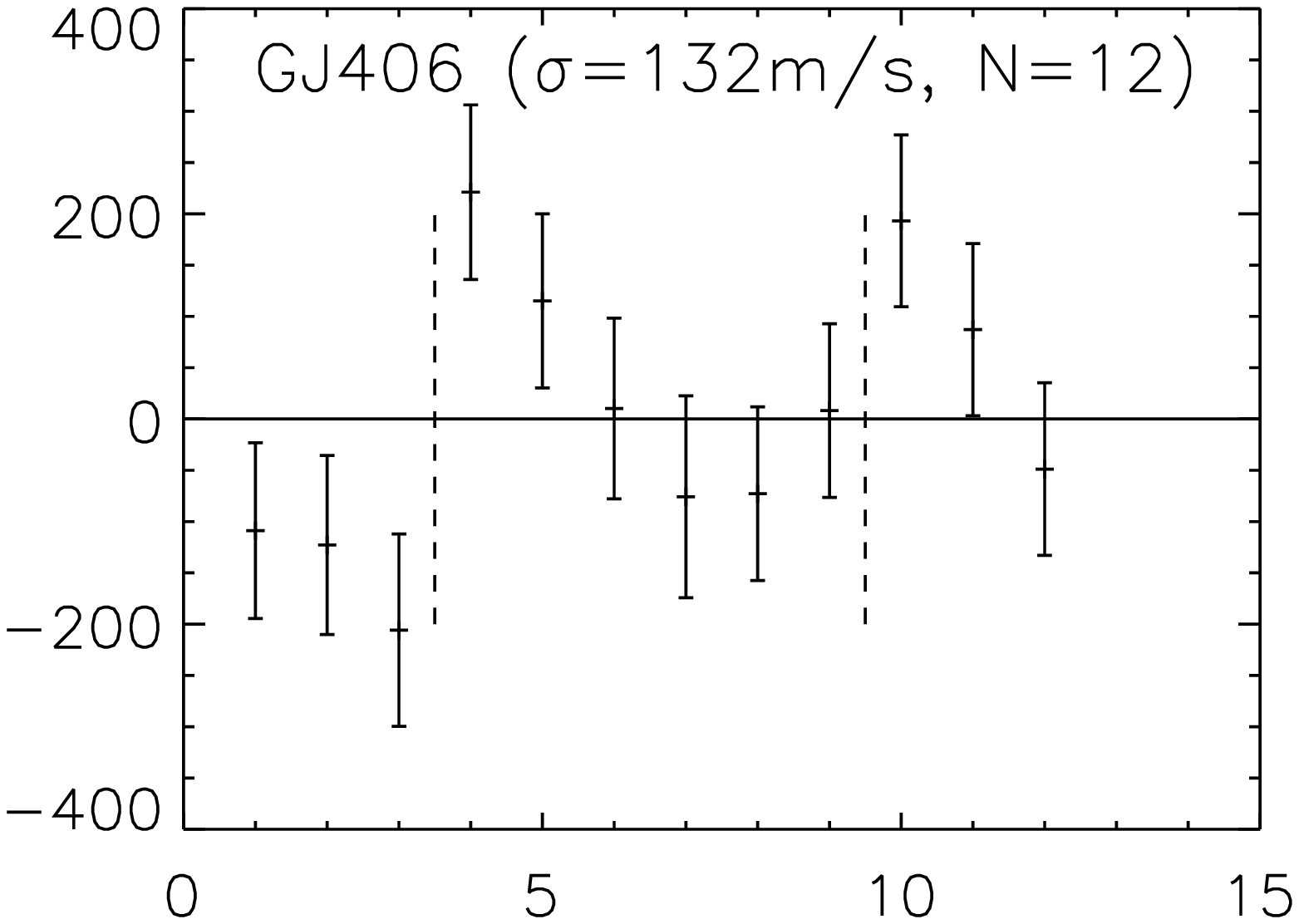}
                \includegraphics[scale=.30,angle=0]{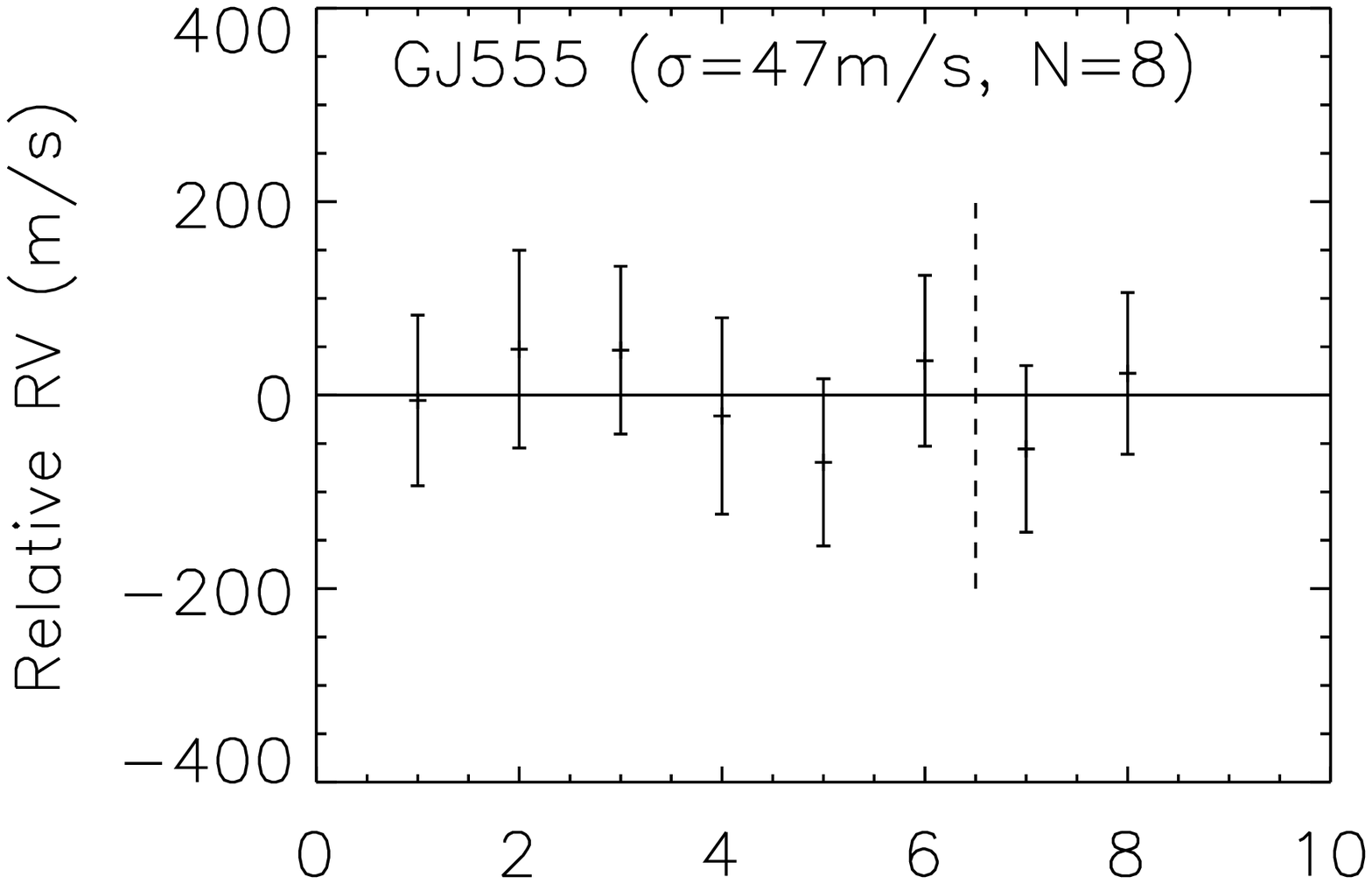}       
                \includegraphics[scale=.30,angle=0]{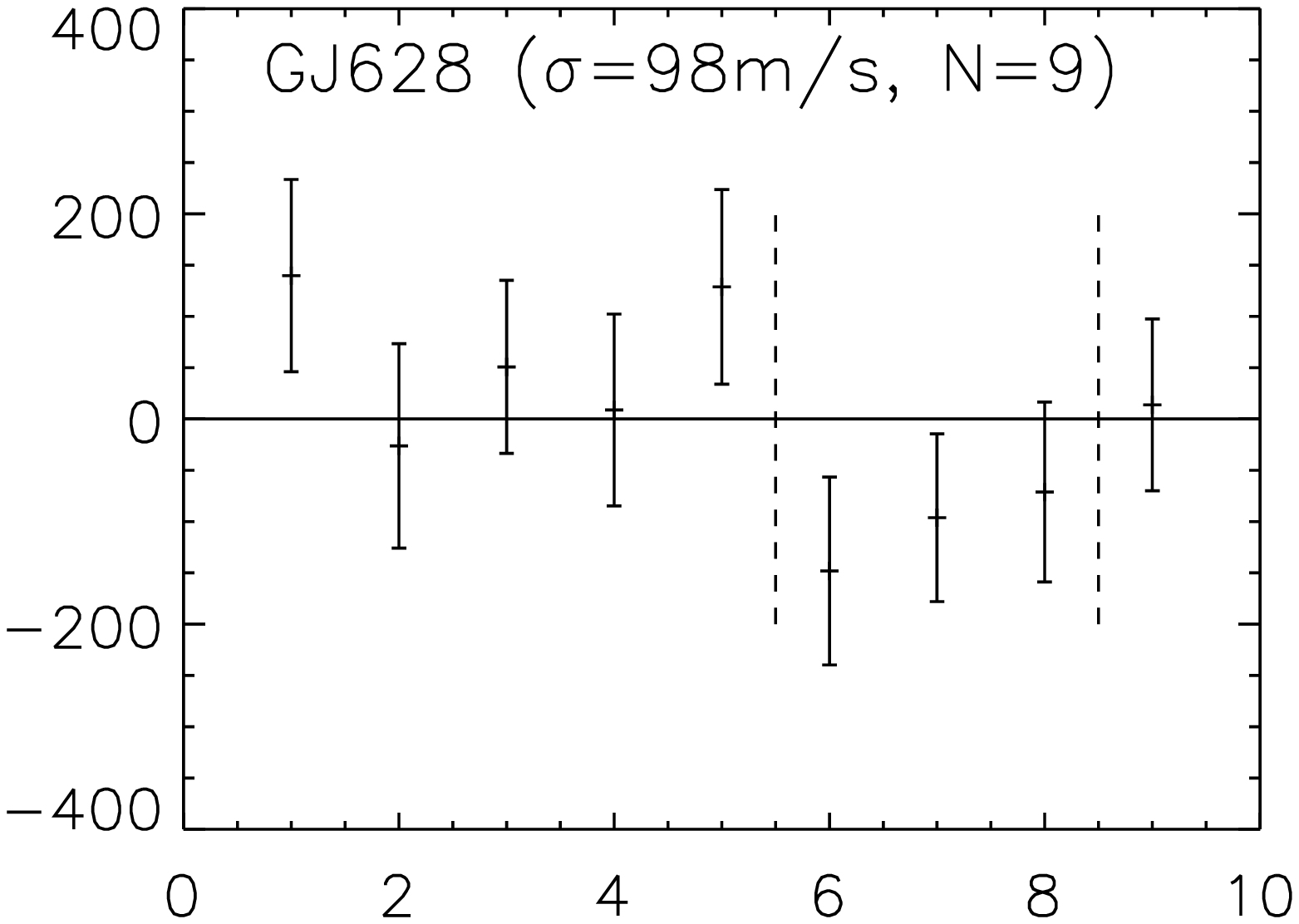}
                \includegraphics[scale=.30,angle=0]{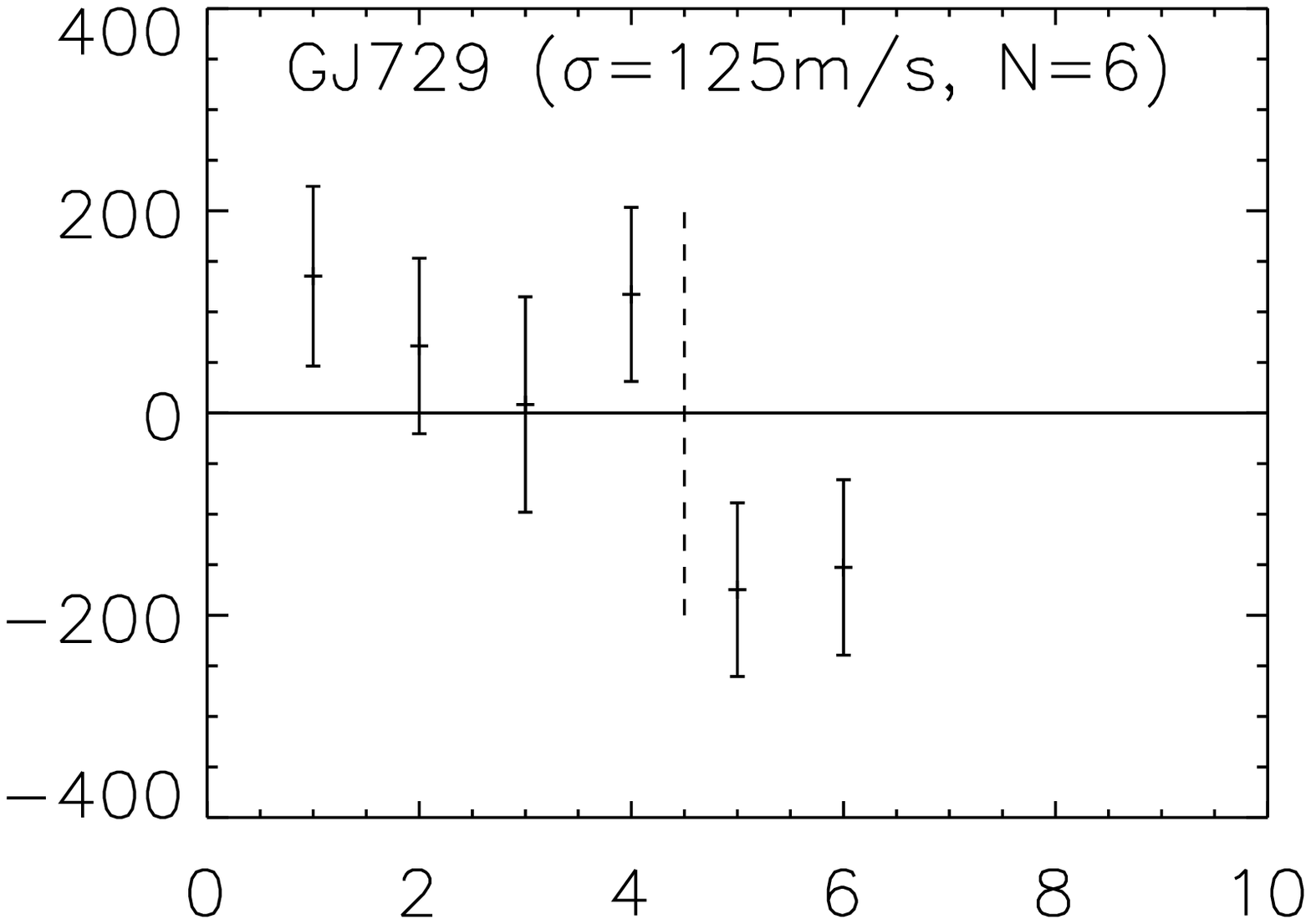}
                \includegraphics[scale=.30,angle=0]{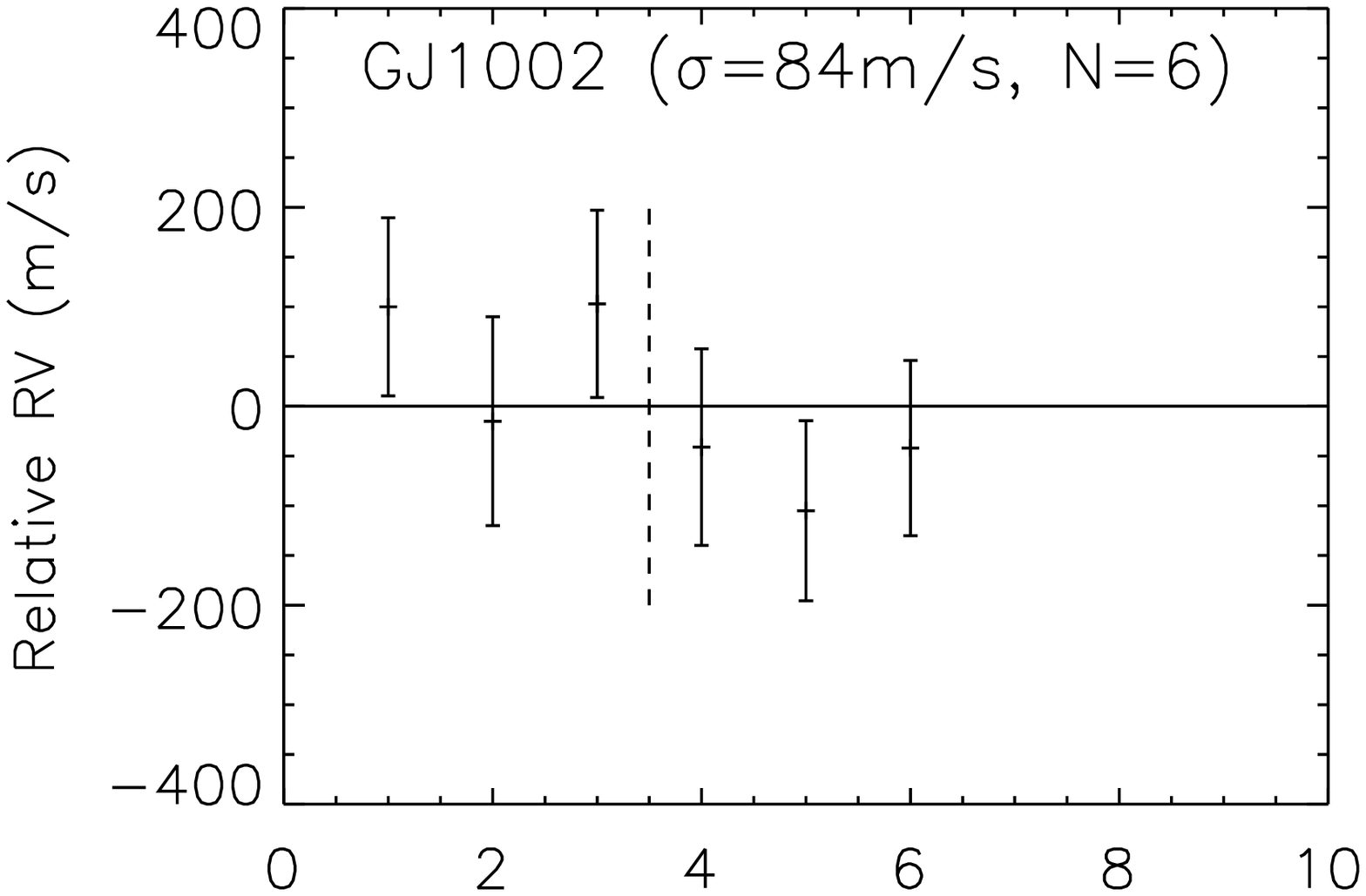} 
                \includegraphics[scale=.30,angle=0]{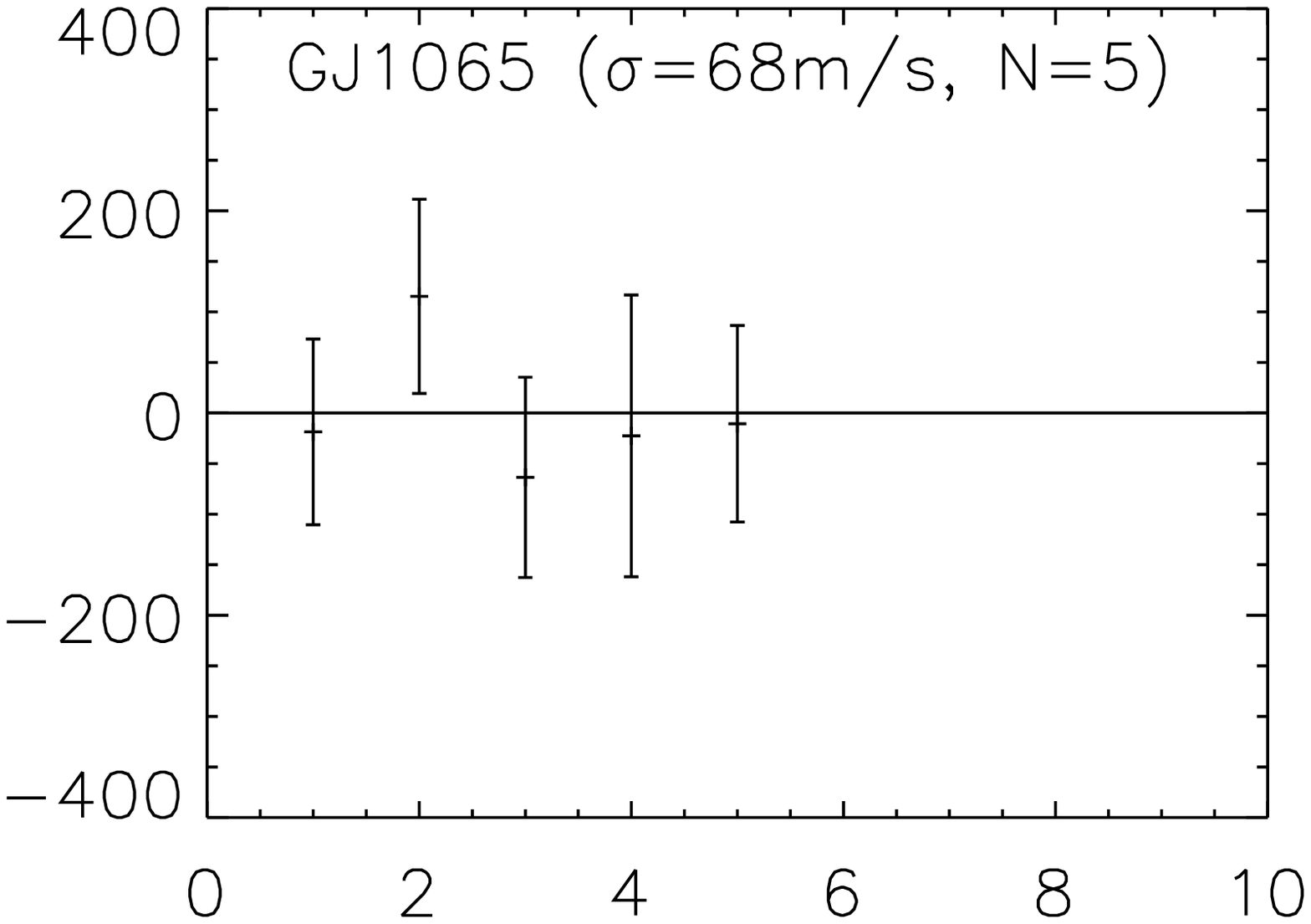}   
                \includegraphics[scale=.30,angle=0]{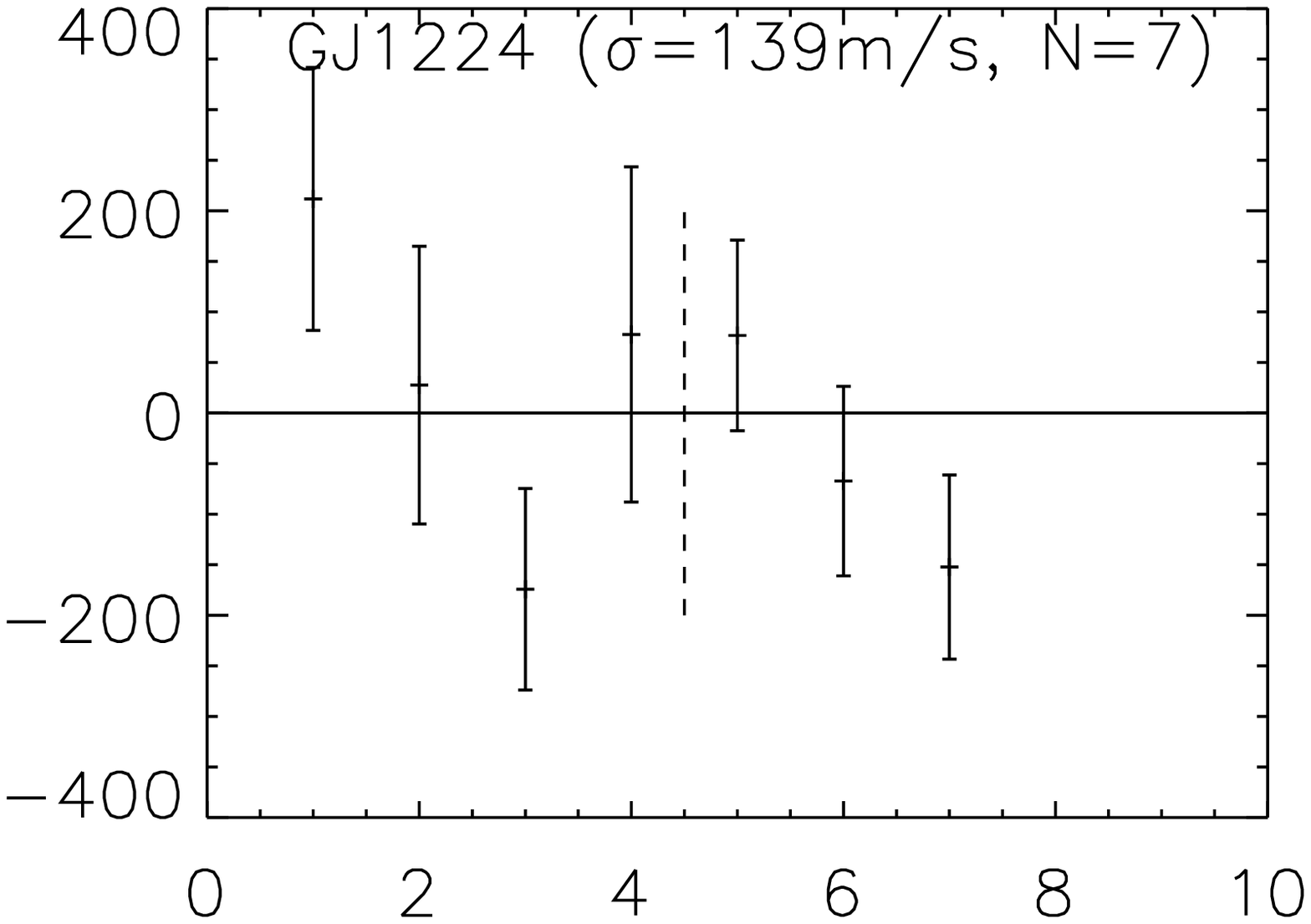}  
                \includegraphics[scale=.30,angle=0]{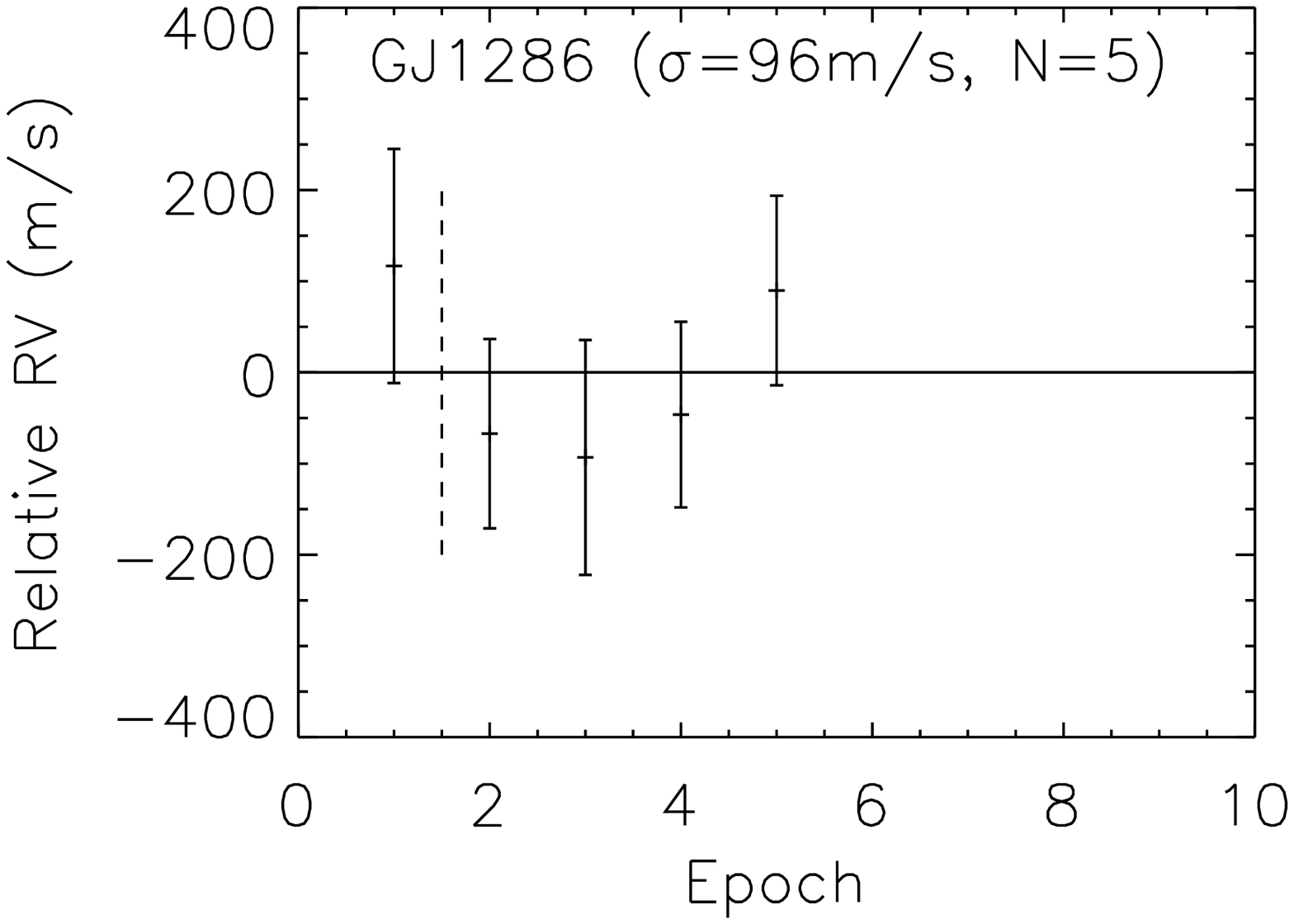} 
                \includegraphics[scale=.30,angle=0]{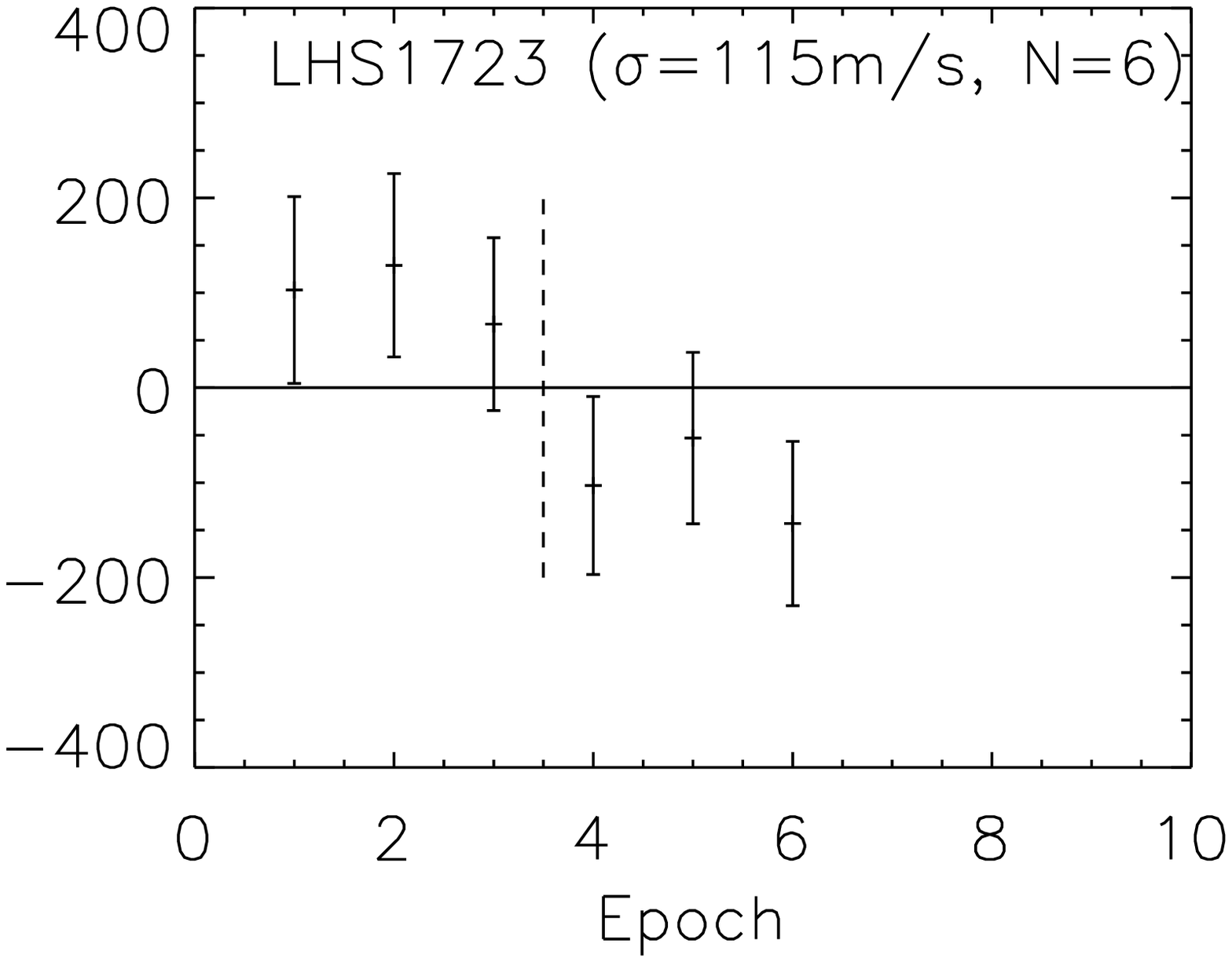}
                \includegraphics[scale=.30,angle=0]{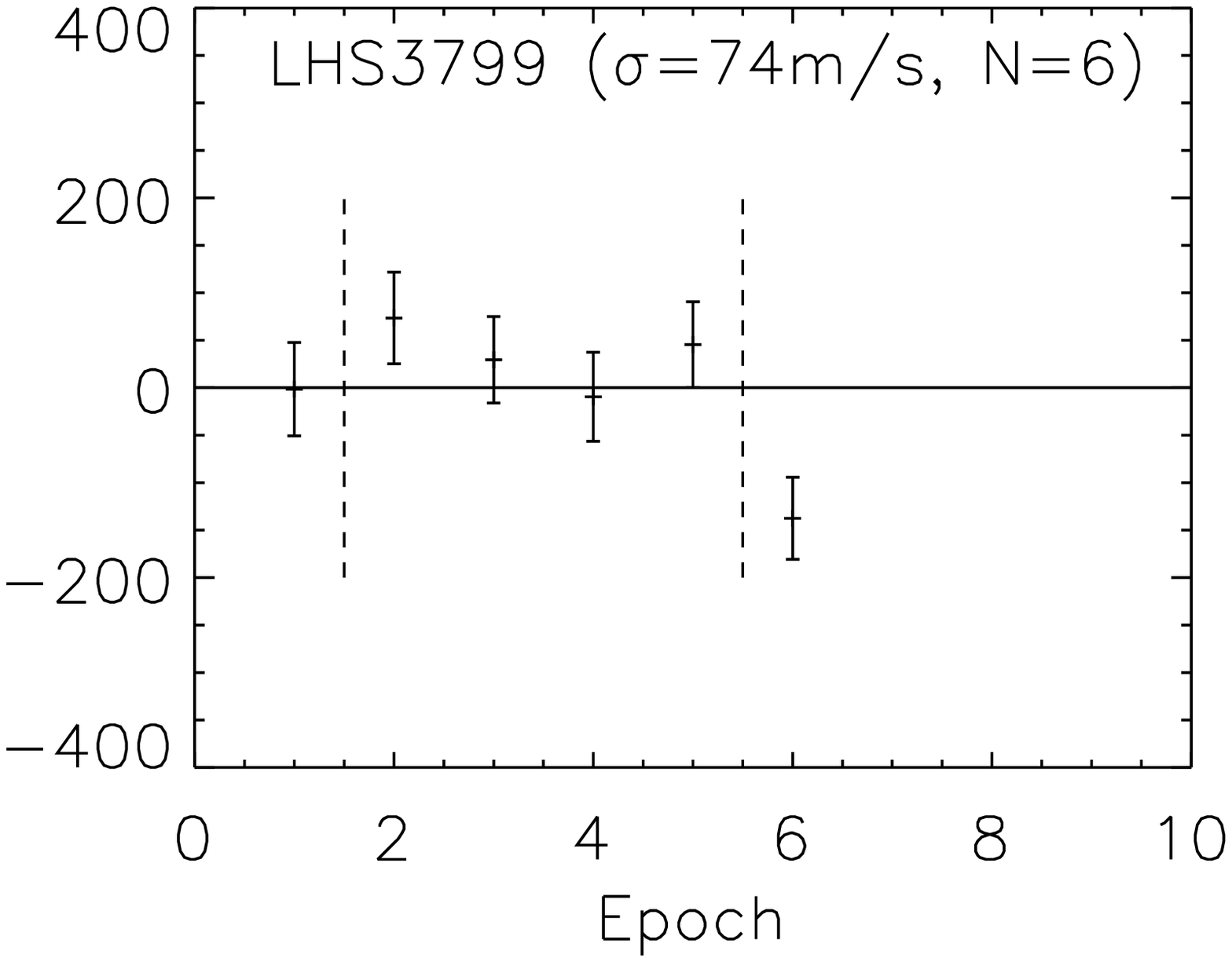}
\caption{Nightly averaged relative radial velocity measurements are plotted by epoch (see Table~\ref{table:rv} for a list of JD).  Vertical dotted lines are used to indicate long time spans between different observing runs ($\sim$ 1 to 3 weeks).  After completing periodogram tests on this RV data to seach for periodicity, we find no indication of companions around these stars.}
\label{fig:rvfig}
\end{subfigure}
\end{center}
\end{figure}

Our observed dispersions for these 12 stars range from 47 m s$^{-1}$ to 139 m s$^{-1}$.  Our average observed dispersion is 99 m s$^{-1}$, with a standard deviation of 27 m s$^{-1}$.  Figure~\ref{fig:rvfig} shows the RV measurements for all epochs for each of the target stars.  The median RV error for high SNR spectra of the 10 slowly rotating stars ($v$sin$i$ $<$ 3.0 km s$^{-1}$) is 88 m s$^{-1}$.  This is dominated by the instrumental error (73 m s$^{-1}$), which suggests a limiting precision of $\sim$90 m s$^{-1}$ for high SNR, slowly rotating mid to late M-dwarfs.  We note that some of the instrumental uncertainty could be a consequence of our modeling prescription; a more sophisticated approach may yield better results.  We also note that a precision of 58 m s$^{-1}$ has been reported for multiple epoch measurements of the M0 star GJ 281 using CSHELL \citep{Crockett2011}.  These precisions are nevertheless considerably better than the design specs for CSHELL \citep{1990SPIE.1235..131T, 1993SPIE.1946..313G}, especially considering the small wavelength coverage, and are credited to the talents of the instrument team.


\section{Astrometry}
\subsection{Observations}
 
All optical astrometric observations were made using the 0.9-m telescope at CTIO.  The astrometry program began as an NOAO Surveys Program in 1999 August and continued from 2003 February as part of the SMARTS (Small and Moderate Aperture Research Telescope System) Consortium.  Stars have been intermittently added to the observing list since 1999 and stars discussed here continue to be observed.  The Cerro Tololo Interamerican Observatory Parallax Investigation (CTIOPI) program was originally designed to measure accurate parallaxes of nearby stars.  We are now using the same data and techniques to look for perturbations that remain in our astrometric signal after solving for the parallactic motion and the proper motion of our targets.  The presence of a periodic perturbation in our data might signify that our star is orbiting around a common center of mass with an unseen companion.  To do this, we use the same instrumental setup as that used for the photometry frames.  We observe each star in one of the \textit{V}, \textit{R}, or \textit{I} filters.  Stars are observed through the filter that gives the strongest reference field, while not compromising the counts of the target star (the filter and number of reference stars used are given in Table~\ref{table:ast}).  Strong reference fields that give the most precise parallax measurement include 5 to 12 reference stars that are  bright (peak counts greater than 1000), close on the chip to the target star, and in a configuration that surrounds the target.  We require all stars used as reference stars to have a minimum of 100 counts.  Our reduction routine accounts for plate scaling and rotation effects, but ignores higher-order terms  \citep[astigmatism, coma, chromatic aberration; see][]{Jao2005}.  

Exposure times are set such that the target star or in some cases a very close bright reference star does not saturate.  Our maximum exposure time is 600 seconds.  We aim to obtain exposure times of at least 30 seconds for every star, although this is not always possible for our brightest stars.  

Frames are only collected under seeing conditions better than 2.4$\arcsec$, determined by the FWHM of the stars in the field to be used in our reduction.  Also, the target star and its reference stars must have an ellipticity less than 20$\%$, in order to determine the centroid of the stars with the most accuracy and to eliminate frames with possible tracking/guiding errors.  The guider is typically used for any exposure times longer than 300 seconds. 

Eight of the twelve stars discussed here are observed astrometrically with the \textit{V} filter.  During the course of observations, the first \textit{V} filter was cracked and replaced by another \textit{V} filter.  The use of the second \textit{V} filter from 2005 February to 2009 July causes a few milli-arcseconds (mas) offset in astrometric residuals of known singles from other techniques.  In 2009 July, we switched back to the original \textit{V} filter, as the minor crack on the filter edge does not affect the data acquired on our central quarter region of the CCD.  Using data from both filters gives the same parallax measurement, but with slightly higher errors \citep{Subasavage2009, 2010AJ....140..897R} and the average residual deviation for the stars is still less than 4 mas for our 12 stars (see \S7.3).  Therefore, we choose to use data obtained in both filters to maximize the time coverage.

\subsection{Astrometric Reductions}
 
We correct all centroids of reference stars and the target star for differential chromatic refraction \citep[DCR;][]{Jao2005}.  We measure accurate positions using the SExtractor Centroiding algorithm from  \citet{1996A&AS..117..393B} and use the Gaussfit program  \citep{Jefferys1987} to simultaneously solve for the parallax relative to the reference stars and proper motion on all available data \citep[for more details see][]{Jao2005}.  If after running our Gaussfit program, reference stars are found to have proper motions greater than 0\farcs05$/$yr or a parallax greater than 5 mas based on photometric parallaxes, then those stars are rejected as reference stars.  To obtain the absolute parallax of our target star, we must correct for the parallactic motion of the references stars as these stars are not infinitely far away.  We use photometric parallaxes and accurate \textit{V}, \textit{R} and \textit{I} photometry described in \S4 to correct our relative parallax to the absolute parallax value \citep{Jao2005}.  All the frames are used to fit the parallactic orbit to the star.  However, only nights with 2 or more good images are used in order to calculate the astrometric signal that remains after correcting for the parallactic motion and proper motion of our target.

\subsection{Astrometric Results}
In Table~\ref{table:ast}, we list the number of seasons the target has been observed, the number of parallax frames, the start and end dates of observations, the time duration of the observations, the number of reference stars, the relative parallax ($\pi$ rel), the correction to absolute parallax ($\pi$ corr), the absolute parallax ($\pi$ abs), the proper motion amplitude ($\mu$), the proper motion position angle (P.A.), and the tangential velocity (v$_{tan}$) for the 12 targets in our sub-sample.  
We do note that the parallax correction to absolute for GJ 628 was much larger (4.8$\sigma$) than typical.  We expect that this is a consequence of the reference stars being highly reddened, which will bias distance estimates.  Rather than using this number, we use the average of the corrections to absolute from 221 stars previously published using the CTIOPI pipeline \citep{Jao2005,2006AJ....132.2360H, 2007ApJ...669L..45G, Subasavage2009,2010AJ....140..897R, Jao2011, 2011AJ....142..104R, 2011ApJ...729L..26V}.

All of our measured parallaxes are within 3 sigma of the weighted average of previously published values, except GJ 300.  Our parallax value of GJ 300 supersedes that in \citet{2006AJ....132.2360H}, because we now have roughly twice as many frames over twice the time span, and now use an improved centroiding technique.

After solving for the parallactic and proper motions, the average residual deviations for all 12 stars is 2.74 mas in right ascension (R.A.) and is 3.36 mas in declination (Decl.).  The average residual deviations range from 0.86 mas to 4.85 mas in R.A. and from 1.54 mas to 5.53 mas in Decl. (see Table~\ref{table:astresults}).  Residual deviations are calculated by taking the standard deviation of the absolute values of nightly mean positions.  We also report the statistical uncertainty (henceforth referred to as the mean error) for each star, which ranges from 2.39 mas to 6.18 mas in R.A. and from 2.65 mas to 6.88 mas in Decl.  We calculate the error on a single night by taking the standard deviation of the offsets from zero for the frames, typically five, taken on each star during a night.  Then, we calculate the mean error for the star by taking the average of all the nightly errors.  The mean errors for both R.A. and Decl. are also listed Table~\ref{table:astresults}.  On average our errors are 1.5 times larger than residual deviations.  We take this as indication that we are slightly over estimating our errors.  For all 12 stars, the astrometric signals that remain after correcting for the parallactic and proper motions are plotted over time in Figure~4, split into R.A. and Decl.


\section{Companion Detection Limits}
To identify companions around these 12 stars, we search for hidden periodic signal in the data using Lomb-Scargle periodograms\citep{1982ApJ...263..835S}.  For each star, we search for companions with periods between 2 to 100 days for our RV data and between 300 and 3000 days for our astrometric data (in R.A. and Decl. separately).  Using the IDL program \textit{scargle.pro}, no frequencies have powers that exceed the 1$\sigma$ false alarm probability.  We conclude that there are no periodic signals in either our RV or astrometric data.  From this, we assume our stars are single stars within our detection limits.

To set limits on the presence of companions, we perform Monte Carlo simulations to determine the minimum object mass that we would have been able to detect in our data given our measurement errors and observing cadence.   
\subsection{Radial Velocity Limits}
As reported in \S6.4, these stars all have constant RVs to within 139 m s$^{-1}$.  To determine the minimum mass companion we would expect to detect, we simulate 1,000,000 circular orbits for each star allowing the inclinations and companion masses to vary at orbital periods between 0.5 and 100 days.  For each star, we use the stellar masses calculated in \S2 and allow the stellar masses to fluctuate within 20$\%$ to account for possible errors in the mass estimates.   We assume circular orbits, which is likely appropriate for companions to old stars with short periods \citep[$\leq$ 10 days;][]{2004RMxAC..22...95L}. The cosine of inclination ranges from 0 to 90 degrees.  After creating randomly orientated circular orbits, we then extract predicted radial velocity values at the Julian Dates we observed from these simulated orbits.  Then, we compare these extracted radial velocity values to our determined RV errors at their respective Julian Date (see Table~\ref{table:rv}) to determine a $\chi$$^2$ value.  Then, we use this $\chi$$^2$ value as an input for the IDL routine, \textit{mpchitest.pro} \citep{2009ASPC..411..251M}, to determine which orbits we would have been able to detect with a 3$\sigma$ confidence.  Figures~5-16 provide illustrations of the RV Monte Carlo simulation results for our stars.  The plots are shaded to display the fraction of objects that would have been detected with a 3$\sigma$ confidence in each period and mass increment bins.  The sizes of the mass and period bins are 0.10 M$_{JUP}$ and 1 day, respectively.  The illustrations show that most Jupiter-mass companions with periods less than 30 days would be detected.  Although some Jupiter-mass companions could be detected at longer periods, the sensitivity becomes strongly dependent on the observing cadence relative to the orbital ephemeris.  Aliasing is apparent in our RV data.

To interpret our data collectively, we follow the same method as above, except that we determine the companion mass that we would have been able to detect at selected orbital periods of 3 days, 10 days, 30 days, and 100 days.  We allow our select period to range from $\pm$10$\%$ to mitigate the bias caused by aliasing and give a better estimate of the masses of the systems we would be able to detect.  The minimum companion mass we would have been able to detect is given in increments of 0.5 M$_{JUP}$.  On average for our 12 stars, these simulations suggest that with 3$\sigma$ confidence, we would have been able to detect 90$\%$ of companions with a mass of $\sim$1 M$_{JUP}$ in 3 day orbits, $\sim$1.5 M$_{JUP}$ in 10 day orbits, $\sim$3 M$_{JUP}$ in 30 day orbits and $\sim$7 M$_{JUP}$ in 100 day orbits.  The individual detection limits for each star are provided in Table~\ref{table:90}.  GJ 1065 is only observed over a baseline of $\sim$13 days, hence the limits on the mass of a companion that we would be able to detect around this star are significantly larger for an orbital period of 100 days than the rest of stars in this sub-sample.  

Exoplanets with longer orbital periods can exhibit large eccentricities.  In our orbital period range of 0.5 to 100 days, there are 1356 exoplanets known to date\footnote[10]{Numbers are from exoplanet.eu as of October 27, 2014.}.  Of those, close to 5$\%$ have eccentricities listed above 0.2 and less than 1$\%$ of exoplanets have eccentricities above 0.6.  The planet with highest eccentricity in this period range is Kepler 419b \citep[e=0.85;][]{2014ApJ...791...89D}.  To be inclusive of planets with non-zero eccentricities that have orbital periods between 30 and 100 days, we performed our simulations again allowing the eccentricity to vary from 0.0 to 0.9. In these cases, we find that the minimum companion mass we could detect with our data with a 90$\%$ detection rate increases by a factor of three when the eccentricity varies uniformly over this range, compared to when the eccentricities are set to zero.

\subsection{Astrometric Limits}
Similar to the Monte Carlo simulations performed for the RV orbits, we simulate 1,000,000 photocentric orbits for each star allowing the inclinations, companion masses and eccentricities to change.  For each star, we use the stellar masses calculated in \S2 and allow the stellar masses to fluctuate within 20$\%$.   We allow the inclination to vary between 0 to 90 degrees and the eccentricity to vary between 0 and 0.9.  We allow the longitude of periastron, $\omega$, and the longitude of the ascending node, $\Omega$, to vary from 0 to 180 degrees.  We use the weighted distance measurements determined from the literature and from our parallax measurement in \S7.3.  We allowed the orbital periods to range from 2 to 8 years to represent our observing cadence and temporal baseline.  We then use these simulated orbits to extract astrometric positions at our observed Julian Dates.  We compare these extracted astrometric positions to our determined astrometric errors (see Table~\ref{table:astresults}) to determine a $\chi$$^2$ value.  Then, we use this $\chi$$^2$ value as an input for the IDL routine, \textit{mpchitest.pro} \citep{2009ASPC..411..251M}, to determine which orbits we would have been able to detect with a 3$\sigma$ confidence.  Figures~5-16 display the astrometric Monte Carlo simulation results.  The plots are shaded to include the fraction of objects that would have been detected with a 3$\sigma$ confidence in each period and mass increment bin of 0.10 M$_{JUP}$ and 0.1 years, respectively.  

To obtain a quantitative estimate of our detection limits, we again perform 1,000,000 Monte Carlo simulations.  We keep the same parameters as described above, except we search for companions at selected orbital periods of 2 years, 4 years, 6 years, and 8 years.  We give the minimum companion mass we would have been able to detect in increments of 0.5 M$_{JUP}$.  On average for our 12 stars, these simulations suggest that with 3$\sigma$ confidence, we would have been able to detect 90$\%$ of companions with masses of $\sim$18.5 M$_{JUP}$ in 2 year orbits, $\sim$11.5 M$_{JUP}$ in 4 year orbits, $\sim$9 M$_{JUP}$ in 6 year orbits and $\sim$7.5 M$_{JUP}$ in 8 year orbits for all of our stars (see Table~\ref{table:90}).

\subsection{Additional Detection Limits from the Literature}
\subsubsection{Optical RV Limits}
To place the strictest limits on excluding companions for the stars in this sub-sample, we also include the results for eleven of the stars that have been previously observed with high dispersion optical spectroscopy to search for planetary companions.  \citet{2012MNRAS.424..591B} present 8 epochs of GJ 1286 with a root mean square scatter of 22.1 m s$^{-1}$ over 1.1 days.  With that precision and short time cadence, \citet{2012MNRAS.424..591B} would only be sensitive to planets with periods on the order of a few days.  The remaining ten stars with previous RV measurements are included in the HARPS M-dwarf survey \citep{Bonfils2013}.  Of those ten, the authors present Monte Carlo results to exclude the presence of companions around eight of the stars, as the other two stars have fewer than four epochs of data.  On average for the eight stars, the authors are able to exclude with a 99$\%$ confidence the presence of companions with masses greater than 0.2 M$_{JUP}$ in 10 to 100 day orbital periods and greater than 1.3 M$_{JUP}$ in an eight year orbit \citep{Bonfils2013}.  We find no previous high dispersion spectroscopy data for G 99-49; thus our data place the strictest companion limits for this star.  
\subsubsection{High Resolution Imaging Limits}
To include companions beyond several AU, we include high resolution results for companions.  AO imaging or HST imaging to search for stellar and brown companions has been completed for 11 out of the 12 stars \citep{2012AJ....144...64D, 1994ApJ...428..797N, 2001AJ....121.2189O, 2010PASP..122.1195T}.  GJ 1065 has not been imaged with high resolution techniques. For 10 of the 12 stars, \citet{2001AJ....121.2189O} would have been able to detect any stellar companions at separations greater than 10 AU and most old ($<$5 Gyr) brown dwarfs with masses more than 40 M$_{JUP}$ between 40 and 120 AU.  \citet{2012AJ....144...64D} have further constrained the spatial and mass regime of companions around 7 of the 12 stars in the sub-sample.  \citet{2012AJ....144...64D} adopt 0.04 M$_{SUN}$ ($\sim$42 M$_{JUP}$) at 3 Gyr as the minimum mass detectable with their HST/NICMOS snapshot high resolution imaging survey, and find no L dwarf companions in the separation range of 5 to 70 AU and no T dwarf companions  ($\sim$0.05-0.08 M$_{SUN}$) to these M-dwarfs in the separation range of 10 to 70 AU.  

More targeted surveys have pushed the detection limits of companions down to the planetary range with the use of near infrared AO \citep{2007ApJ...670.1367L, 2005ApJ...623.1141L, 2005ApJ...625.1004M, 2000AJ....119..906S}.  \citet{2005ApJ...625.1004M} find that GJ 628 has no companions more massive than 5 M$_{JUP}$ at 4 AU and no companions more massive than 10 M$_{JUP}$ at 1 AU.  \citet{2007ApJ...670.1367L} also rule out the presence of companions more massive than 5 M$_{JUP}$ in the separation range of 25 to 50 AU around GJ 628.  These two studies imply that we can also find massive planets between 1 and 50 AU around nearby M-dwarfs via imaging techniques, assuming the luminosities from models are correct.

\subsubsection{Common Proper Motion Limits}
To extend our search to inlcude wide companions ($>$ 100 AU), we include common proper motion searches for companions.  \citet{2002AJ....123.2027H} have searched 10 of the 12 targets and find no stellar or brown dwarf companions at wide separations ($\sim$100$-$1400 AU) around these stars.  With a limiting J band magnitude of $\sim$16.5 mag, \citet{2002AJ....123.2027H} would have been able to find $\sim$40 M$_{JUP}$ for an assumed age of 5 Gyr at a distance of 5.8 pc.
\section{Summary}
In recent years, we have gained a better statistical understanding on how common Jupiter-size and Earth-size planets are around stars from the \textit{Kepler} mission that photometrically monitored approximately 150,000 stars in its first 3.5 years  \citep[e.g.][]{2010ApJ...713L.109B, 2012ApJS..201...15H, 2013ApJ...767...95D, 2013ApJ...766...81F, 2014arXiv1406.7356M}.  Of those, 3,000 are M-stars that are brighter than 16th magnitude in the Kepler passband \citep[Kp$<$16][]{2010ApJ...713L.109B}.  As \textit{Kepler} is magnitude limited, most M-stars being monitored are early M-dwarfs.  \textit{Kepler} has discovered many planets around these stars and even around mid M-dwarfs, like the M4 dwarf Kepler 42  \citep{2012ApJ...747..144M, 2014ApJS..213....5M}.  However \textit{Kepler} and the subsequent \textit{K2} mission \citep{2014PASP..126..398H} will not able to survey all of our faint stellar neighbors and thereby determine an unbiased measurement of the frequency of planets around the latest M-dwarfs.  With the current limited number of mid to late M-dwarfs that have been surveyed for companions by ground based telescopes and \textit{Kepler}, we cannot concretely say whether planets even exist around these stars, as the latest known star with an exoplanet is the M4.5 dwarf GJ 1214 \citep{1995AJ....110.1838R, 2009Natur.462..891C}.  Therefore, the New Worlds New Horizons 2010 Astronomy Decadal Survey has stated the need to develop new innovative reduction techniques and instrumentation to explore the lowest mass stars on the main sequence, as these stars are often too faint and chromospherically active for current ground based optical companion surveys and the \textit{Kepler} mission.  Motivated by this, we have identified and characterized a volume-limited survey of mid to late M-dwarfs and present our infrared radial velocity and astrometric programs, which are both viable methods to discover Jupiter-mass and brown dwarf companions.

At the start of this study, we identified a volume-limited equatorial sample of 60 mid M-dwarfs that extends out to 10 pc, which are observable from most facilities in the northern and southern hemispheres.  Since this time, two of these stars have been found to have spectroscopic companions \citep[GJ 867BD, LHS 1610AB;][]{2014AJ....147...26D, Bonfils2013}.  For the remaining 58 isolated stars in the sample, we provide new \textit{V}, \textit{R} and \textit{I} photometry and new low dispersion optical ($6000 - 9000$\,\AA) spectra, from which we measure the H$\alpha$ equivalent widths, $Na\,I$ indices and spectral types.  We determine that 35 stars are emission-line stars, and note the trend of later type M-stars being more likely to be emission-line stars than earlier type M-stars.  

For a smaller sub-sample of 12 stars, we present rotational velocities and absolute radial velocities.  Ten of these 12 stars have $v$sin$i$ values below our detection limit (3 km s$^{-1}$), while G 99-49 and GJ 729 have $v$sin$i$ values of 5.8$\pm$0.3 km s$^{-1}$ and 3.8$\pm$0.6 km s$^{-1}$, respectively.   For these 12 stars our observed RV dispersion is 99 m s$^{-1}$ with a standard deviation of 27 m s$^{-1}$.  We also demonstrate the achievable RV precision using our technique is $\sim$90 m s$^{-1}$ over timescales from 13 days to 5 years.  Our spectroscopic results indicate that on average we would have detected companions with masses greater than 1.5 M$_{JUP}$ in 10 day orbital periods and greater than 7 M$_{JUP}$ in 100 day orbital periods.  

Over baselines of 9$-$13 years, we provide astrometry with typical residuals (after determining parallaxes and proper motions) of $\sim$3 mas.  This allows us to exclude the presence of companions with masses greater than 11.5 M$_{JUP}$ with orbital periods of 4 years and 7.5 M$_{JUP}$ with orbital periods of 8 years.  Although we do not detect companions around any of the stars in our sub-sample, these results do show that we could easily detect brown dwarfs with wide orbits around low mass stars.

An ensemble result of this work is that all 12 stars studied here with both infrared spectroscopy and astrometry and in the literature with various other techniques are found to be single stars without stellar, brown dwarf, or giant planet companions within the respective detection limits of the studies.  These results provide a first step in the process of vetting our nearest neighbors for planet searches.  We suggest these slowly rotating, single stars would be prime targets for upcoming Earth mass RV planet searches using new instruments \citep[e.g. ISHELL with use of a gas cell, SPIROU and CARMENES][]{2012SPIE.8446E..2CR, 2012SPIE.8446E..4ER, 2010ASPC..430..521Q} coming online, as these instruments should be sensitive enough to search for Earth-mass planets around mid to late M-dwarfs.

\acknowledgements 
The authors would like to thank Guillem Anglada-Escud{\'e}, Jeremy Jones, John Rayner, Robert Parks, Peter Plavchan, Angelle Tanner and our anonymous referre for intellectual insight.  Also, the authors would like to note that the astrometric data were obtained by the RECONS team through the NOAO Surveys Program and the SMARTS Consortium.  The infrared spectroscopic data were obtained from the Infrared Telescope Facility, which is operated by the University of Hawaii under Cooperative Agreement no. NNX-08AE38A with the National Aeronautics and Space Administration, Science Mission Directorate, Planetary Astronomy Program.  The authors are grateful for the telescope time granted to them.  This publication makes use of data products from the Two Micron All Sky Survey, which is a joint project of the University of Massachusetts and the Infrared Processing and Analysis Center/California Institute of Technology, funded by the National Aeronautics and Space Administration and the National Science Foundation.  This work was supported by the National Science Foundation through a Graduate Research Fellowship to Cassy Davison.  This project was funded in part by NSF/AAG grant no. 0908018 and NSF grants AST05-07711 and AST09-08402.

{\it Facilities:} \facility{IRTF (CSHELL)}, \facility{CTIO:2MASS}, \facility{CTIO:0.9m}.

\onecolumn

\begin{landscape}
\begin{deluxetable}{lcclcccrrrccccc}
\tabletypesize{\footnotesize}
\tablecolumns{14}
\setlength{\tabcolsep}{0.06in} 
\tablewidth{0pt}
\tablecaption{Effectively Single Equatorial ($\pm$30$^\circ$ Decl.) mid M-dwarfs within 10 pc  }    
\tablehead{
\colhead{Name}  & \colhead{R.A. J2000.0} &\colhead{Decl. J2000.0}   &\colhead{\hspace{-0.5cm}Parallax$^a$} &\colhead{SpType}  & \colhead{Mass$^b$} & \colhead{M$_V$} &\colhead{\textit{V$_J$}} &\colhead{\textit{R$_{KC}$}} &\colhead{\textit{I$_{KC}$}}&\colhead{$\#$ nts. } &\colhead{\textit{J}$^c$} & \colhead{\textit{H}$^c$} &\colhead{\textit{K$_S$}$^c$} &\colhead{$v$sin$i$$^d$}    \\ 
                &                        &                        &\colhead{\hspace{-0.5cm}(mas)}      &                                     & \colhead{(M$_{\odot}$)} & \colhead{(mag)} &\colhead{(mag)} &\colhead{(mag)} &\colhead{(mag)}& &\colhead{(mag)} & \colhead{(mag)} &\colhead{(mag)} &\colhead{(km s$^{-1}$)}  }

\startdata
GJ 1002              &  00 06 43.2  & $-$07 32 17.0 & 213.00$\pm$3.60$^{1}$         &   M5.0V & 0.11 & 15.48  & 13.84 &12.21 & 10.21 &3 & 8.23 & 7.79 & 7.44& $<$3.0$^1$ \\  
GJ 54.1              &  01 12 30.6  & $-$16 59 56.3 &269.08$\pm$2.99$^{1,2}$       &   M4.0Ve& 0.13 & 14.30   & 12.15 &10.73 &  8.95 &2 & 7.26 & 6.75 & 6.42& $<$2.5$^2$\\
LHS 1302             &  01 51 04.5  & $-$06 07 04.8 &100.78$\pm$1.89$^3  $         &   M4.5Ve& 0.13 & 14.51  & 14.49 &13.00 & 11.16 &5 &9.41 & 8.84 & 8.55 &  ...\\
GJ 83.1              &  02 00 13.0  & +13 03 07.0   &224.80$\pm$2.90$^1$           &   M4.0Ve& 0.14 & 14.11  & 12.35 &10.95 &  9.18 &3 & 7.51 & 6.97 & 6.65 &  $<$2.5$^2$\\
LHS 1326             &  02 02 16.2  & +10 20 13.7   &112.00$\pm$3.20$^1$           &   M5.0Ve& 0.10 & 15.95  & 15.70 &13.99 & 11.91 &3 & 9.84 & 9.25 & 8.93& $<$4.5$^3$ \\ 
LHS 1375             &  02 16 29.8  & +13 35 13.7   &117.70$\pm$4.00$^1$           &   M5.0Ve& 0.10 & 16.15  & 15.80 &14.14 & 12.01 &3 & 9.87 & 9.31 & 8.98 & 12.4$^4$ \\
GJ 102               &  02 33 37.2  & +24 55 39.2   &102.40$\pm$2.70$^1$           &   M3.5Ve& 0.17 & 13.10   & 13.05 &11.76 & 10.10 &3 & 8.47 & 7.91 & 7.63 &  ... \\
GJ 105B              &  02 36 15.4  & +06 52 19.1   &138.79$\pm$0.43$^{1,2,4}$     &   M3.5V & 0.22 & 12.42  & 11.71 &10.49 &  8.88 &2 & 7.33 & 6.79 & 6.57 & $<$2.5$^2$\\
SO 0253+1652         &  02 53 00.8  & +16 52 53.3   &259.41$\pm$0.89$^{3,5}$       &   M7.0V & 0.08 & 17.21  & 15.14 &13.03 & 10.65 &3 & 8.39 & 7.88 & 7.59 &   10.0$^5$\\ 
LP 771-095A          &  03 01 51.1  & $-$16 35 30.7 &143.81$\pm$2.49$^{3}$         &   M2.5V & 0.25 & 12.01  & 11.22 &10.07 &  8.66 &4 & 7.11 & 6.56 & 6.29 &  5.5$^6$\\
GJ 1057              &  03 13 23.0  & +04 46 29.3   &117.10$\pm$3.50$^1$           &   M4.5V & 0.13 & 14.28  & 13.94 &12.45 & 10.62 &3 & 8.78 & 8.21 & 7.83 & $<$2.2$^1$\\
GJ 1065              &  03 50 44.3  & $-$06 05 41.7 &105.40$\pm$3.20$^1$           &   M3.5V & 0.18 & 12.93  & 12.82 &11.60 & 10.04 &3 & 8.57 & 8.00 & 7.75 & 4.0$^6$\\
GJ 166C              &  04 15 21.7  & $-$07 39 17.4 &200.65$\pm$0.23$^2$           &   M4.0Ve& 0.19 & 12.75  &11.24 &  9.99 &  8.31 &3 & 6.75 & 6.28 & 5.96 &  5.0$^8$\\
LP 655-048           &  04 40 23.3  & $-$05 30 08.3 &105.50$\pm$3.20$^6$           &   M6.0Ve& 0.08 & 17.92  &17.80 & 15.73 & 13.37 &5 &\hspace{-0.1cm}10.66 & 9.99 & 9.55 & 16.5$^9$\\ 
LHS 1723             &  05 01 57.4  & $-$06 56 46.5 &187.92$\pm$1.26$^3$           &   M4.0Ve& 0.15 & 13.57  &12.20 & 10.86 &  9.18 &4 & 7.62 & 7.07 & 6.74 &  $<$3.2$^7$\\
GJ 203               &  05 28 00.2  & +09 38 38.1   &102.60$\pm$2.09$^{1,2}$       &   M3.0V & 0.19 & 12.52  &12.46 & 11.27 &  9.78 &3 & 8.31 & 7.84 & 7.54 &4.0$^6$\\
GJ 213               &  05 42 09.3  & +12 29 21.6   &171.50$\pm$1.00$^{1,2,7}$     &   M3.5V & 0.19 & 12.71  &11.54 & 10.32 &  8.68 &2 & 7.12 & 6.63 & 6.39 & $<$2.5$^2$\\
G 99-49              &  06 00 03.5  & +02 42 23.6   &190.77$\pm$1.86$^{1,3}$       &   M3.5Ve& 0.19 & 12.71  &11.31 & 10.04 &  8.43 &6 & 6.91 & 6.31 & 6.04 & 7.4$^7$\\
GJ 232               &  06 24 41.3  & +23 25 58.6   &119.40$\pm$2.30$^1$           &   M4.0V & 0.15 & 13.55  &13.16 & 11.86 & 10.21 &3 & 8.66 & 8.16 & 7.91 & $<$3.1$^7$\\
GJ 1093              &  06 59 28.7  & +19 20 57.7   &128.80$\pm$3.50$^1$           &   M5.0Ve& 0.11 & 15.49  &14.94 & 13.25 & 11.24 &4 & 9.16 & 8.55 & 8.23 & $<$2.8$^1$\\
GJ 273               &  07 27 24.5   &+05 13 32.8   &266.23$\pm$0.66$^{1,2,7}$     &   M3.0V & 0.25 & 12.01  & 9.88 &  8.68 &  7.14 &3 & 5.71 & 5.22 & 4.86 & 2.5$^2$\\
GJ 283B              &  07 40 19.2  & $-$17 24 45.0 &109.45$\pm$0.51$^{1,8,9}$     &   M6.5Ve& 0.16 & 13.26  &13.06 & 12.89 & 12.72 &4 &\hspace{-0.1cm}10.16 & 9.63 & 9.29 & ...\\  
GJ 285               &  07 44 40.2  & +03 33 08.8   &167.19$\pm$2.05$^{1,2}$       &   M4.0Ve& 0.23 & 12.31  &11.19 &  9.91 &  8.22 &4 & 6.58 & 6.01 & 5.70 & 4.5$^{10}$\\
GJ 1103              &  07 51 54.7  & $-$00 00 11.8 &114.00$\pm$3.30$^1$           &   M4.5V & 0.15 & 13.54  &13.26 & 11.89 & 10.19 &3 & 8.50 & 7.94 & 7.66 & ...\\ 
GJ 299       	     &  08 11 57.6  & +08 46 22.1   &146.30$\pm$3.10$^1$           &   M3.5V & 0.15 & 13.69  &12.86 & 11.57 &  9.91 &3 & 8.42 & 7.93 & 7.66 &3.0$^7$\\
GJ 300               &  08 12 40.9  & $-$21 33 06.8 &125.78$\pm$0.97$^3$           &   M3.5V & 0.20 & 12.65  &12.15 & 10.85 &  9.22 &3 & 7.60 & 6.96 & 6.71 & $<$3.0$^6$\\
GJ 1111              &  08 29 49.3  & +26 46 33.7   &275.80$\pm$3.00$^1$           &   M6.0Ve& 0.09 & 17.16  &14.96 & 12.89 & 10.59 &3 & 8.24 & 7.62 & 7.26 &8.1$^7$\\ 
LHS 2090             &  09 00 23.6  & +21 50 05.4   &156.87$\pm$2.67$^3$           &   M6.0Ve& 0.09 & 17.09  &16.11 & 14.12 & 11.84 &3 & 9.44 & 8.84 & 8.44 & 20.0$^3$\\ 
LHS 2206       	     &  09 53 55.2  & +20 56 46.0   &108.69$\pm$2.06$^{3,10}$      &   M4.0Ve& 0.14 & 14.20  &14.02 & 12.63 & 10.85 &3 & 9.21 & 8.60 & 8.33 & 16.5$^3$\\
LHS 292              &  10 48 12.6  & $-$11 20 08.2 &220.30$\pm$3.60$^1$           &   M6.5Ve& 0.08 & 17.50  &15.78 & 13.63 & 11.25 &3 & 8.86 & 8.26 & 7.93 & 3.0$^1$\\ 
GJ 402               &  10 50 52.0  & +06 48 29.2   &145.67$\pm$3.17$^{1,2}$       &   M4.0V & 0.21 & 12.53  &11.71 & 10.43 &  8.84 &3 & 7.32 & 6.71 & 6.37 &  $<$2.5$^2$\\
GJ 406		     &  10 56 28.9  & +07 00 53.2   &419.10$\pm$2.10$^1$           &   M5.0Ve& 0.09 & 16.69  &13.58 & 11.64 &  9.44 &2 & 7.09 & 6.48 & 6.08 & $<$3.0$^1$\\ 
GJ 447        	     &  11 47 44.4  & +00 48 16.4   &298.14$\pm$1.37$^{1,2}$       &   M4.0V & 0.16 & 13.52  &11.15 &  9.79 &  8.13 &3 & 6.51 & 5.95 & 5.65 &  $<$2.5$^2$\\
GJ 1154              &  12 14 16.5  & +00 37 26.4   &119.40$\pm$3.50$^1$           &   M4.5Ve& 0.14 & 14.03  &13.64 & 12.17 & 10.31 &2 & 8.46 & 7.86 & 7.54 & 5.2$^1$\\
GJ 1156              &  12 18 59.4  & +11 07 33.9   &152.90$\pm$3.00$^1$           &   M4.5Ve& 0.12 & 14.87  &13.95 & 12.33 & 10.38 &2 & 8.53 & 7.88 & 7.57 & 9.2$^7$\\
GJ 486               &  12 47 56.6  & +09 45 05.0   &119.58$\pm$2.64$^2$           &   M4.0V & 0.28 & 11.79  &11.40 & 10.21 &  8.67 &2 & 7.20 & 6.67 & 6.36 & $<$2.5$^2$\\
GJ 493.1             &  13 00 33.5  & +05 41 08.1   &123.10$\pm$3.50$^1$           &   M4.5Ve& 0.15 & 13.85  &13.40 & 12.02 & 10.26 &2 & 8.55 & 7.97 & 7.66 &16.8$^7$ \\
GJ 555               &  14 34 16.8  & $-$12 31 10.3 &               160.78$\pm$1.98$^{1,2,11}$    &   M3.5V & 0.22 & 12.37  &11.34 & 10.06 &  8.44 &3 & 6.84 & 6.26 & 5.94 &$<$2.5$^2$\\
LHS 3003             &  14 56 38.3  & $-$28 09 47.4 &152.49$\pm$2.02$^{1,8,12}$     &   M7.0Ve& 0.08 & 17.99  &17.07 & 14.92 & 12.54 &5 & 9.97 & 9.32 & 8.93 &5.0$^8$\\ 
GJ 609               &  16 02 51.0  & +20 35 21.8   &100.30$\pm$3.10$^1$           &   M4.0V & 0.20 & 12.59  &12.58 & 11.32 &  9.70 &2 & 8.13 & 7.65 & 7.37 & $<$3.0$^7$\\
GJ 628               &  16 30 18.1  & $-$12 39 45.4 &234.38$\pm$1.50$^{1,2}$       &   M3.0V & 0.26 & 11.92  &10.07 &  8.89 &  7.37 &3 & 5.95 & 5.37 & 5.08 &1.5$^{10}$\\
GJ 643               &  16 55 25.3  & $-$08 19 20.8 &154.96$\pm$0.52$^{1,8,13,14}$                &   M3.0V & 0.19 & 12.72  &11.77 & 10.55 &  9.01 &3 & 7.56 & 7.06 & 6.72 &$<$2.7$^7$\\
GJ 644C              &  16 55 35.3  & $-$08 23 40.1 &154.96$\pm$0.52$^{1,8,13,14}$                &   M7.0Ve& 0.08 & 17.80  &16.85 & 14.64 & 12.25 &3 & 9.78 & 9.20 & 8.82 & 9.0$^1$\\ 
GJ 1207              &  16 57 05.7  & $-$04 20 56.0 &115.26$\pm$1.50$^3$           &   M3.5Ve& 0.19 & 12.56  &12.25 & 10.99 &  9.43 &5 & 7.97 & 7.44 & 7.12 & 10.7$^6$\\
GJ 699               &  17 57 48.5  & +04 41 36.2   &545.51$\pm$0.29$^{1,2, 15}$    &   M3.5V & 0.17 & 13.17  & 9.49 &  8.27 &  6.70 &1 & 5.24 & 4.83 & 4.52 &$<$2.5$^2$ \\
GJ 1224              &  18 07 32.9  & $-$15 57 47.0 &132.60$\pm$3.70$^1$           &   M4.0Ve& 0.14 & 14.09  &13.48 & 12.08 & 10.31 &3 & 8.64 & 8.09 & 7.83 &$<$3.0$^{10}$ \\
GJ 1230B             &  18 41 09.8  & +24 47 19.5   &120.90$\pm$7.20$^1$           &   M4.5Ve& 0.19 & 12.74  &12.33 & 10.97 &  9.26 &4 & 8.86 & 8.0 & 7.77 &$<$7.1$^1$\\
GJ 729               &  18 49 49.4  & $-$23 50 10.4 &337.22$\pm$1.97$^{1,2}$       &   M3.5Ve& 0.17 & 13.14  &10.50 &  9.26 &  7.68 &3 & 6.22 & 5.66 & 5.37 & 4.0$^2$\\
GJ 752B              &  19 16 57.6  & +05 09 02.2   &171.20$\pm$0.50$^{1,2,16,17}$ &   M8.0Ve& 0.07 & 18.62  &17.45 & 15.21 & 12.78 &4 & 9.91 & 9.23 & 8.77 & 6.5$^1$\\  
GJ 1235              &  19 21 38.7  & +20 52 02.8   &100.10$\pm$3.50$^1$           &   M4.0Ve& 0.16 & 13.47  &13.47 & 12.12 & 10.46 &2 & 8.80 & 8.22 & 7.94 & ...\\
GJ 1256              &  20 40 33.6  & +15 29 57.2   &102.00$\pm$2.20$^1$           &   M4.0V & 0.16 & 13.51  &13.47 & 12.10 & 10.37 &3 & 8.64 & 8.08 & 7.75 &  $<$6.5$^{11}$\\
LP 816-060           &  20 52 33.0  & $-$16 58 29.0 &175.03$\pm$3.40$^2$           &   M3.0V & 0.19 & 12.72  &11.50 & 10.25 &  8.64 &3 & 7.09 & 6.52 & 6.20 &  $<$6.5$^{11}$\\
G 188-038            &  22 01 13.1  & +28 18 24.9   &111.70$\pm$1.73$^{1,2}$       &   M3.5Ve& 0.23 & 12.29  &12.05 &10.77  & 9.16  &2 & 7.64 & 7.04 & 6.78 & 35.1$^6$\\
LHS 3799             &  22 23 07.0  & $-$17 36 26.1 &134.40$\pm$4.90$^1$           &   M4.5Ve& 0.14 &  13.94 &13.30 &11.87  & 10.04 &5 & 8.24 & 7.64 & 7.32 &  $<$6.5$^{11}$\\
LP 876-010           &  22 48 04.5  & $-$24 22 07.5 &132.07$\pm$1.19$^{18}$        &   M4.0Ve& 0.17 &  13.19 &12.59 & 11.31 &  9.61 &3 & 8.08 & 7.53 & 7.21 &  22.0$^{12}$\\
GJ 896A              &  23 31 52.2  & +19 56 14.3   &159.88$\pm$1.53$^{1,2,19}$     &   M3.5Ve& 0.33 & 11.32  &10.30 &  9.13 &  7.66 &2 & 6.16 & 5.57 & 5.33 & 10.0$^8$\\
GJ 896B              &  23 31 52.6  & +19 56 13.9   &159.88$\pm$1.53$^{1,2,19}$     &   M4.0Ve& 0.16 &  13.42 &12.40 & 11.04 &  9.28 & 2& 7.10 & 6.56 & 6.26  &15.0$^8$\\
GJ 1286              &  23 35 10.5  & $-$02 23 20.8 &138.30$\pm$3.50$^1$           &   M5.0Ve& 0.11 &  15.43 &14.73 &13.10  & 11.10 &3 & 9.15 & 8.51 & 8.18 & $<$5.7$^1$\\

\enddata
\tablenotetext{a}{When multiple parallax references are listed, the reported value here is the weighted means for each system.  Parallax References: (1) \citet{1995gcts.book.....V}.  (2) \citet{2007A&A...474..653V};  (3) \citet{2006AJ....132.2360H};  (4) \citet{1996AJ....111..492I};  (5) \citet{2009AJ....137..402G};  (6) \citet{2012ApJ...758...56S};  (7) \citet{2008AJ....136..452G};  (8) \citet{2005AJ....130..337C};  (9) \citet{Subasavage2009};  (10) \citet{2010A&A...514A..84S};  (11) \citet{Jao2005};  (12) \citet{1996MNRAS.281..644T};  (13) \citet{1999A&A...341..121S};  (14) \citet{1998A&AS..133..149M};  (15) \citet{1999AJ....118.1086B};  (16) \citet{1995AJ....110.3014T};  (17) \citet{2009ApJ...700..623P};  (18) \citet{2013AJ....146..154M};  (19) \citet{1996AJ....112.2300W}.  This table only includes previously published parallax measurements.  New astrometric measurements from this paper are given in Table 6.}     
\tablenotetext{b}{Masses were determined using the mass-luminosity relationship of \citet{1999ApJ...512..864H} for stars with M$_v$ $>$ 12.89 mag, and the relationship of \citet{1993AJ....106..773H} for brighter stars.  Typically errors using these relations are close to 20$\%$.}
\tablenotetext{c}{All \textit{J}, \textit{H}, \textit{K$_s$} magnitudes are from the 2MASS All Sky Catalogue of point sources from \citet{Skrutskie2006}.}   
\tablenotetext{d}{$v$sin$i$ References: (1) \citet{2003ApJ...583..451M};  (2) \citet{2010AJ....139..504B}; (3) \citet{2009ApJ...704..975J};  (4) \citet{2014MNRAS.439.3094B}; (5) \citet{2012ApJS..203...10T};  (6) \citet{2013arXiv1310.5820R};  (7) \citet{1998A&A...331..581D};  (8) \citet{2005MNRAS.358..105J};  (9) \citet{2010ApJ...710..924R};  (10) \citet{2007ApJ...656.1121R};  (11) \citet{Bonfils2013};  (12) \citet{2013AJ....146..154M}.}

\label{table:sample}

\end{deluxetable}
\end{landscape}

\begin{landscape}
\begin{deluxetable}{lccllccccccccc}
\tabletypesize{\footnotesize}
\tablecolumns{14}
\tablewidth{0pt}
\tablecaption{Close-separation ($<$4$\arcsec$) mid M-Dwarf Equatorial ($\pm$30$^\circ$ Decl.) Multiples within 10 pc\label{table:binaries} }
\tablehead{
\colhead{Name}  & \colhead{R.A. J2000.0} &\colhead{Decl. J2000.0}   &\colhead{\hspace{-0.6cm}Parallax$^a$} &\colhead{SpType$^b$}&  \colhead{M$_V$$^c$} &\colhead{\textit{V}$^d$} &\colhead{\textit{R}$^d$} &\colhead{\textit{I}$^d$} &\colhead{\textit{J}$^d$} & \colhead{\textit{H}$^d$} &\colhead{\textit{K$_S$}$^d$} &Configuration$^e$   \\ 
                &                        &                        &\colhead{\hspace{-0.6cm}(mas)}      &                          & \colhead{(mag)} &\colhead{(mag)} &\colhead{(mag)} &\colhead{(mag)} &\colhead{(mag)} & \colhead{(mag)} &\colhead{(mag)} & } 
\startdata
GJ 1005AB            &  00 15 27.7  & $-$16 07 56.0 &168.42$\pm$0.89$^{1,2,3,4}$     &   M4.0VJ$^1$   &  13.75 & 12.62$^1$  &  11.46$^1$  &10.05$^1$  & 7.22 & 6.71 & 6.39 & AB   \\
GJ 2005ABC           &  00 24 44.2  & $-$27 08 25.2 &129.71$\pm$2.43$^{1,5}$         &   M5.5VJ$^2$   &  15.98 & 15.42$^2$  &  13.71$^2$  &11.56$^2$  & 9.25 & 8.55 & 8.24 & ABC \\
GJ 65AB              &  01 39 01.5  & $-$17 57 01.8 &373.70$\pm$2.70$^1$             &   M5.5V$^1$    &  14.92 & 12.06$^2$  &  10.40$^2$  & 8.34$^2$  & 6.28  & 5.69  & 5.34  & AB  \\
GJ 105C              &  02 36 04.7  &   +06 53 14.8 &138.79$\pm$0.43$^{1,6,7}$       &   M7.0V$^3$    &    ... &  ...       &    ... &  ...  &  ... & ...  & ...  & AC-B \\
LP 771-096BC$^f$     &  03 01 51.4  & $-$16 35 36.1 &143.81$\pm$2.49$^{6,8}$         &   M3.50VJ$^4$  &  12.16 & 11.37$^3$  &  10.13$^3$ & 8.58$^3$    & 7.29 & 6.77 & 6.50 & A-BC$^{f}$\\
LHS 1610AB           &  03 52 41.7  & +17 01 05.7   &100.88$\pm$2.05$^8$             &   M4.0VJ$^4$   &  13.87 & 13.85$^3$  &  12.42$^3$ & 10.66$^3$  & 8.93 & 8.38 & 8.05 & AB\\
GJ 190AB             &  05 08 35.1  & $-$18 10 19.4 &107.57$\pm$2.08$^{1,6}$         &   M3.5VJ$^2$   &  10.48 & 10.32$^4$  &   9.17$^4$ & 7.67$^4$   & 6.17 & 5.59 & 5.31 & AB\\
GJ 234AB             &  06 29 23.4  & $-$02 48 50.3 &244.44$\pm$0.92$^{1,2,9}$       &   M4.5VJ$^4$   &  13.06 & 11.12$^4$  &   9.78$^4$ & 8.08$^4$   & 6.38 & 5.75 & 5.49 & AB\\
LTT 17993AB          &  07 36 25.1  &   +07 04 43.2 &116.60$\pm$0.97$^8$             &   M4.5VJ$^2$   &  13.58 & 13.25$^3$  &  11.81$^3$ & 9.97$^3$  & 8.18 & 7.61 & 7.28 & AB\\
GJ 1116AB            &  08 58 15.2  &   +19 45 47.1 &191.20$\pm$2.50$^1$             &   M5.5VJ$^1$   &  15.06 & 13.65$^5$  &  11.97$^5$ & 9.83$^5$  &  7.79 & 7.24 & 6.89 & AB\\
LTT 12352AC          &  08 58 56.3  & $+$08 28 25.9 &147.66$\pm$1.98$^8$             &   M3.5VJ$^4$   &  11.77 & 10.92$^3$  &   9.67$^3$ & 8.05$^3$   & 6.51 & 5.97 & 5.69 & AC-B \\
GJ 473AB             &  12 33 16.3  &   +09 01 26.0 &227.90$\pm$4.60$^1$             &   M5.5VJ$^1$   &  14.28 & 12.49$^4$  &  10.93$^4$ & 8.97$^4$  & 7.00 & 6.40 & 6.04 & AB \\
GJ 569BC             &  14 54 29.4  &   +16 06 08.9 &100.62$\pm$1.28$^{1,6}$         &   M8.5VJ$^5$   &  ...   &  ...       &  ...       & ...   &  ... &  ... & ...  & A-BC \\
GJ 644ABD            &  16 55 29.6  & $-$08 19 55.3 &154.96$\pm$0.52$^{1,2,5,10}$    &   M2.5VJ$^1$   &  9.98  &  9.03$^4$  &  7.94$^4$  & 6.57$^4$  & 5.27 & 4.78 & 4.40 & ABD-C-GJ 643\\
GJ 695BC             &  17 46 25.1  &   +27 43 01.4 &120.32$\pm$0.16$^{1,6}$         &   M3.5VJ$^2$   &  10.26 &  9.86$^6$  &  8.70$^6$  & 7.25$^6$  & 5.77 & 5.17 & 4.95 & AD-BC \\ 
GJ 1230AC            &  18 41 09.8  &   +24 47 14.4 &120.90$\pm$7.20$^1$             &   M4.5VJ$^1$   &  12.57 & 12.16$^5$  &  10.82$^5$ & 9.07$^5$  & 7.53 & 6.91 & 6.62 & AC-B \\
GJ 791.2AB           &  20 29 48.3  &   +09 41 20.2 &112.90$\pm$0.30$^{1,6}$         &   M4.5VJ$^5$   &  13.34 & 13.08$^4$  &  11.73$^4$ & 9.98$^4$  & 8.23 & 7.67 & 7.31 & AB \\ 
GJ 829AB             &  21 29 36.8  &   +17 38 35.9 &149.01$\pm$1.69$^{1,6}$         &   M3.4VJ$^{1}$ &  11.17 & 10.30$^7$  &   9.15$^7$ & 7.70$^7$  & 6.25 & 5.74 & 5.45 & AB \\
GJ 831AB             &  21 31 18.6  & $-$09 47 26.5 &128.21$\pm$2.05$^{1,6}$         &   M4.5VJ$^1$   &  12.58 & 12.04$^4$  &  10.74$^4$ & 9.04$^4$  & 7.32 & 6.70 & 6.38 & AB \\       
GJ 866ABC            &  22 38 33.7  & $-$15 17 57.3 &289.50$\pm$4.40$^1$             &   M5.0VJ$^{1}$ &  14.68 & 12.37$^2$  &  10.70$^2$ & 8.64$^2$  & 6.55 & 5.95 & 5.54 & ABC \\
GJ 867BD             &  22 38 45.3  & $-$20 36 51.9 &113.37$\pm$1.04$^{1,6,11}$      &   M3.5VJ$^7$   &  11.75 & 11.45$^4$  &  10.29$^4$ & 8.78$^4$  & 7.34 & 6.82 & 6.49& AC-BD\\
\enddata

\tablenotetext{a}{When multiple parallax references are listed, the reported value here is the weighted mean for each system.  Parallax References: (1) \citet{1995gcts.book.....V}; (2)  \citet{1999A&A...341..121S};  (3) \citet{2010A&A...514A..84S};  (4) \citet{1998AJ....116.1440H};  (5) \citet{2005AJ....130..337C};  (6) \citet{2007A&A...474..653V};  (7) \citet{1996AJ....111..492I};  (8) \citet{2006AJ....132.2360H};  (9) \citet{2003AJ....125.1530G};  (10) \citet{1998A&AS..133..149M};  (11) \citet{2014AJ....147...26D}. }   
\tablenotetext{b} {J represents a joint spectral type when two or more stars cannot be deconvolved into their individual components.  Spectral type References:  (1) \citet{1994AJ....108.1437H}; (2) \citet{1995AJ....110.1838R}; (3) \citet{2000AJ....120.2082G}; (4); \citet{2006AJ....132.2360H}; (5) \citet{1991ApJS...77..417K}; (6) \citet{2014AJ....147...26D}.}
\tablenotetext{c}{Using our photometric measurements and the parallax data, the absolute magnitude errors range from 0.03 to 0.08 mag.}

\tablenotetext{d}{We give joint photometry of the close binary stars as these close systems cannot be deconvolved into their individual components.  All stars with spectral type denoted by the letter J also have joint photometry.  \textit{V}, \textit{R} and \textit{I} References: (1) \cite{2010AJ....140..897R};  (2) \citet{1991AJ....101..662B};  (3) \citet{2006AJ....132.2360H};  (4) \citet{1990A&AS...83..357B};  (5) \citet{1996AJ....112.2300W};  (6) this paper;  (7) \citet{1991AJ....102.1795W}.  All J, H, K$_s$ magnitudes are from the 2MASS All Sky Catalogue of point sources from \citet{Skrutskie2006}.}       
\tablenotetext{e}{System Configuration.  We indicate widely ($>$4$\arcsec$) separated pairs with a hyphen.  There is no spacing between components with separations less than 4$\arcsec$ (e.g. AB).}
\tablenotetext{f}{\textrm{L}P 771-096BC is a distant companion to LP 771-095, which is labeled A under this configuration.}

\end{deluxetable}

\end{landscape}



\begin{deluxetable}{lcccc}
\tabletypesize{\footnotesize}
\tablecolumns{14}
\tablewidth{0pt}
\tablecaption{Optical Spectroscopic Measurements for Effectively Single Equatorial mid M-dwarfs within 10 pc \label{table:halpha} }
\tablehead{
\colhead{Name} & \colhead{SpType} & \colhead{Date} & \colhead{H-$\alpha$} & \colhead{$Na\,I$}  \\
 & & \colhead{YYYYMMDD} & \colhead{EW \AA} & \colhead{Index}    } 
\startdata
GJ 1002    & M5.0V      &    19931031 &    -0.1  &  1.31  \\
           &            &    20041001 &     0.2  &  1.36  \\
\hline						   							 
GJ 54.1    & M4.0Ve     &    19931031 &    -1.3  &  1.25  \\
           &            &    20041002 &    -1.1  &  1.30  \\
\hline						   							 
LHS 1302   & M4.5Ve     &    20031206 &    -3.6  &  1.27  \\
           &            &    20061209 &    -3.4  &  1.25  \\
\hline						   							 
GJ 83.1    & M4.0Ve     &    19911113 &     0.0  &  1.20  \\
           &            &    20041001 &    -2.1  &  1.28  \\
\hline						   							 
LHS 1326   & M5.0Ve     &    20031206 &    -1.4  &  1.33  \\
           &            &    20061208 &    -0.8  &  1.32  \\
\hline						   							 
LHS 1375   & M5.0Ve     &    20031205 &    -4.1  &  1.33  \\
           &            &    20061208 &    -2.9  &  1.32  \\
\hline						   							 
GJ 102     & M3.5Ve     &    20031206 &    -3.1  &  1.22  \\
\hline						   							 
GJ 105B    & M3.5V      &    19931029 &     0.3  &  1.12  \\
           &            &    20041002 &     0.2  &  1.18  \\
\hline						   							 
SO 0253+1652 & M7.0V     &    20031205 &     0.8  &  1.37 \\
             &           &    20061206 &    -0.4  &  1.39  \\
\hline						   							 
LP 771-095A & M2.5V     &    20031206 &     0.0  &  1.16  \\
\hline						   							 
GJ 1057    & M4.5V      &    20031206 &    -0.3  &  1.26  \\
\hline						   							 
GJ 1065    & M3.5V      &    20020402 &     0.0  &  1.09  \\
           &            &    20061207 &     0.0  &  1.21  \\
\hline						   							 
GJ  166C   & M4.0Ve     &    20031206 &    -5.3  &  1.21  \\
\hline						   							 
LP 655-048 & M6.0Ve     &    20040311 &   -17.8  &  1.35  \\
\hline						   							 
LHS 1723   & M4.0Ve     &    20031206 &    -1.5  &  1.25  \\
\hline						   							 
GJ 203     & M3.0V      &    19980208 &     0.2  &  1.13  \\
           &            &    20061208 &     0.1  &  1.15  \\
\hline						   							 
GJ 213     & M3.5V      &    19911113 &     0.3  &  1.12  \\
           &            &    20040312 &     0.0  &  1.14  \\
\hline						   							 
G 99-49    & M3.5Ve     &    19930314 &    -3.4  &  1.20  \\
           &            &    20031206 &    -4.6  &  1.22  \\
\hline						   							 
GJ 232     & M4.0V      &    19900122 &     0.5  &  1.17  \\
           &            &    20040312 &     0.0  &  1.22  \\
\hline						   							 
GJ 1093    & M5.0Ve     &    19930314 &    -0.9  &  1.29  \\
           &            &    20031206 &    -2.1  &  1.33  \\
\hline						   							 
GJ 273     & M3.0V      &    19900122 &     0.4  &  1.11  \\
           &            &    20040312 &     0.1  &  1.15  \\
\hline
GJ 283B    & M6.5Ve    &    19930316 &     1.7  &  1.33   \\
           &           &    20031205 &    -0.5  &  1.40   \\
           &           &    20061209 &    -0.1  &  1.38   \\
\hline						   						       
GJ 285     & M4.0Ve    &    19930315 &    -9.4  &  1.20  \\
           &           &    20041003 &    -7.7  &  1.24  \\
\hline						   						       
GJ 1103    & M4.5V     &    20040312 &     0.2  &  1.20  \\
\hline						   						       
GJ 299     & M3.5V     &    19930315 &     1.6  &  1.19  \\
           &           &    20050131 &     0.1  &  1.24  \\
\hline						   						       
GJ 300     & M3.5V     &    19930315 &     0.8  &  1.20  \\
           &           &    20041003 &    -0.2  &  1.25  \\
\hline						   						       
GJ 1111    & M6.0Ve    &    19951202 &    -4.4  &  1.40  \\
           &           &    20031206 &    -3.8  &  1.37  \\
           &           &    20090203 &    -4.0  &  1.50  \\
           &           &    20100301 &    -7.6  &  1.43  \\
GJ 1111    & M6.0Ve    &    20100305 &    -8.0  &  1.47  \\
\hline						   						       
LHS 2090   & M6.0Ve    &    20020402 &   -15.8  &  1.16  \\
           &           &    20061209 &    -7.9  &  1.33  \\
\hline						   						       
LHS 2206   & M4.0Ve    &    20020401 &    -4.4  &  1.14  \\
           &           &    20061209 &    -3.3  &  1.28  \\
\hline						   						       
LHS 292    & M6.5Ve    &    19901123 &    -0.8  &  1.37  \\
           &           &    20031207 &    -2.4  &  1.43  \\
           &           &    20060531 &    -9.5  &  1.36  \\
\hline						   						       
GJ 402     & M3.0V     &    20090203 &    -0.2  &  1.26  \\
           &           &    20100305 &     0.0  &  1.28  \\
           &           &    20110517 &     0.3  &  1.21  \\
\hline						   						       
GJ 406     & M5.0Ve    &    19951203 &   -16.1  &  1.37  \\
           &           &    20030717 &   -11.6  &  1.33  \\
           &           &    20090505 &    -8.0  &  1.38  \\
\hline						   						       
GJ 447     & M4.0V     &    19930315 &     1.0  &  1.21  \\
           &           &    20040608 &     0.2  &  1.23  \\
           &           &    20090505 &    -0.1  &  1.26  \\
\hline						   						       
GJ 1154    & M4.5Ve    &    20040313 &    -6.6  &  1.29  \\
\hline						   						       
GJ 1156    & M4.5Ve    &    19930314 &    -5.6  &  1.25  \\
           &           &    20040608 &    -6.5  &  1.27  \\
\hline						   						       
GJ 486     & M4.0V     &    20040312 &     0.3  &  1.16  \\
\hline						   						       
GJ 493.1   & M4.5Ve    &    20040312 &    -4.9  &  1.27  \\
\hline						   						       
GJ 555     & M3.5V     &    19930314 &     0.6  &  1.18  \\
           &           &    20040608 &     0.5  &  1.20  \\
\hline						   						       
LHS 3003   & M7.0Ve    &    19930317 &   -28.5  &  1.29  \\
           &           &    20040311 &    -7.5  &  1.40  \\
           &           &    20060527 &   -19.4  &  1.35  \\
           &           &    20090505 &     0.0  &  1.32  \\
           &           &    20110727 &     2.4  &  1.00  \\
\hline						   						       
GJ 609     & M4.0V     &    20090506 &    -0.2  &  1.16  \\
           &           &    20110727 &     0.1  &  1.16  \\
\hline						   						       
GJ 628     & M3.0V     &    19930315 &     0.9  &  1.13  \\
           &           &    20040608 &     0.0  &  1.17  \\
\hline						   						       
GJ 643     & M3.0V     &    20040608 &     0.0  &  1.19  \\
\hline						   						       
GJ 644C    & M7.0Ve    &    19950812 &   -11.6  &  1.44  \\
           &           &    20060525 &    -7.1  &  1.42  \\
           &           &    20060526 &   -10.4  &  1.40  \\
\hline						   						       
GJ 1207    & M3.5Ve    &    20020402 &    -3.9  &  1.11  \\
           &           &    20060526 &    -4.9  &  1.17  \\
           &           &    20090505 &    -7.2  &  1.27  \\
\hline						   						       
GJ 699     & M3.5V     &    19950814 &     0.0  &  1.22  \\
           &           &    20041001 &     0.0  &  1.21  \\
           &           &    20060527 &     0.6  &  1.17  \\
\hline						   						       
GJ 1224    & M4.0Ve    &    19931101 &    -4.6  &  1.25  \\
           &           &    20040608 &    -4.3  &  1.26  \\
\hline						   						       
GJ 1230B   & M4.5Ve    &    19931101 &    -0.1  &  1.26  \\
           &           &    20040930 &    -1.2  &  1.29  \\
\hline						   						       
GJ 729     & M3.5Ve    &    19931101 &    -2.4  &  1.20  \\
GJ 729     & M3.5Ve    &    20040929 &    -2.5  &  1.25  \\
\hline						   						       
GJ 752B    & M8.0Ve    &    19950812 &    -2.6  &  1.31  \\
           &           &    20030715 &    -6.1  &  1.33  \\
           &           &    20040930 &    -6.2  &  1.29  \\
           &           &    20060525 &    -6.9  &  1.27  \\
\hline						   						       
GJ 1235    & M4.0Ve    &    20031011 &    -0.5  &  1.24  \\
           &           &    20060526 &     0.4  &  1.21  \\
\hline						   						       
GJ 1256    & M4.0V     &    20031011 &    -0.2  &  1.27  \\
\hline						   						       
LP 816-060 & M3.0V     &    20020401 &     0.5  &  1.05  \\
           &           &    20031207 &     0.1  &  1.19  \\
\hline						   						       
G 188-038  & M3.5Ve    &    20031012 &    -5.8  &  1.20  \\
\hline						   						       
LHS 3799   & M4.5Ve    &    20031207 &    -3.7  &  1.30  \\
           &           &    20060525 &    -3.8  &  1.31  \\
           &           &    20060531 &    -4.3  &  1.27  \\
\hline						   						       
LP 876-010 & M4.0Ve    &    20041002 &    -6.5  &  1.24  \\
           &           &    20060525 &    -4.0  &  1.21  \\
           &           &    20090726 &    -3.3  &  1.23  \\
\hline						   						       
GJ 896A    & M3.5Ve    &    19931031 &    -6.2  &  1.17  \\
           &           &    20041002 &    -5.4  &  1.17  \\
           &           &    20110517 &    -6.9  &  1.17  \\
\hline						   						       
GJ 896B    & M4.0Ve    &    19931031 &    -4.9  &  1.20  \\
           &           &    20110517 &    -6.6  &  1.17  \\
\hline						   						       
GJ 1286    & M5.0Ve    &    19931031 &    -0.7  &  1.32  \\
           &           &    20041001 &    -0.7  &  1.35  \\
\enddata

\end{deluxetable}


\begin{deluxetable}{cccc}
\tabletypesize{\footnotesize}
\tablecolumns{14}
\tablewidth{0pt}
\tablecaption{Radial Velocity Measurements }
\tablehead{
\colhead{Name}  & \colhead{HJD $-$ 2,400,000} &\colhead{Radial Velocity}  &\colhead{SNR}  \\ 
                &                 &\colhead{m s$^{-1}$}       &               }
\startdata
G 99$-$049                      &           54787.56*  & 30171$\pm$ 92  &  200  \\
		       		& 	    54788.44   & 30077$\pm$ 96  &  258  \\
	  			&           54790.51*  & 30171$\pm$ 87  &  254   \\
  				& 	    54791.47*  & 30132$\pm$ 118  &  202  \\
				& 	    55146.52   & 30244$\pm$ 122  &  119  \\
		       		& 	    55149.50   & 30352$\pm$ 105  &  140  \\
		       		& 	    55151.53   & 30317$\pm$ 102  &  215  \\
		       		& 	    55154.56   & 30057$\pm$ 90  &  208   \\
		       		& 	    55638.31   & 30144$\pm$ 95  &  176  \\
		       		& 	    55641.23   & 30130$\pm$ 104  &  145   \\
		       		& 	    55642.27   & 30294$\pm$ 86  &  245   \\

\hline 

GJ 300                          &           54790.60* & 9064$\pm$ 84  &   199  \\     			 	       
               		      	&           54791.65* & 9198$\pm$ 97  &   103  \\				       
  				&   	    54963.27 & 8882$\pm$ 85  &   178   \\				       
          			&           54964.29 & 9237$\pm$ 109  &    89   \\				       
          		        &           55151.64 & 8922$\pm$ 96  &   118   \\				       
	                     	&           55154.63 & 9061$\pm$ 83 &   184   \\				       
	                     	&           55317.25 & 9047$\pm$ 83  &   190   \\				       
        	             	&           55321.30 & 8994$\pm$ 81  &   181   \\				       
	                     	&           55644.36 & 9042$\pm$ 84  &   176   \\
				&	    56671.60 & 8981$\pm$ 85  &   154    \\				       
\hline 	             	              	      	         					   
GJ 406	                     	&           54962.33 & 19429$\pm$ 81  &  212   \\			   
	                     	&           54963.38 & 19415$\pm$ 83  &  193   \\			   
			   	&           54964.38 & 19332$\pm$ 89  &  148   	\\			   
			      	&           55320.41 & 19759$\pm$ 81  &  199    	\\			   
		           	&           55321.40 & 19653$\pm$ 80  &  204    \\			   
	                     	&           55323.41 & 19548$\pm$ 84  &  166   	\\
	                     	&           55330.33 & 19462$\pm$ 94  &   87       \\	   
	                     	&           55331.30 & 19465$\pm$ 80  &  239  	\\
				&	    55335.34 & 19546$\pm$ 80  &  162    \\
				&	    55637.46 & 19731$\pm$ 80  &  257     \\
    				&	    55643.32 & 19625$\pm$ 79  &  180      \\
				&	    55645.39 & 19489$\pm$ 80  &  273      \\
					    			   
\hline 	             	              	      	            							       
GJ 555	                     	&           54959.43 & $-$1417$\pm$ 84  &       214    	\\		       
                    	 	&           54960.44 & $-$1364$\pm$ 98  &	116    	\\		       
  				&      	    54961.41 & $-$1365$\pm$ 82  &	197    \\		       
         			&           54962.44 & $-$1433$\pm$ 98  &	238   	\\		       
		 	      	&           54963.47 & $-$1481$\pm$ 99  &	124   \\		       
	                     	&           54964.50 & $-$1376$\pm$ 84  &	166   	\\		       
	                     	&           55320.52 & $-$1467$\pm$ 81  &	199   	\\
				&	    55355.44 & $-$1389$\pm$ 78  &	226     \\       
\hline 	             	              	    	        	         							       
GJ 628	                     	&           54959.40 & $-$20969$\pm$ 89  & 139   \\				       
		               	&           54960.46 & $-$21135$\pm$ 96  & 132    \\	
				&	    54962.46 & $-$21058$\pm$ 80  & 168    \\			       
			  	&  	    54963.47 & $-$21100$\pm$ 89  & 181   \\				       
			      	&   	    54964.51 & $-$20980$\pm$ 91  & 149    \\				       
			      	&           55320.53 & $-$21257$\pm$ 87  & 178   \\				       
	                     	&           55331.36 & $-$21205$\pm$ 77  & 258   \\				         
	                     	&           55335.45 & $-$21180$\pm$ 83  & 180   \\	
				&	    56174.24 & $-$21095$\pm$ 79  & 229    \\			       	       		  	      	       	        	 

GJ 729	                     	&           54959.56  & $-$10109$\pm$ 84 &   242   \\		       
	                  	&           54960.55  & $-$10178$\pm$ 82 &   257  \\		       
			 	&           54961.54  & $-$10236$\pm$ 97 &   198  	\\		       
		                &           54962.63  & $-$10127$\pm$ 109 &   275   	\\		       
			      	&           55335.51  & $-$10419$\pm$ 81 &   274  	\\		       
	                     	&           55392.46* & $-$10397$\pm$ 81 &   228   	\\		       
\hline 	             	              	       	            							       
GJ 1002	                     	&           54787.41* & $-$39770$\pm$ 85 &    132  \\			       
	                     	&           54788.35  & $-$39886$\pm$ 101 &    162  \\			       
			  	&   	    54790.37* & $-$39768$\pm$ 90 &    207 \\			       
		          	&           55146.44  & $-$39912$\pm$ 95 &    177  \\
			      	&           55150.38  & $-$39976$\pm$ 86 &    113  \\			       
			      	&           55151.39  & $-$39913$\pm$ 84 &    150 \\			       
\hline 				      
GJ 1065	   	                  &           56259.53* & $-$9118$\pm$ 87 &   162    \\
	                          &           56260.55  & $-$8984$\pm$ 92 &   135     \\
				  &           56270.46  & $-$9163$\pm$ 95 &   145   	\\
		                  &           56271.43  & $-$9122$\pm$ 137 &   82     \\		       
			          &           56272.45  & $-$9110$\pm$ 93 &   151  \\		       
        	         	
\hline                                                          						       
	             	              	       	             							       
GJ 1224	   	                 &           54961.51 & $-$32425$\pm$ 127 &   84      \\
                                 &           54962.52 & $-$32609$\pm$ 135 &   122    \\
			         &           54963.52 & $-$32811$\pm$ 96 &   163   \\
		                 &           54964.59 & $-$32559$\pm$ 164 &   100    \\		       
		                 &           55320.56 & $-$32560$\pm$ 90 &   157  \\		       
        	         	 &   	     55321.52 & $-$32704$\pm$ 90 &   160   	\\		       
        	                 &           55335.49 & $-$32789$\pm$ 87 &   175   	\\		       
\hline 	       	 	  	      	                							       
GJ 1286	   	                 &           54787.37* & $-$40687$\pm$ 125 &        74  	\\		       
	   	                 &           55146.37  & $-$40871$\pm$ 100 &	  126 	\\		       
				 &   	     55149.29  & $-$40897$\pm$ 126 &	  76   	\\                       
	 	                 &           55151.36  & $-$40850$\pm$ 98  &	  109  	\\		       
			         &           55154.40  & $-$40714$\pm$ 100 &	  138 	\\		       

\hline 	         		    	      	                							       
LHS 1723	              	&           54787.51* & 42533$\pm$ 94 &   200   \\		       
	   	                &           54788.40  & 42559$\pm$ 93 & 	 149   \\		       
			 	&           54790.47* & 42497$\pm$ 987 & 	 188  	\\		       
		           	&           55149.47  & 42327$\pm$ 90 &	 157   	\\		       
		        	&           55151.50  & 42377$\pm$ 86 &	 180  	\\		       
	                     	&           55154.49  & 42287$\pm$ 82 &	 217   \\		       
\hline 	             	              	      	            						       
LHS 3799                   	&           54790.28* & $-$1643$\pm$ 104 &         158   \\		       
			   	&           55146.32  & $-$1568$\pm$ 92 &	 129   	\\		       
		          	&           55149.24  & $-$1612$\pm$ 83  &	 188   	\\		       
			      	&           55150.27  & $-$1651$\pm$ 85  &	 173   	\\	       
			      	&           55151.27  & $-$1596$\pm$ 90  &	 143   \\		       
                             	&           55392.58  & $-$1779$\pm$ 88  &	 193    \\      
\enddata

\label{table:rv}

\tablenotetext{*}{No telluric standards were observed on this night.  }

\end{deluxetable}

\begin{deluxetable}{lcrccccc}
\tabletypesize{\footnotesize}
\tablecolumns{14}
\tablewidth{0pt}
\tablecaption{Spectroscopic Results}
\tablehead{
\colhead{Name}   & \colhead{Model T$_{eff}$} &\colhead{Abs. RV$^a$}             &\colhead{N}                 & \colhead{$\Delta$ time}   &\colhead{$\sigma$$_{RV}$} &\colhead{ $v$sin$i$$^b$ }  \\ 
                               &  \colhead{(K) }           &\colhead{(m s$^{-1}$)}       &  \colhead{($\#$ nights) }  & \colhead{(days)}       &\colhead{(m s$^{-1}$)}    &\colhead{(km s$^{-1}$)}   }
\startdata
G 99-49    & 3200     & 30190     & 11 &  855         &  98   &        5.8$\pm$0.3                \\
GJ 300     & 3200     & 9043      &  10& 1881         &110    &        0.9$\pm$0.7$^1$                \\     
GJ 406 	   & 3000     & 19537     &  12&  638         &132    &        1.6$\pm$0.7$^1$                 \\      
GJ 555     & 3200     & $-$1413   &  8 &  396         &47     &        0.6$\pm$0.7$^1$                 \\     
GJ 628     & 3000     &$-$21109   &  9 & 1215         &  98   &        1.2$\pm$0.7$^1$                 \\
GJ 729     & 3200     &$-$10244   &  6 &  433         &125    &        3.8$\pm$0.6                \\
GJ 1002	   & 3000     &$-$39871   &  6 &  368         & 84    &        1.4$\pm$0.3$^1$                 \\
GJ 1065	   & 3200     &$-$9099   &  5 &   13         &   68  &        1.0$\pm$0.8$^1$                 \\
GJ 1224    & 3000     &$-$32673   &  7 &  374         &  139  &        1.4$\pm$0.4$^1$                 \\
GJ 1286    & 3000     &$-$40803  &  5 &  367         & 96    &        0.7$\pm$0.9$^1$                \\
LHS 1723   & 3000     & 42430     &  6 &  367         &  115  &        1.0$\pm$0.8$^1$                 \\
LHS 3799   & 3000     &$-$1641    &  6 &  602         &  74   &        1.5$\pm$0.5$^1$                 \\				      
\enddata
\tablenotetext{a}{The error on the absolute RV measurements is estimated to be $\sim$ 100 m s$^{-1}$, based on systematic uncertainties.}
\tablenotetext{b}{The $v$sin$i$ values reported are the values used in the best fit model.  Since values below 3 km s$^{-1}$ cannot be confidently measured at our resolution, we adopt a $v$sin$i$ upper limit value of 3 km s$^{-1}$ for these stars.}

\label{table:rvsum}

\end{deluxetable}

\begin{landscape}
\begin{deluxetable}{lccccccccrrrrrc}
\tabletypesize{\footnotesize}
\tablecolumns{15}
\tablewidth{0pt}
\setlength{\tabcolsep}{0.055in} 
\tablecaption{Astrometric Results}
\tablehead{
\colhead{Name} & \colhead{R.A.} & \colhead{Decl.} & \colhead{Filter} &\colhead{N$_{sea}$$^a$}    &\colhead{N$_{frm}$ }     &\colhead{Coverage$^b$} &\colhead{Years$^b$} &\colhead{ N$_{REF}$ }  &\colhead{$\pi$ (rel)}  &\colhead{$\pi$ (corr) } &\colhead{$\pi$ (abs)} &\colhead{ $\mu$  }   &\colhead{P.A.} &\colhead{ V$_{tan}$ } \\ 
               & \colhead{J2000.0} & \colhead {J2000.0} &                  &                       &                         &                   &                 &                       &\colhead{(mas)}        &\colhead{(mas)}         &\colhead{(mas) }      & \colhead{(mas yr$^{-1}$)} &\colhead{ (deg) }&\colhead{(km s$^{-1}$) } \\
\colhead{(1)}   & \colhead{(2) }   &\colhead{(3)}           &\colhead{ (4)  }        &\colhead{(5) }     &\colhead{ (6)  } &\colhead{(7) }         &\colhead{(8) }           &\colhead{(9) }        &\colhead{(10) }       &\colhead{(11) } &\colhead{ (12)} &\colhead{ (13)} & \colhead{ (14)} & \colhead{ (15)} }
\startdata
G 99-49  & 06 00 03.52  & +02 42 23.6   & V & 14s & 374 & 1999.91-2012.94 &  13.04 & 4 & 191.08$\pm$0.97 &  2.52$\pm$1.57 & 193.60$\pm$1.85 &  307.6$\pm$0.3 &  97.1$\pm$0.10 &   7.5  \\
GJ 300   & 08 12 40.88  & $-$21 33 06.8 & V & 14s & 374 & 1999.91-2012.95 &  13.05 & 8 & 121.37$\pm$0.42 &  1.27$\pm$0.26 & 122.64$\pm$0.49 &  698.6$\pm$0.1 & 178.6$\pm$0.01 &   27.0  \\
GJ 406   & 10 56 28.91  & +07 00 53.2   & R & 12s & 139 & 2000.23-2012.27  &  12.04 & 6 & 413.70$\pm$1.61 &  1.46$\pm$0.17 & 415.16$\pm$1.62 & 4694.5$\pm$0.5 & 236.2$\pm$0.01 &  53.6 \\
GJ 555   & 14 34 16.82  & $-$12 31 10.3 & V & 13s & 193 & 2000.14-2012.26  &  12.12 & 6 & 161.15$\pm$1.45 &  0.58$\pm$0.22 & 161.73$\pm$1.47 &  682.5$\pm$0.4 & 331.0$\pm$0.07 & 20.0  \\
GJ 628   & 16 30 18.07  & $-$12 39 45.4 & V &  10s & 141 & 2003.51-2012.58 &   9.07 & 5 & 229.10$\pm$2.23 & 1.43$\pm$0.17 & 230.53$\pm$2.24 & 1191.5$\pm$0.9 & 185.5$\pm$0.07 & 24.7   \\
GJ 729   & 18 49 49.37  & $-$23 50 10.4 & V & 11s & 124 & 1999.62-2012.75 &  13.13 & 7 & 335.64$\pm$1.30 &  3.95$\pm$0.99 & 339.59$\pm$1.63 &  666.3$\pm$0.4 & 106.8$\pm$0.05 &  9.3  \\
GJ 1002  & 00 06 43.19  & $-$07 32 17.0 & R &  9s &  64 & 2003.77-2012.87 &   9.10 & 4 & 205.09$\pm$3.09 &  2.09$\pm$0.16 & 207.18$\pm$3.09 & 2034.5$\pm$0.9 & 203.8$\pm$0.05 & 46.5   \\
GJ 1065  & 03 50 44.29  & $-$06 05 41.7 & V & 10s & 86 & 2003.95-2012.95  &  9.00 & 5 & 99.98$\pm$1.91 &  1.65$\pm$0.24 & 101.63$\pm$1.93 &  1444.8$\pm$0.7 & 198.9$\pm$0.05 &  67.4  \\
GJ 1224  & 18 07 32.85  & $-$15 57 47.0 & I &  10s& 170 & 2003.52-2012.52 &   9.00 & 7 & 125.04$\pm$0.92 &  1.50$\pm$0.50 & 126.54$\pm$1.05&  702.3$\pm$0.4 & 241.0$\pm$0.06 & 26.1    \\
GJ 1286  & 23 35 10.47  & $-$02 23 20.8 & I &  10s& 135 & 2003.52-2012.88 &   9.36 & 5 & 139.19$\pm$1.07 &  2.28$\pm$0.25 & 141.47$\pm$1.10 & 1141.8$\pm$0.4 & 137.8$\pm$0.04 & 38.3   \\
LHS 1723 & 05 01 57.43  & $-$06 56 46.5 & V & 14s & 258 & 1999.81-2012.75 &  12.94 & 4 & 187.19$\pm$0.76 &  1.47$\pm$0.21 & 188.66$\pm$0.79 &  770.8$\pm$0.3 & 226.3$\pm$0.04 &   19.4  \\
LHS 3799 & 22 23 07.00  & $-$17 36 26.1 & V &  9s & 118 & 2003.52-2012.70  &  9.18  & 6 & 137.70$\pm$1.86 &  0.47$\pm$0.14 & 138.17$\pm$1.87&  769.1$\pm$0.4 & 157.7$\pm$0.06 & 26.4   \\
\enddata
\tablenotetext{a}{Number of seasons (N$_{sea}$) counts observing semesters where a dataset was taken, and denotes if coverage was 'c'ontinuous (more than one night of data in all seasons) or 's'cattered.}
\tablenotetext{b}{‘Coverage’ and ‘Years’ run from the first to last epoch.}
\label{table:ast}

\end{deluxetable}
\end{landscape}

\begin{deluxetable}{lcccc}
\tabletypesize{\footnotesize}
\tablecolumns{14}
\tablewidth{0pt}
\tablecaption{Astrometric Residuals and Errors}
\tablehead{
\colhead{Name}  & \colhead{R.A. Res. Dev.} &\colhead{R.A. Mean Err.}             &\colhead{Decl. Res. Dev.}                 &\colhead{Decl. Mean Err.}  \\ 
                &  \colhead{(mas)  }           &\colhead{(mas) }       &  \colhead{(mas) }  &\colhead{(mas) }     }
\startdata  
G 99-49  &  3.15 & 4.70 & 2.72 & 4.72  \\
GJ 300   & 1.60 & 2.58 & 2.35 & 2.77  \\
GJ 406   & 4.85 & 5.18 & 5.53 & 6.60  \\
GJ 555   &  4.56 & 6.18 & 4.35 & 6.88 \\
GJ 628   & 3.74 & 5.30& 4.15 & 6.87  \\
GJ 729   &  3.25 & 3.43 & 4.41 & 4.90  \\
GJ 1002  & 0.86 & 2.39 & 1.54 & 3.35  \\ 
GJ 1065  &  2.01 & 3.26 & 4.45 & 4.23  \\
GJ 1224  & 2.49 & 4.60 & 2.12 & 5.41 \\   
GJ 1286  & 1.99 & 2.50 & 4.28 & 4.83  \\
LHS 1723 & 2.38 & 3.41 & 2.34 & 3.97  \\
LHS 3799 &  2.02 & 2.67 & 2.07 & 2.65  \\
\enddata
\label{table:astresults}

\end{deluxetable}

\begin{deluxetable}{lccccc|cccc}
\tabletypesize{\footnotesize}
\tablecolumns{14}
\tablewidth{0pt}
\tablecaption{Companion Mass with a 90$\%$ Detection Rate}
\tablehead{
& &\multicolumn{4}{c}{Radial Velocity Mass Limits}  & \multicolumn{4}{c}{Astrometric Mass Limits} \\

\colhead{Name} & \colhead{Stellar Mass} & \colhead{3 d } &\colhead{10 d  }  &\colhead{30 d }      &\colhead{100 d }  &\colhead{2 years}  &\colhead{4 years}      &\colhead{6 years } &\colhead{8 years } \\
              &   \colhead{(M$_{SUN}$)}  &   \colhead{(M$_{JUP}$)}    &   \colhead{(M$_{JUP}$)} &   \colhead{(M$_{JUP}$)} &   \colhead{(M$_{JUP}$)} &   \colhead{(M$_{JUP}$)} &   \colhead{(M$_{JUP}$)} &   \colhead{(M$_{JUP}$)} &   \colhead{(M$_{JUP}$)} }
\startdata  

G 99-49 &0.19 & 1.0  & 1.0 &   2.5  &5.5   & 23.5   & 14.5   & 11.0 & 9.0  \\
GJ 300  &0.20 & 1.0  & 1.0 &   2.0  &2.5   & 15.5   &  9.5   & 7.5  & 6.0  \\
GJ 406  &0.09 & 0.5  & 1.0 &   1.0  &2.0   & 7.0    &  4.5   & 3.5  & 3.0  \\
GJ 555  &0.22 & 1.0  & 1.5 &   4.0  &5.0   & 41.0   & 26.0    & 20.0 & 16.5   \\
GJ 628  &0.26 & 1.0  & 1.5 &   2.5  &5.5   & 19.0   & 11.0    & 8.5  & 7.5  \\
GJ 729  &0.17 & 1.0  & 1.0 &   4.0  &4.5   & 9.5    &  6.0    & 4.5  & 4.0        \\
GJ 1002 &0.11 & 1.0  & 1.0 &   3.0  &8.0   & 9.5    &  6.0    & 4.5  & 4.0  \\
GJ 1065 &0.18 & 1.0  & 3.5 &   4.5  &24.0  & 31.5   & 19.0    & 15.5 & 12.5 \\
GJ 1224 &0.14 & 1.0  & 1.5 &   2.5  &5.5   & 25.5   &  15.5   & 12.0 & 10.0  \\
GJ 1286 &0.11 & 1.5  & 1.0 &   3.0  &8.0   & 11.0   &  6.5    & 5.0  & 4.0  \\
LHS 1723&0.15 & 1.0  & 1.0 &   3.5  &10.0  & 13.0   &  8.5    & 6.5  & 5.5  \\
LHS 3799&0.14 & 1.0  & 1.0 &   3.0  &3.0   & 16.0   & 10.0    & 8.0  & 6.5  \\
\enddata

\label{table:90}

\end{deluxetable}



\begin{figure}
\begin{center}
\begin{subfigure}
                
                \includegraphics[scale=.33,angle=90]{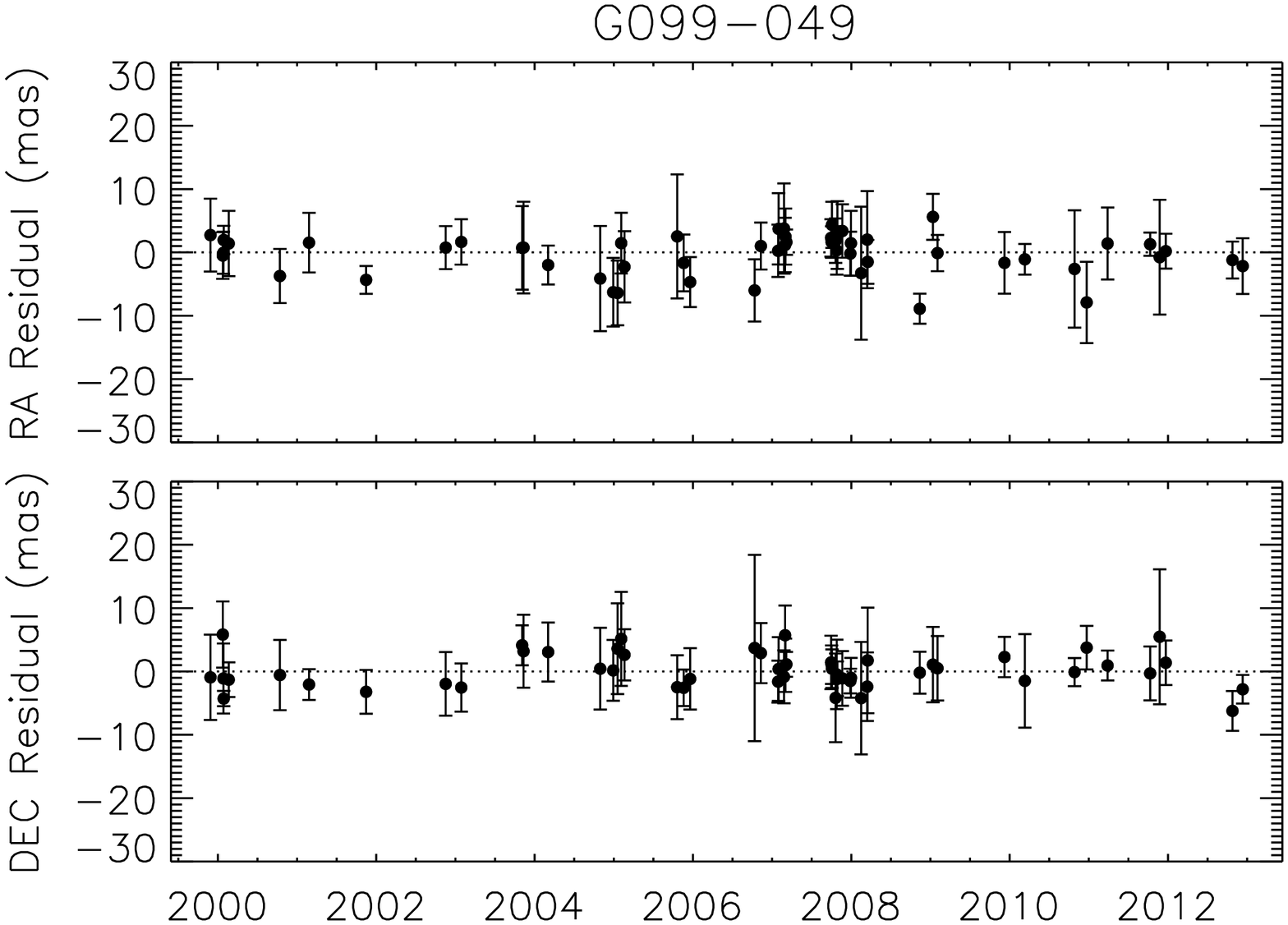}
                \includegraphics[scale=.33,angle=90]{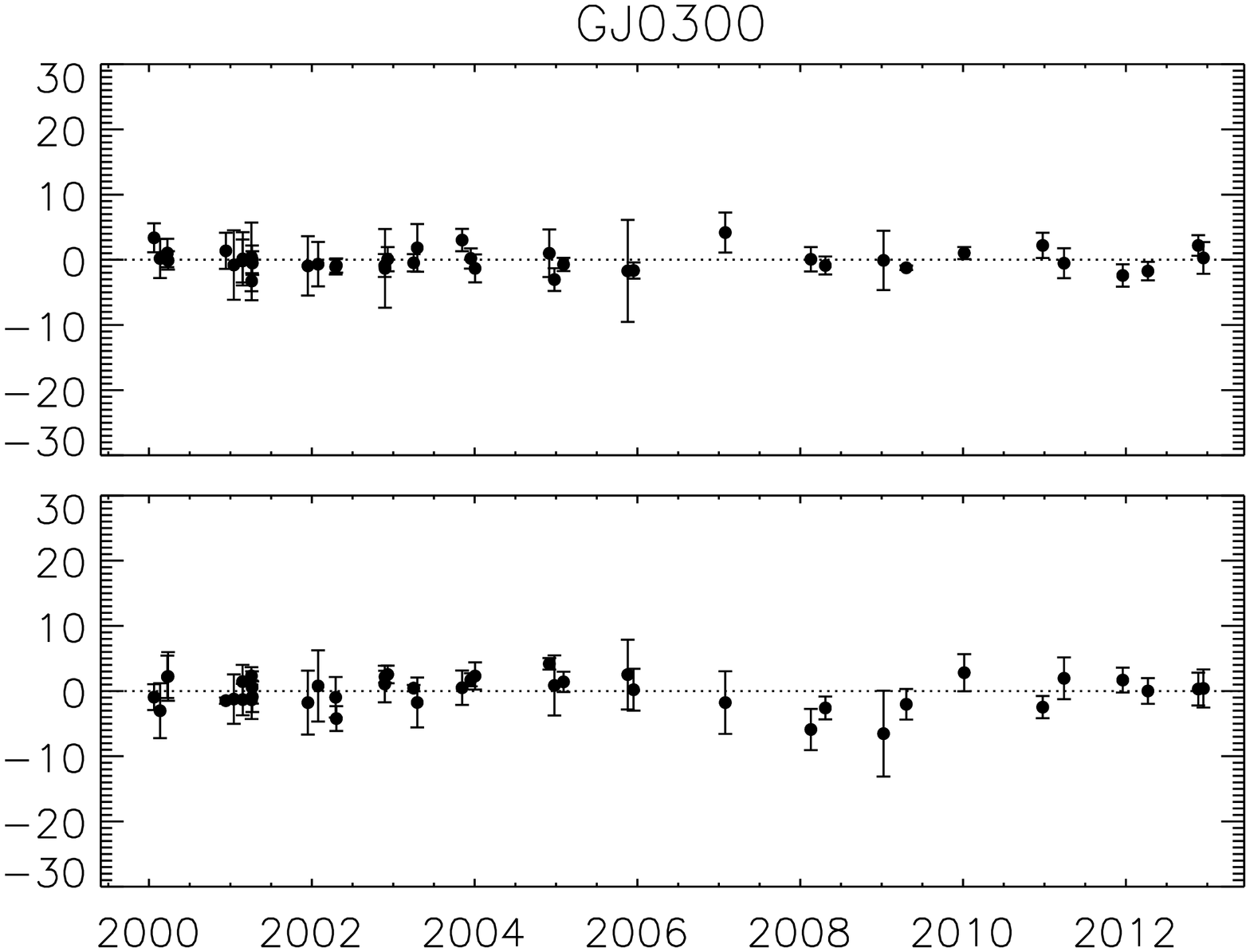}
                \includegraphics[scale=.33,angle=90]{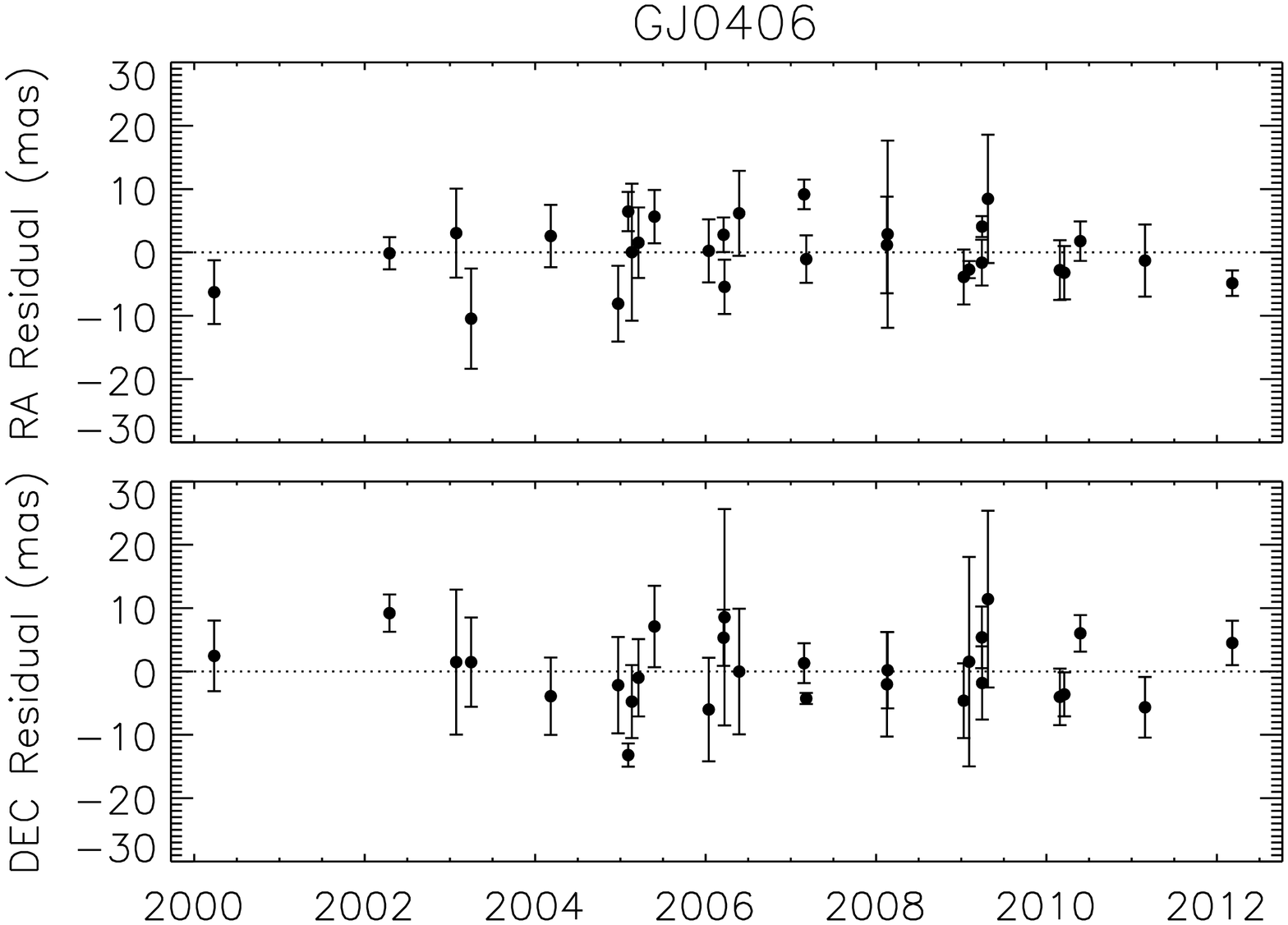}
                \includegraphics[scale=.33,angle=90]{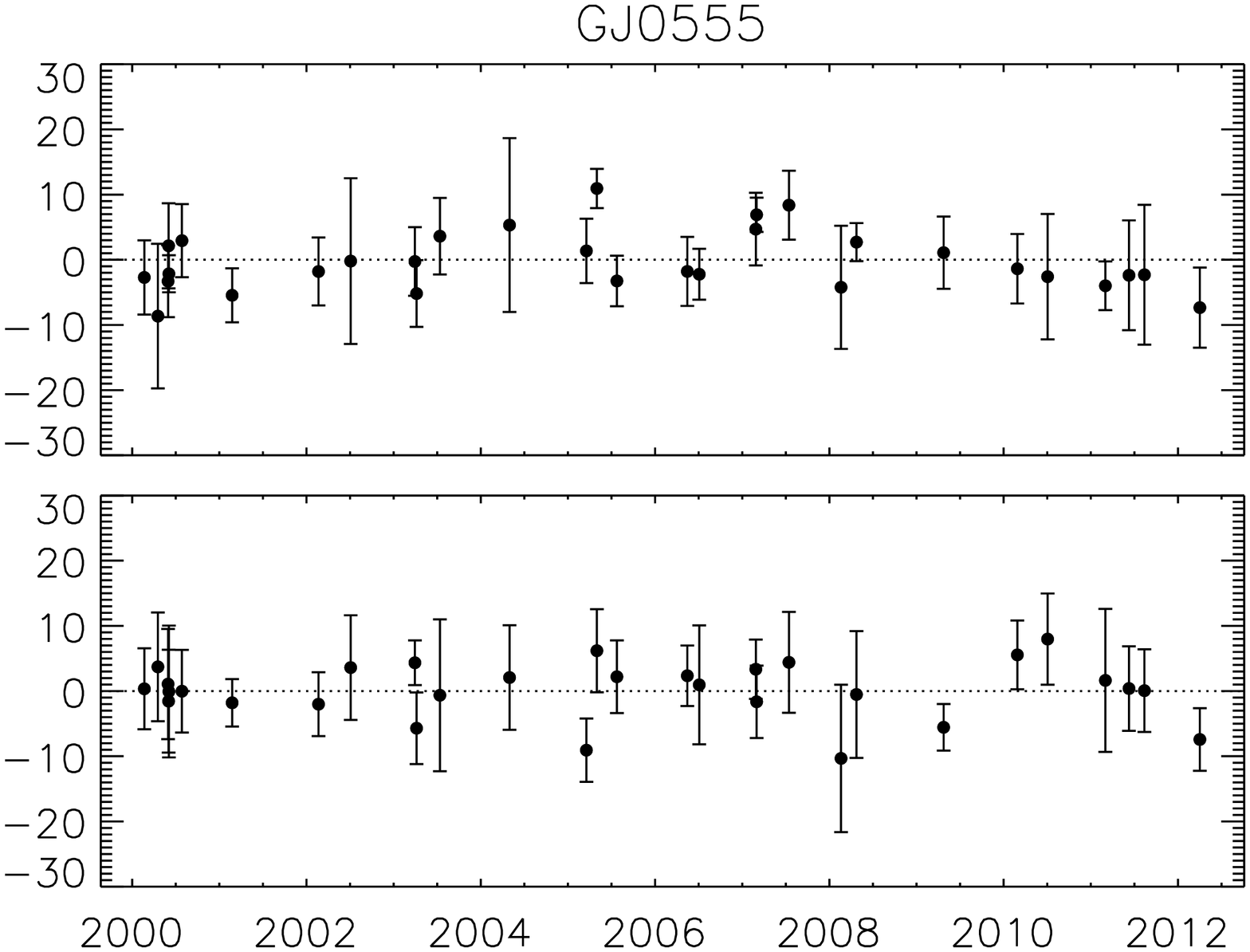}
                \includegraphics[scale=.33,angle=90]{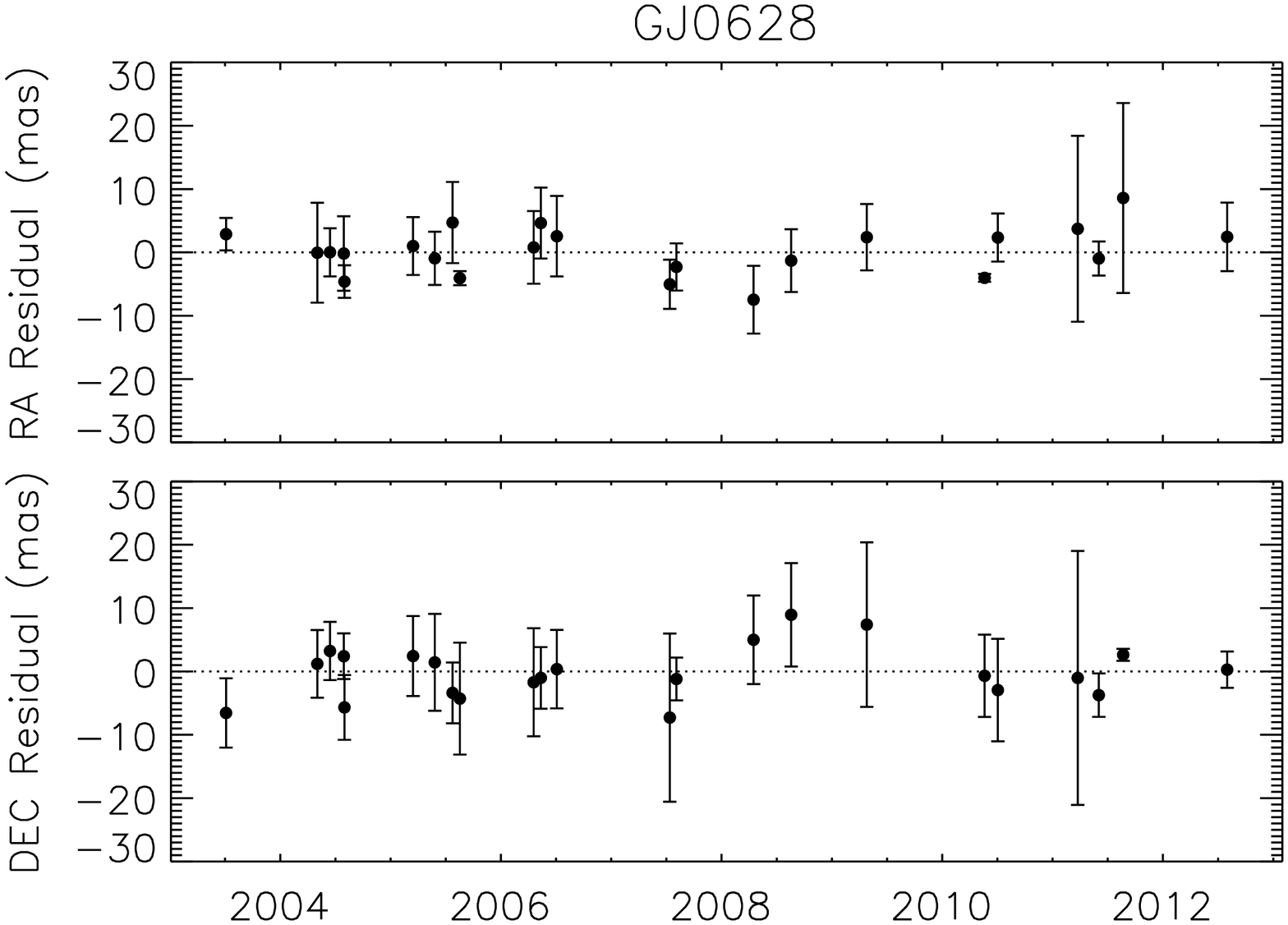}
                \includegraphics[scale=.33,angle=90]{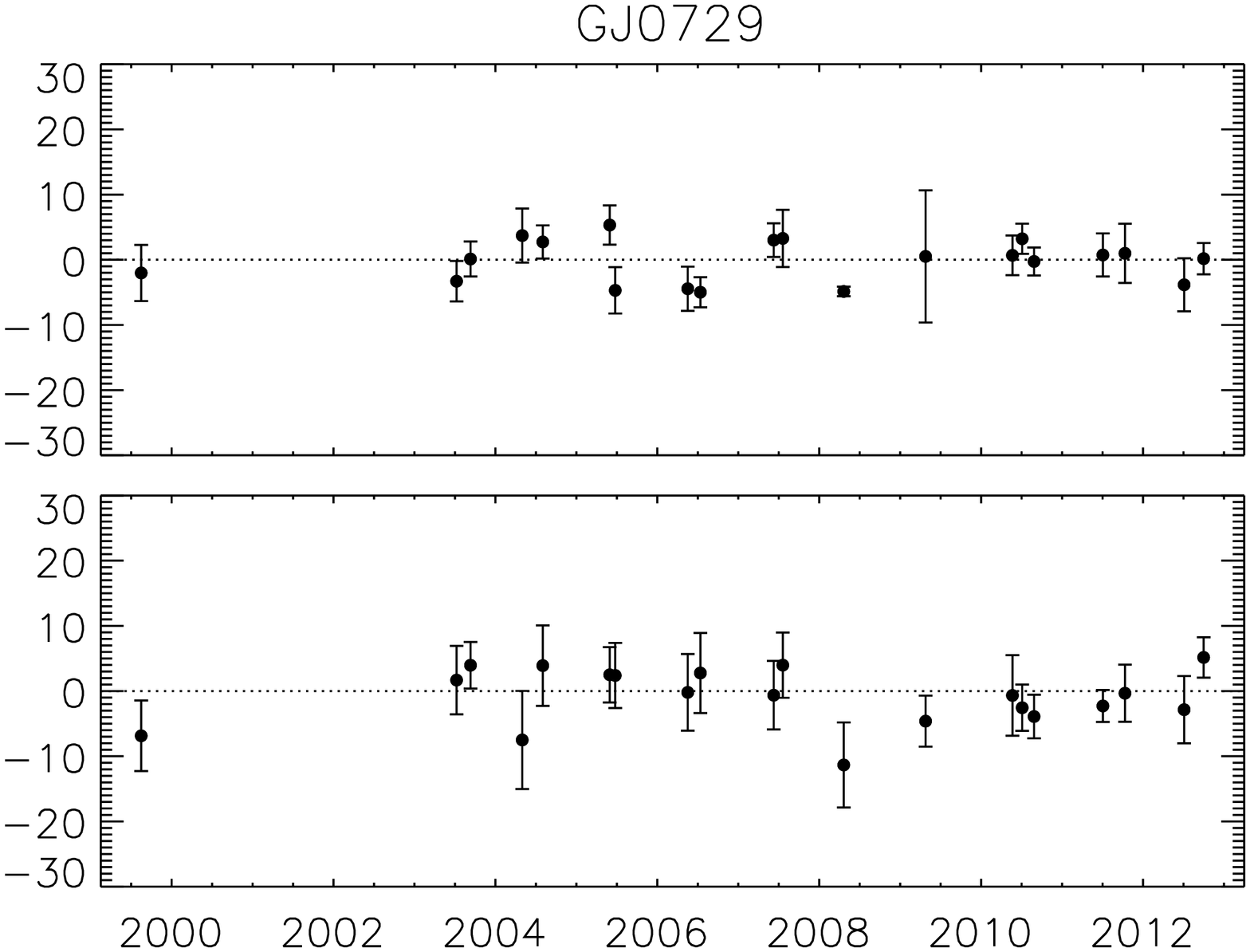}
 \end{subfigure}
\label{fig:astfig}

\end{center}
\end{figure}


\begin{figure}
\begin{center}
\begin{subfigure}
		              
		\includegraphics[scale=.33,angle=90]{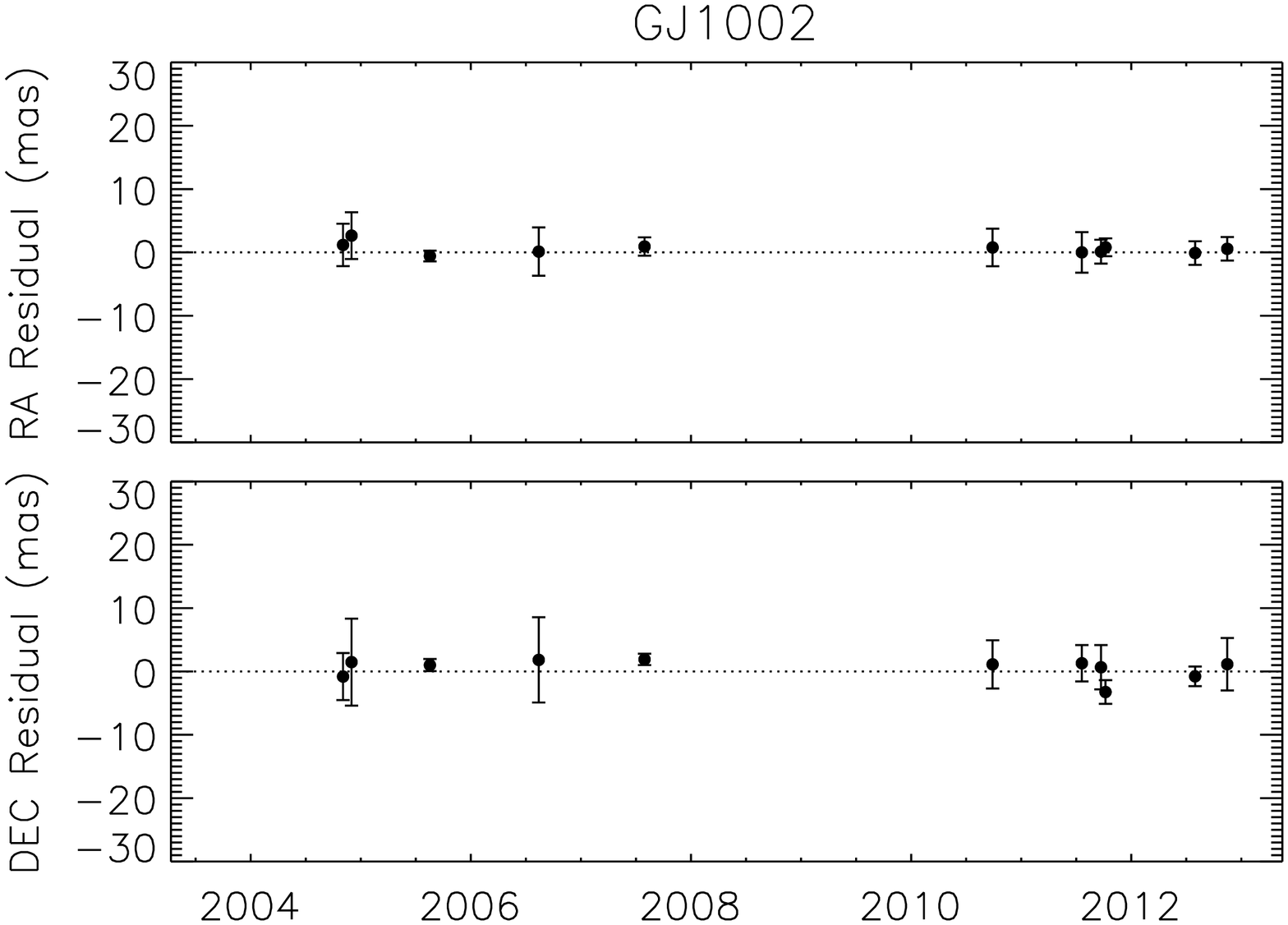}
                \includegraphics[scale=.33,angle=90]{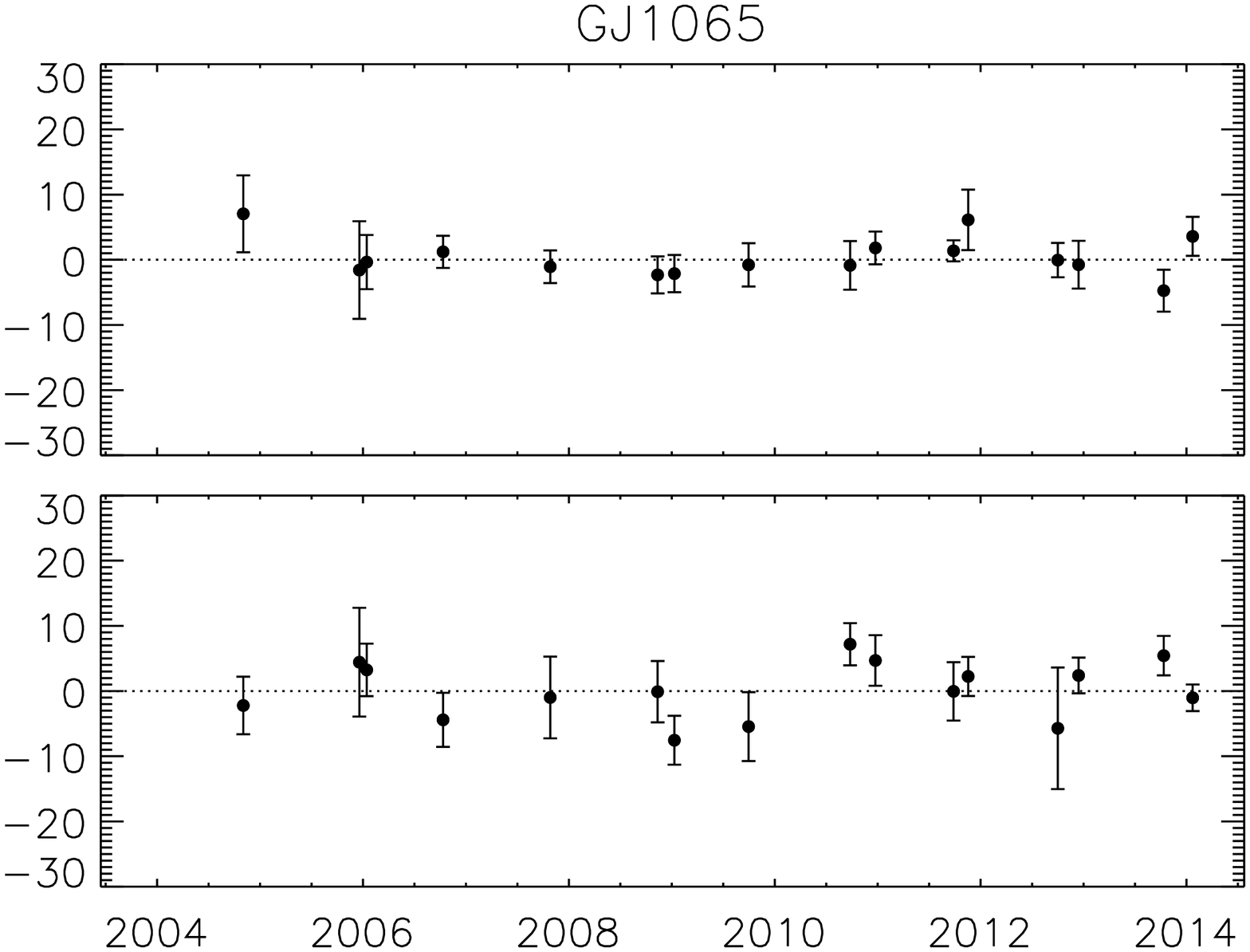}
                \includegraphics[scale=.33,angle=90]{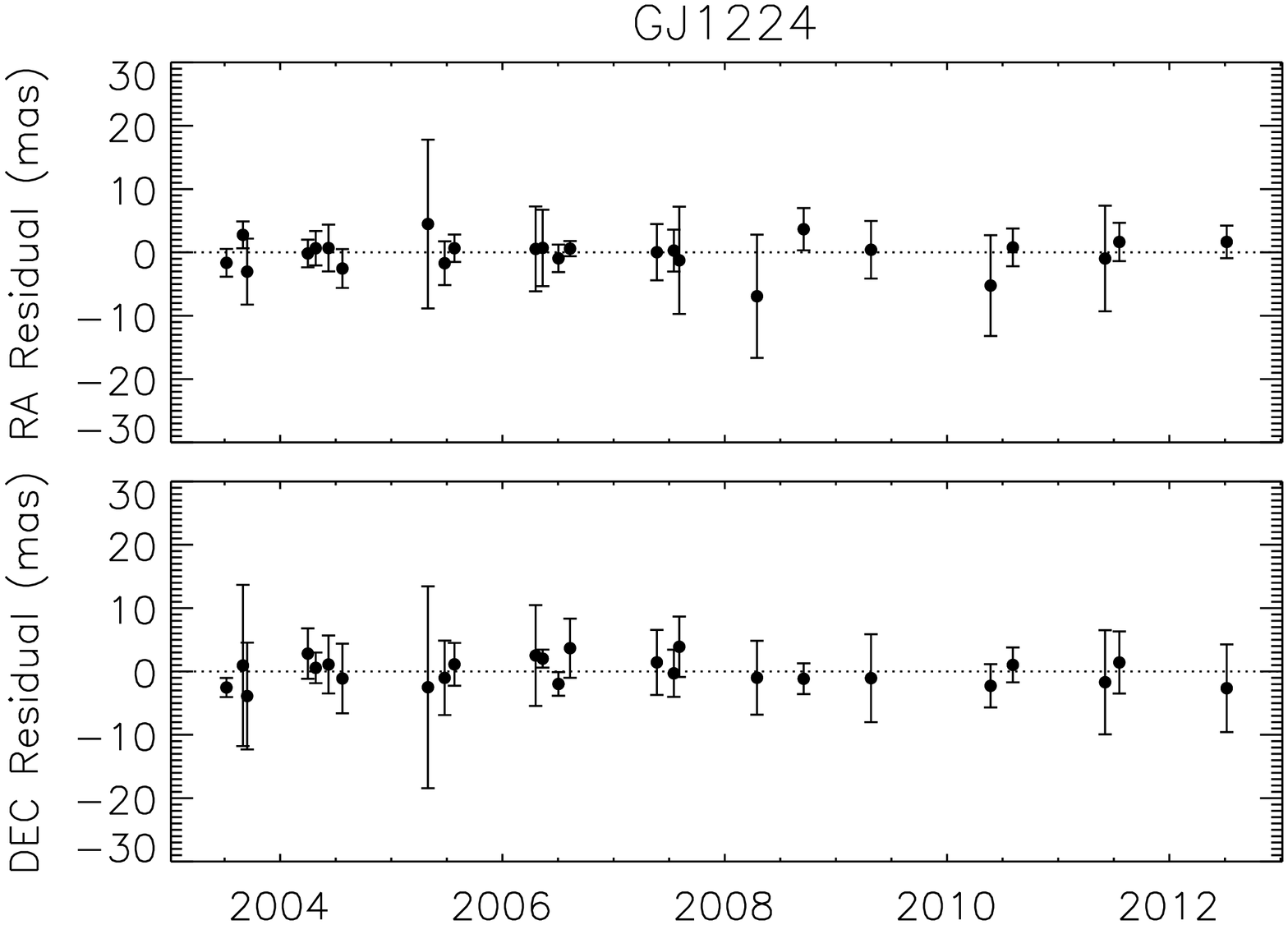}
                \includegraphics[scale=.33,angle=90]{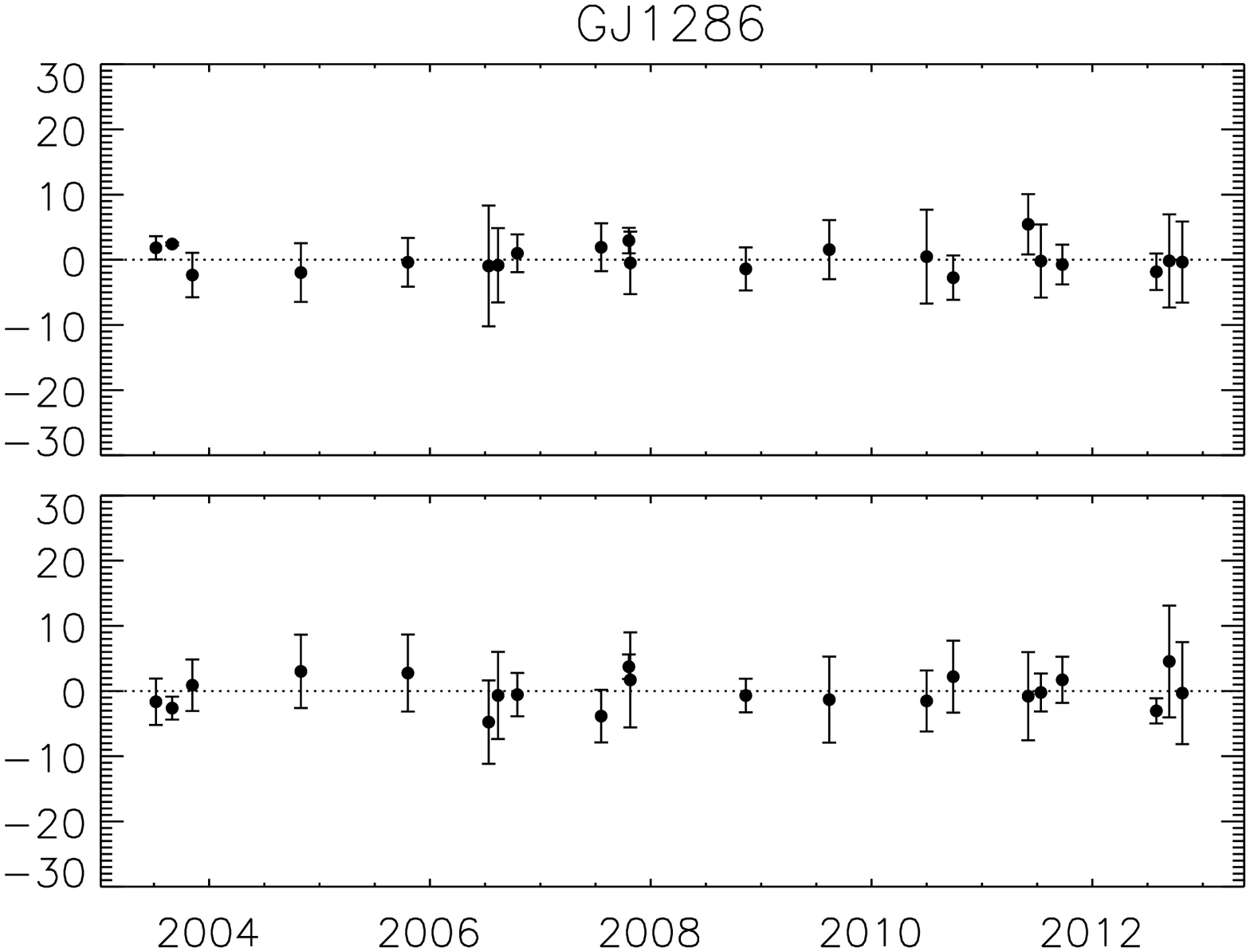} 
                \includegraphics[scale=.33,angle=90]{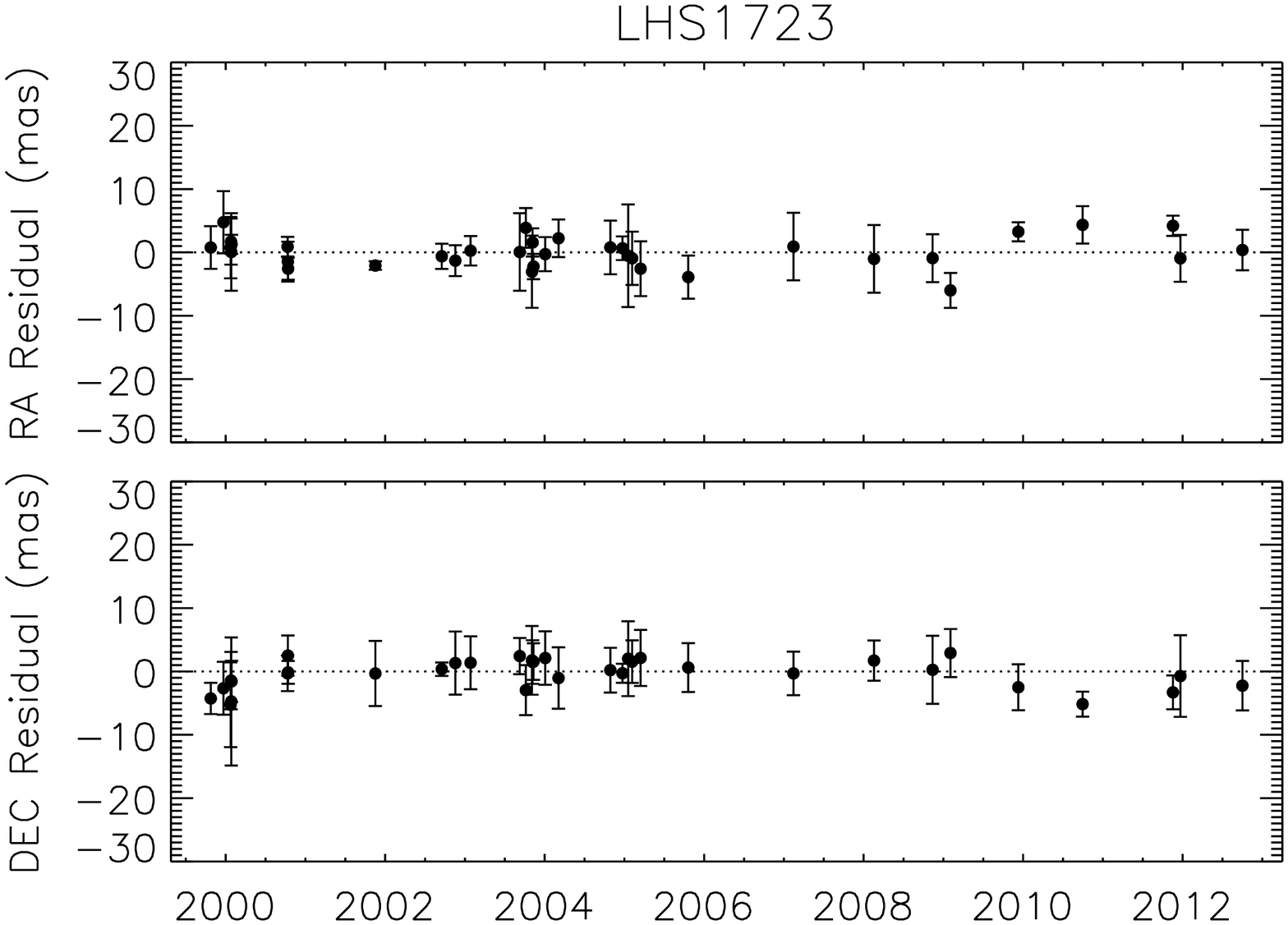} 
		\includegraphics[scale=.33,angle=90]{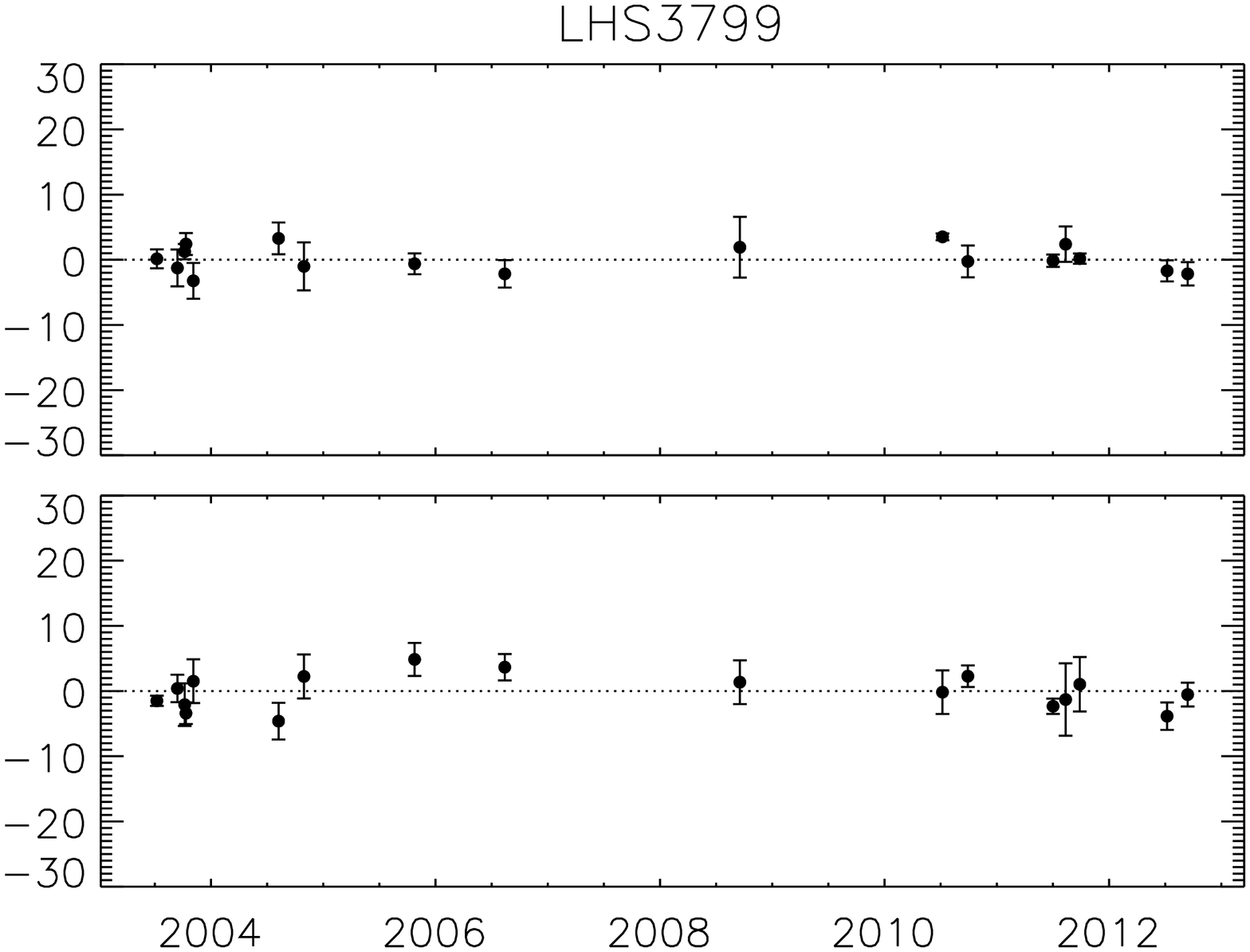}

                          \caption{Relative astrometric measurements are plotted over time for both R.A. and Decl.  After solving for parallactic and proper motion, we see no indications of companions in the astrometric residuals for any of these stars. }

 \end{subfigure}

\end{center}
\end{figure}


\begin{figure}
\begin{center}
\begin{subfigure}
		              
		\includegraphics[scale=.44,angle=0]{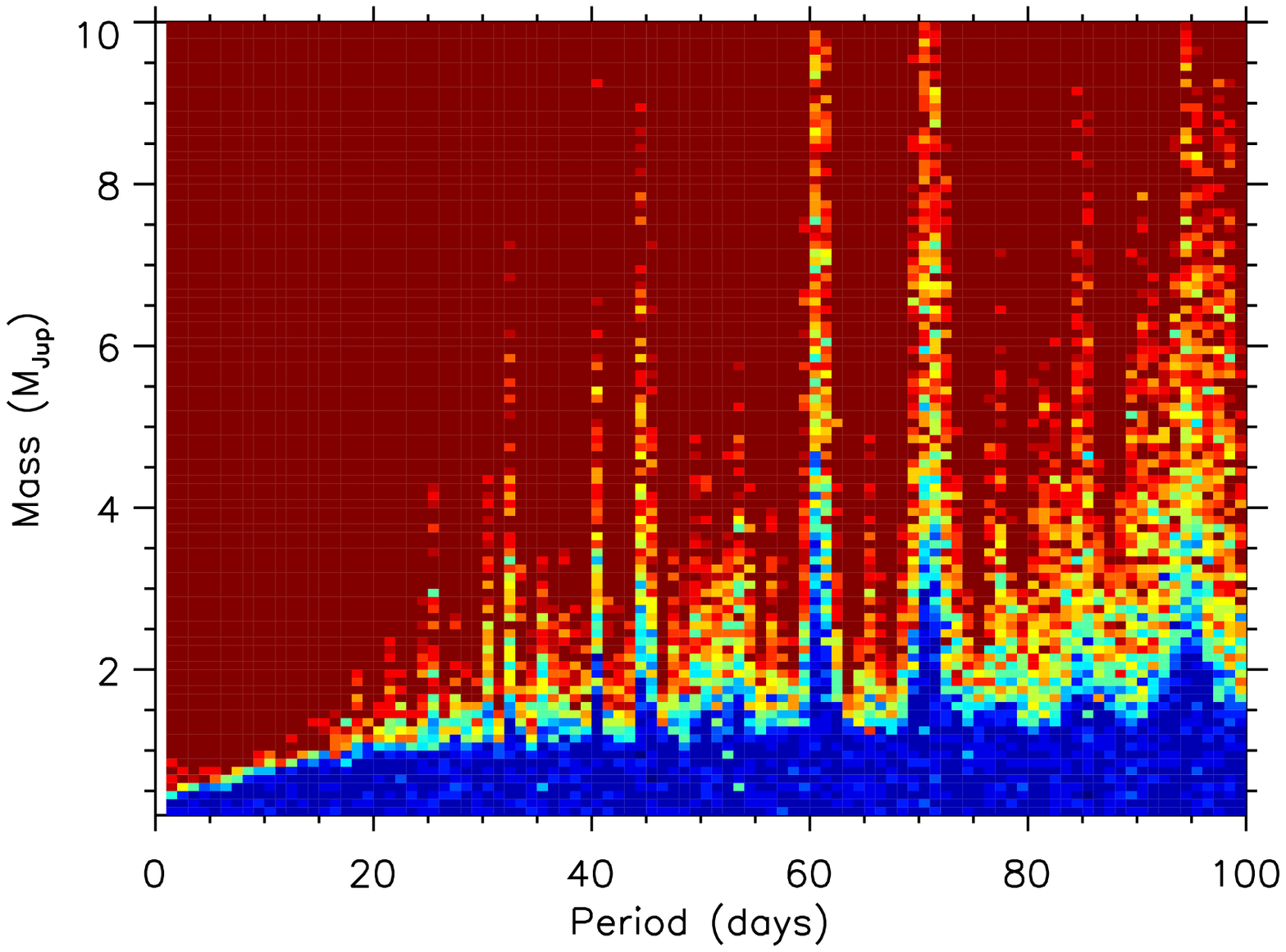}
                \includegraphics[scale=.44,angle=0]{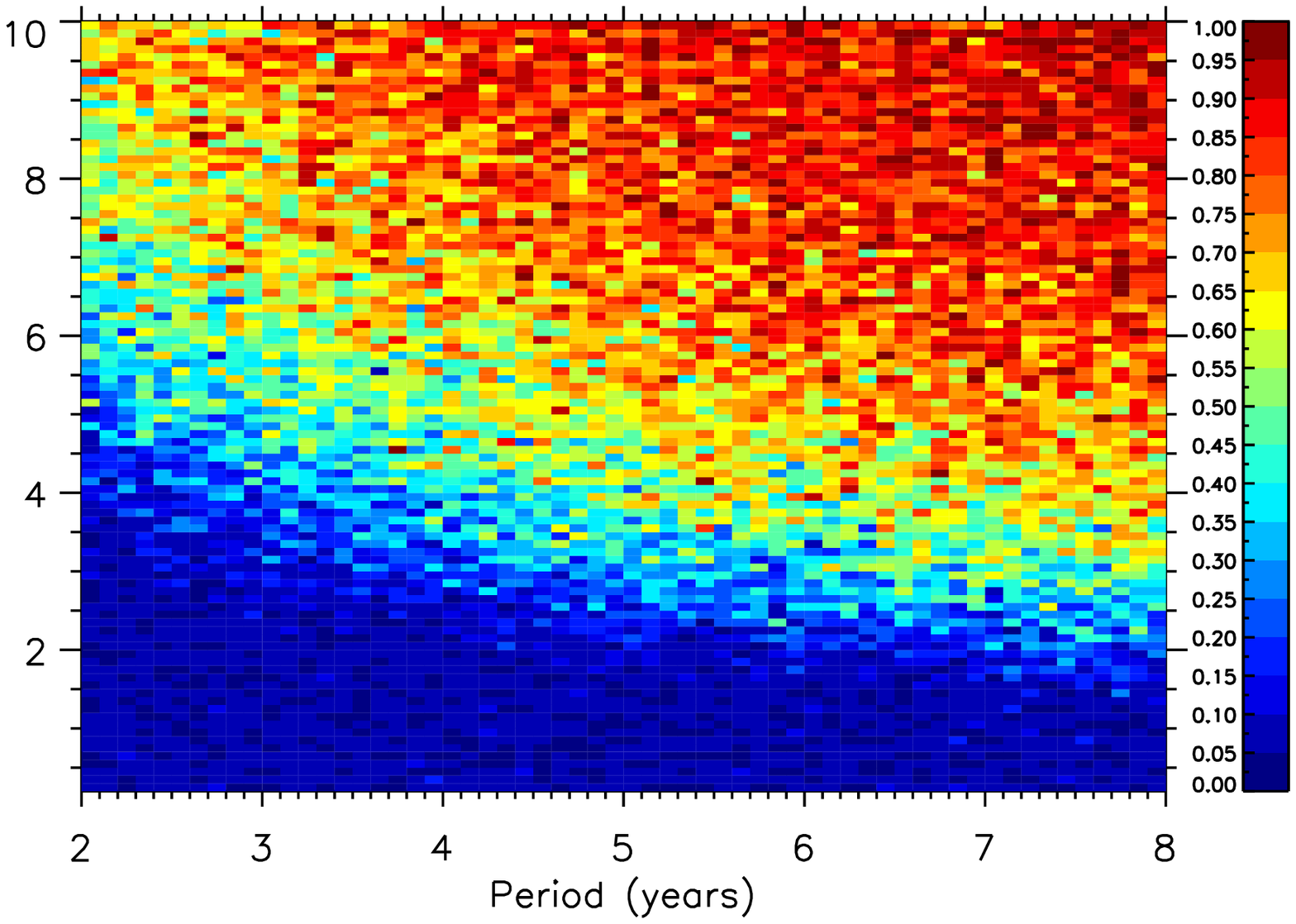}

                          \caption{Fraction of detected companions in a companion mass versus orbital period plot, based on Monte Carlo simulations using the RV data (left panel) and astrometric data (right panel) for G 99-49. }

 \end{subfigure}

\end{center}
\end{figure}



\begin{figure}
\begin{center}
\begin{subfigure}
		              
		\includegraphics[scale=.44,angle=0]{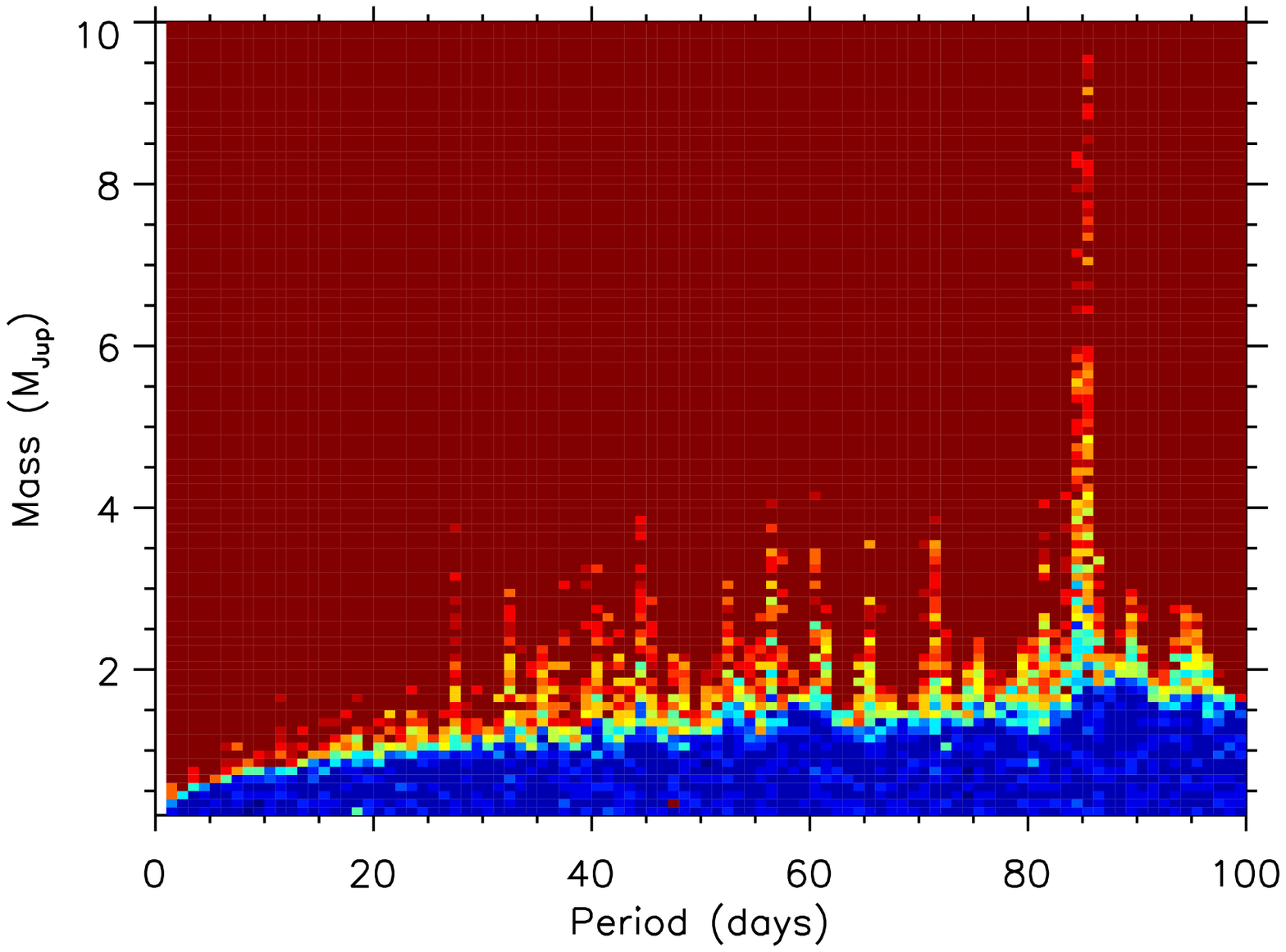}
                \includegraphics[scale=.44,angle=0]{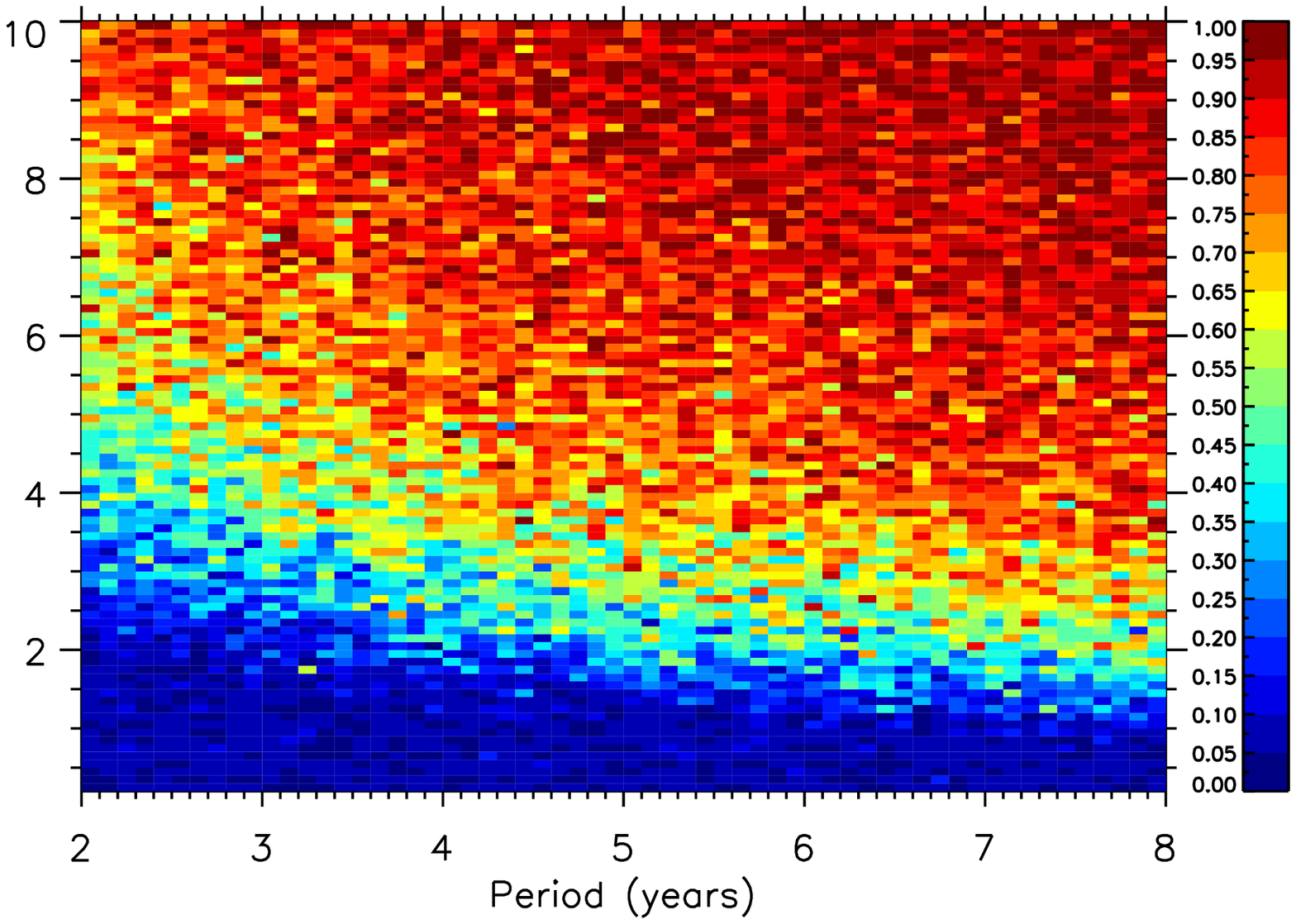}

                          \caption{Fraction of detected companions in a companion mass versus orbital period plot, based on Monte Carlo simulations using the RV data (left panel) and astrometric data (right panel) for GJ 300. }

 \end{subfigure}

\end{center}
\end{figure}


\begin{figure}
\begin{center}
\begin{subfigure}
		              
		\includegraphics[scale=.44,angle=0]{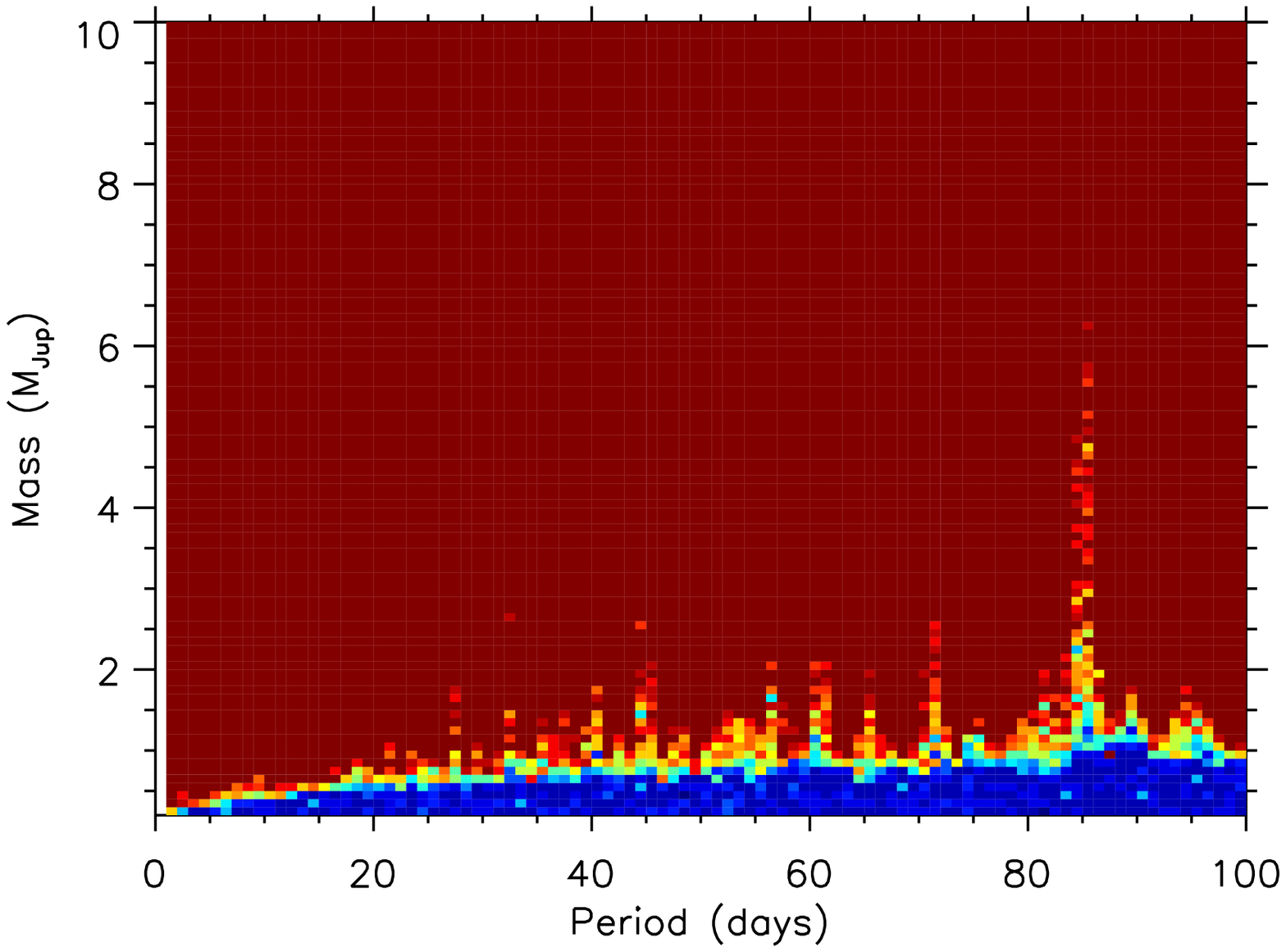}
                \includegraphics[scale=.44,angle=0]{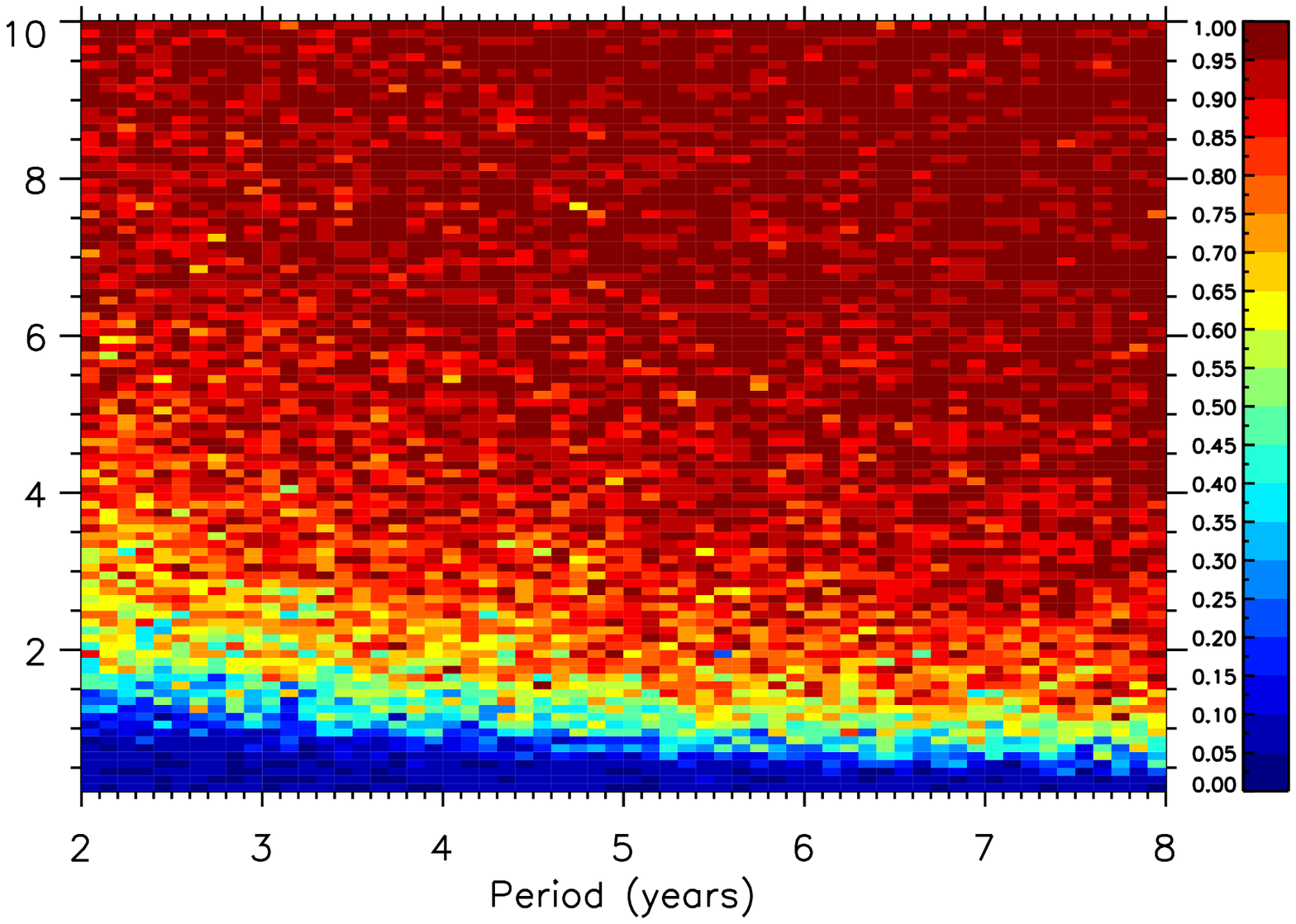}

                          \caption{Fraction of detected companions in a companion mass versus orbital period plot, based on Monte Carlo simulations using the RV data (left panel) and astrometric data (right panel) for GJ 406.  }

 \end{subfigure}

\end{center}
\end{figure}


\begin{figure}
\begin{center}
\begin{subfigure}
		              
		\includegraphics[scale=.44,angle=0]{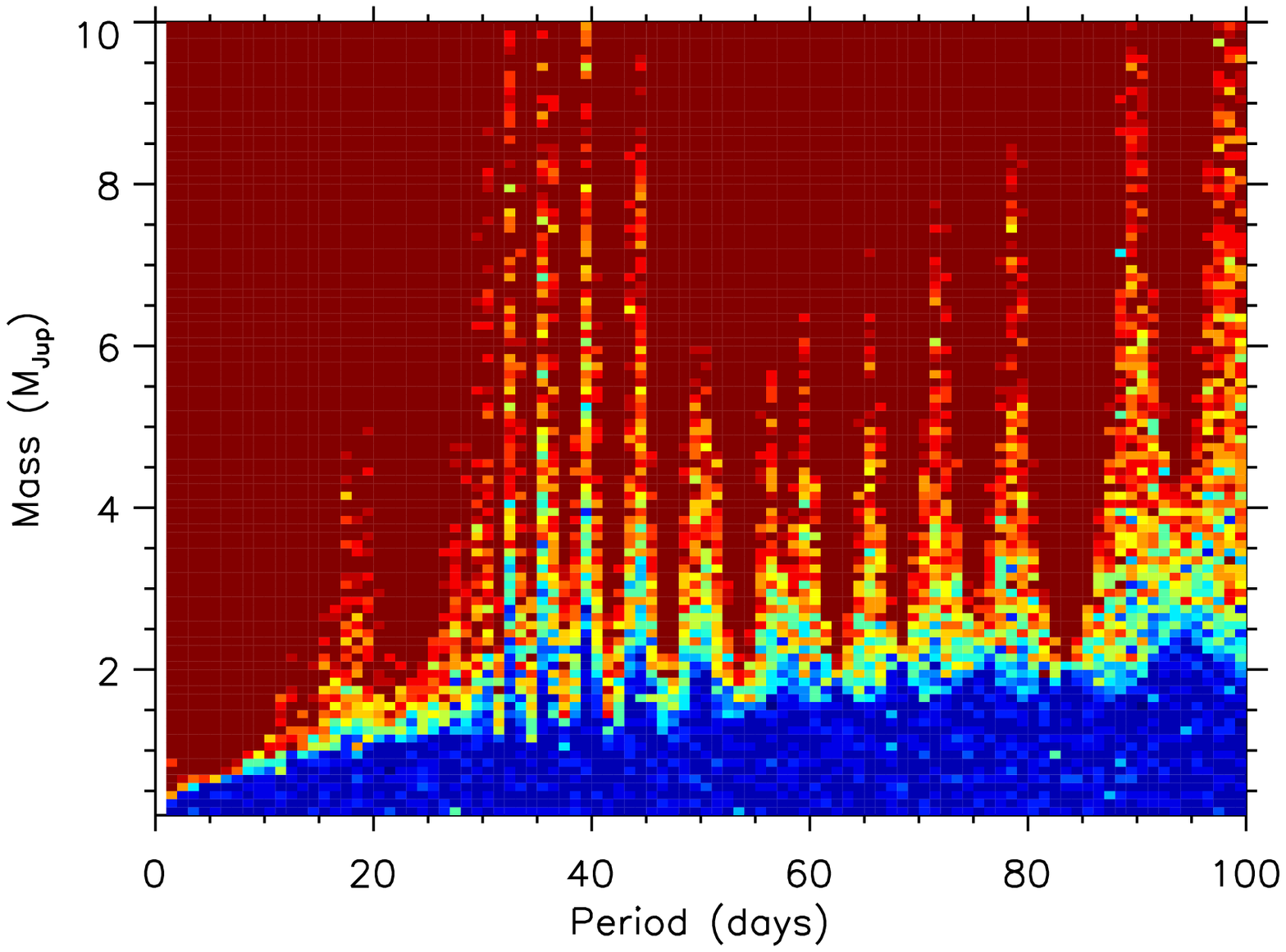}
                \includegraphics[scale=.44,angle=0]{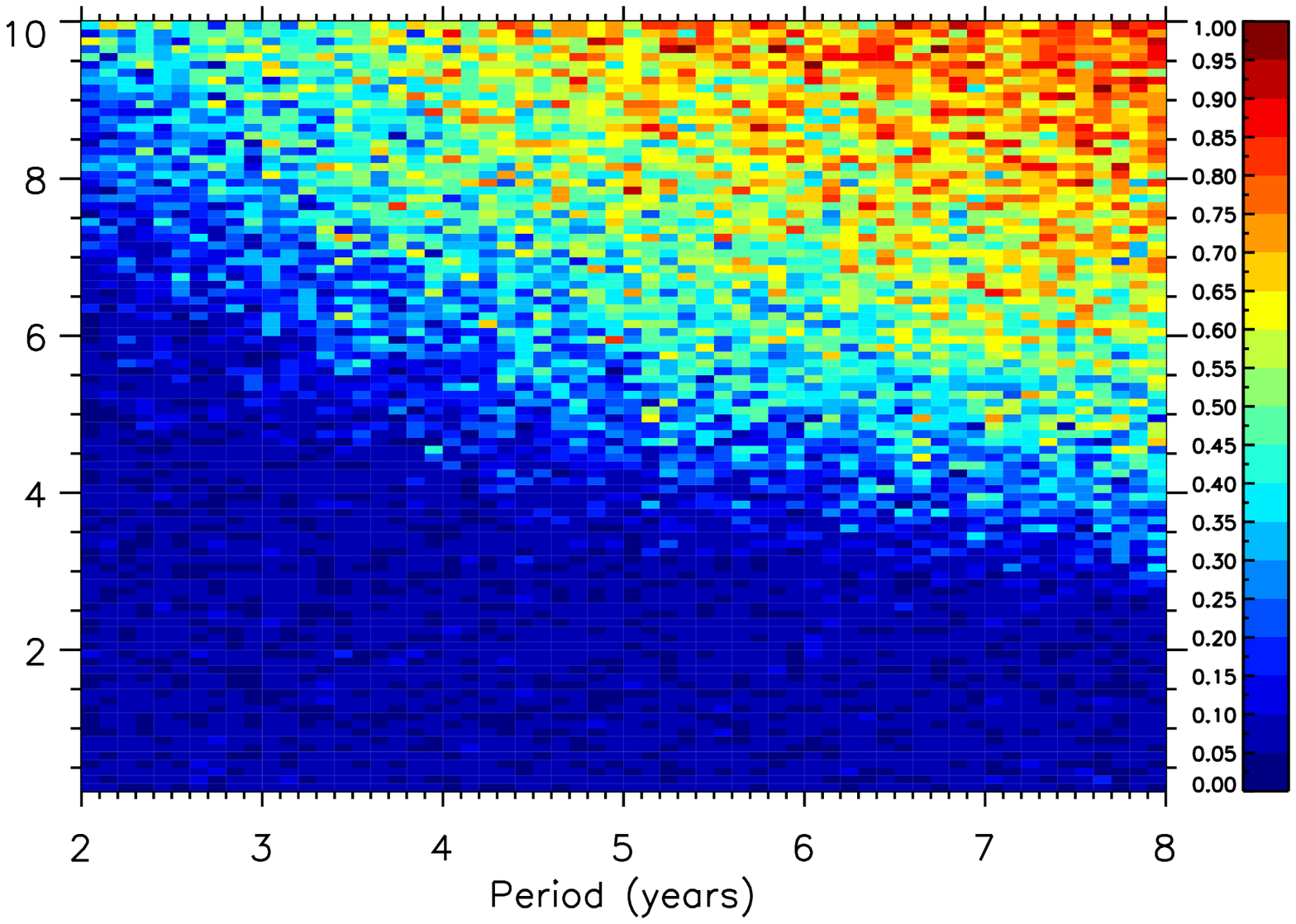}

                          \caption{Fraction of detected companions in a companion mass versus orbital period plot, based on Monte Carlo simulations using the RV data (left panel) and astrometric data (right panel) for GJ 555. }

 \end{subfigure}

\end{center}
\end{figure}


\begin{figure}
\begin{center}
\begin{subfigure}
		              
		\includegraphics[scale=.44,angle=0]{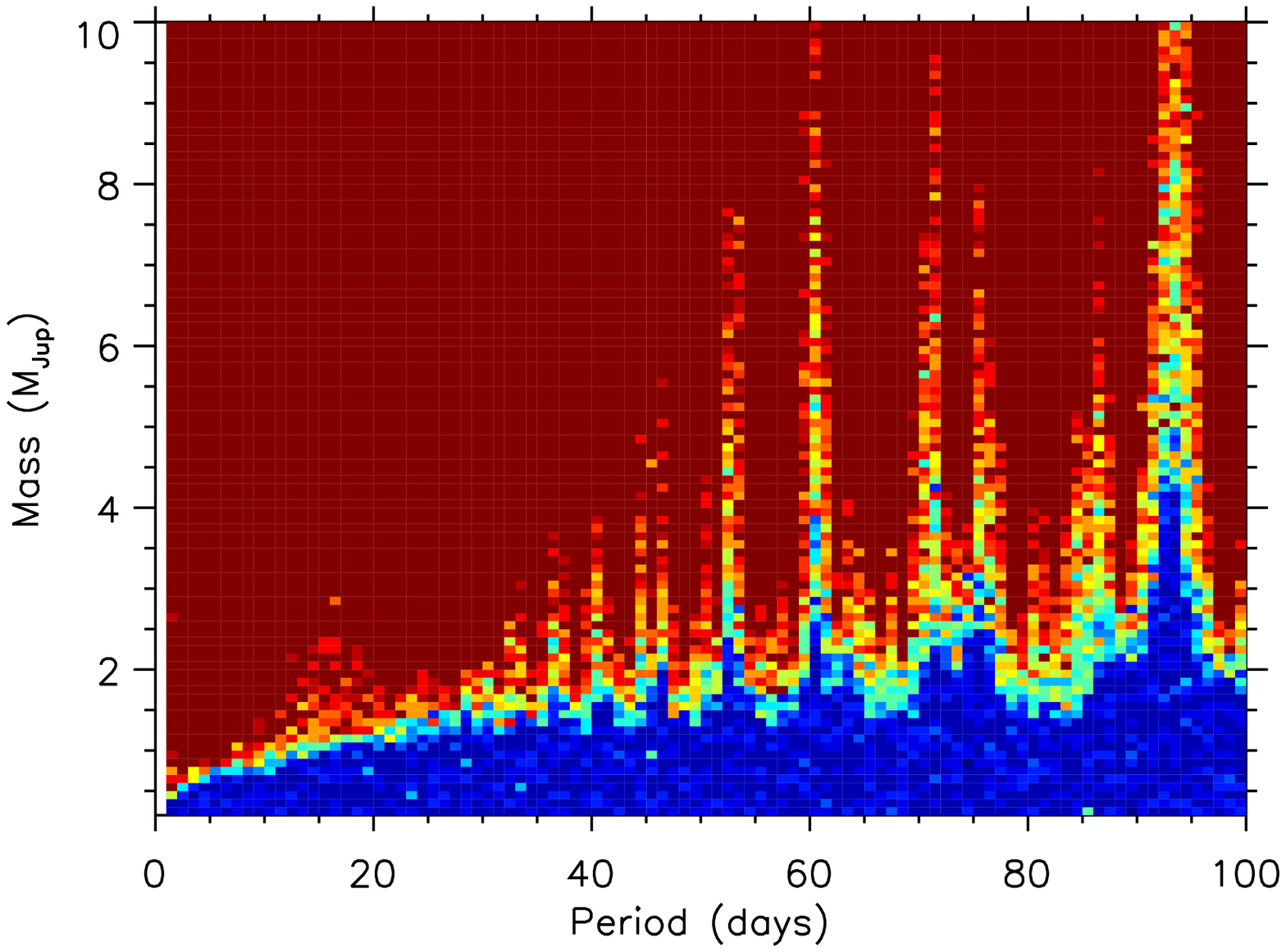}
                \includegraphics[scale=.44,angle=0]{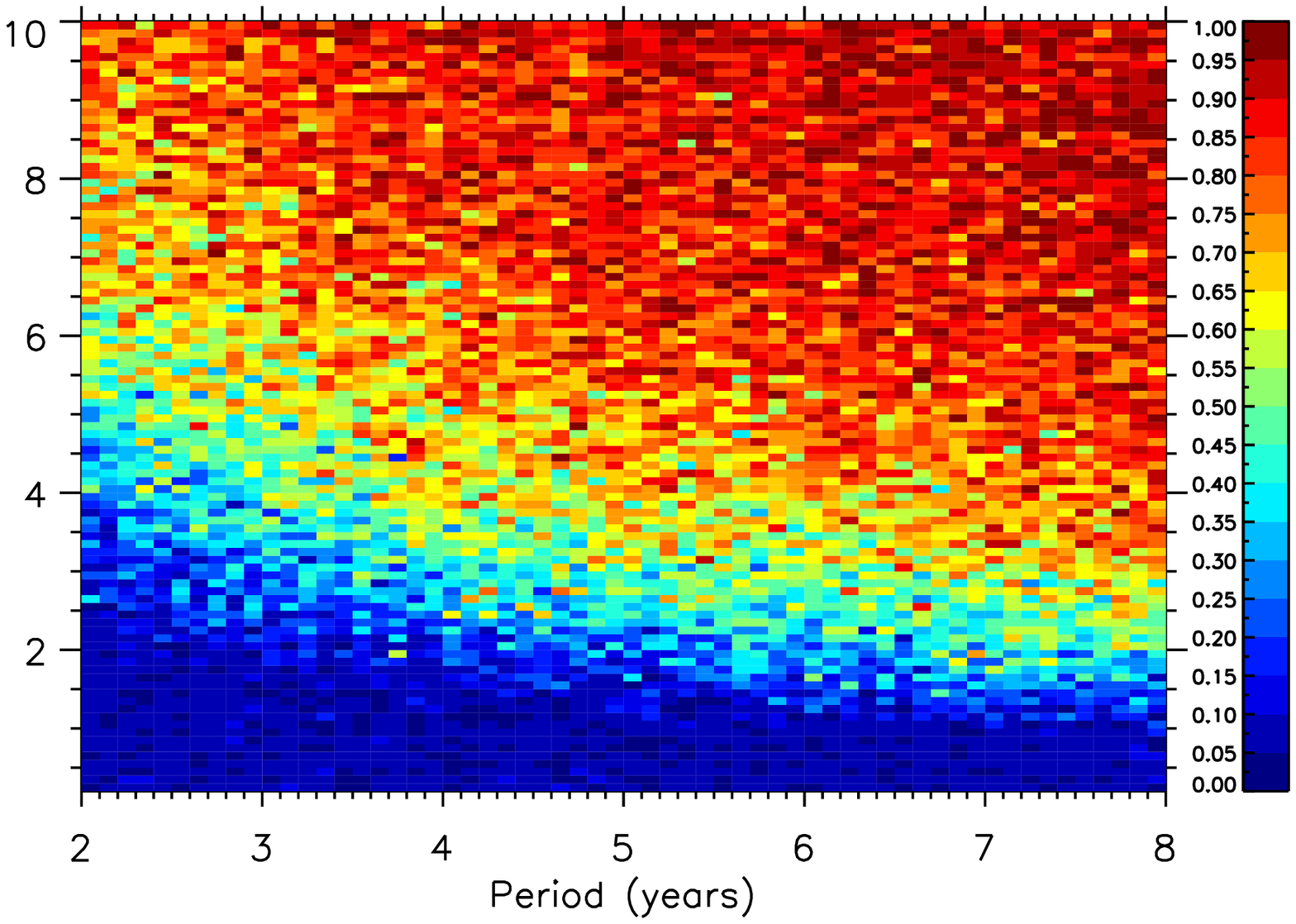}

                          \caption{Fraction of detected companions in a companion mass versus orbital period plot, based on Monte Carlo simulations using the RV data (left panel) and astrometric data (right panel) for GJ 628.  }

 \end{subfigure}

\end{center}
\end{figure}


\begin{figure}
\begin{center}
\begin{subfigure}
		              
		\includegraphics[scale=.44,angle=0]{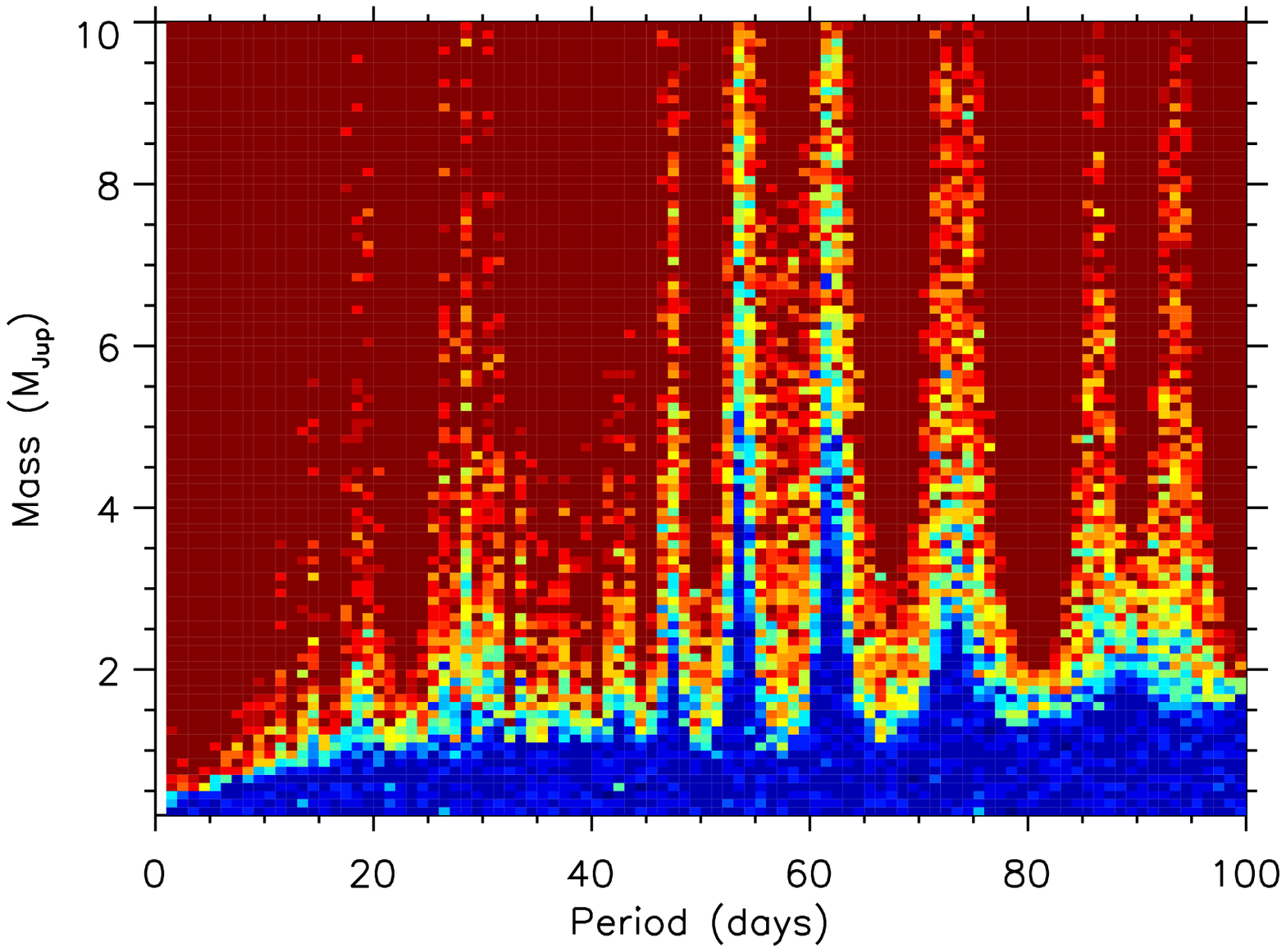}
                \includegraphics[scale=.44,angle=0]{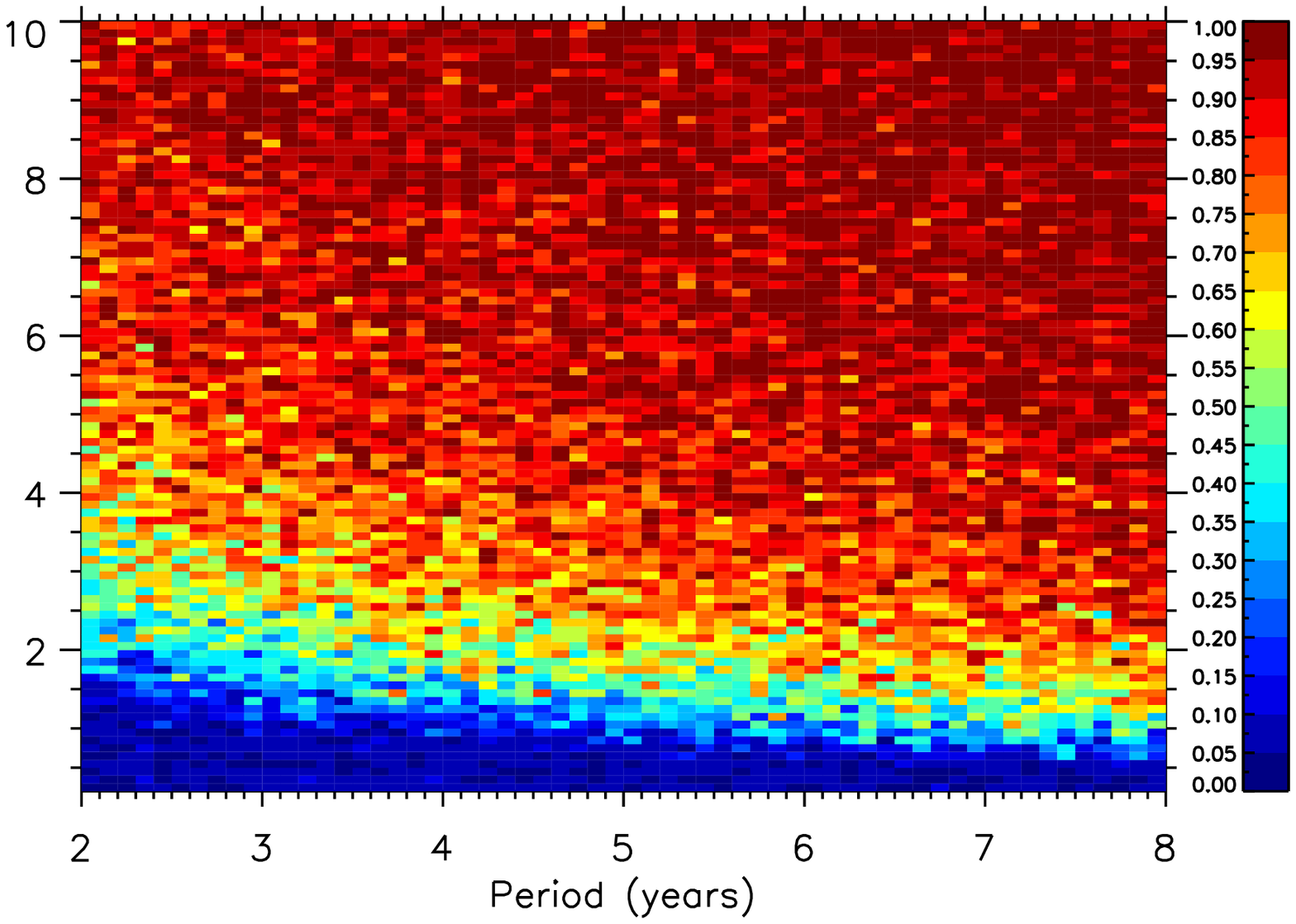}

                          \caption{Fraction of detected companions in a companion mass versus orbital period plot, based on Monte Carlo simulations using the RV data (left panel) and astrometric data (right panel) for GJ 729. }

 \end{subfigure}

\end{center}
\end{figure}


\begin{figure}
\begin{center}
\begin{subfigure}
		              
		\includegraphics[scale=.44,angle=0]{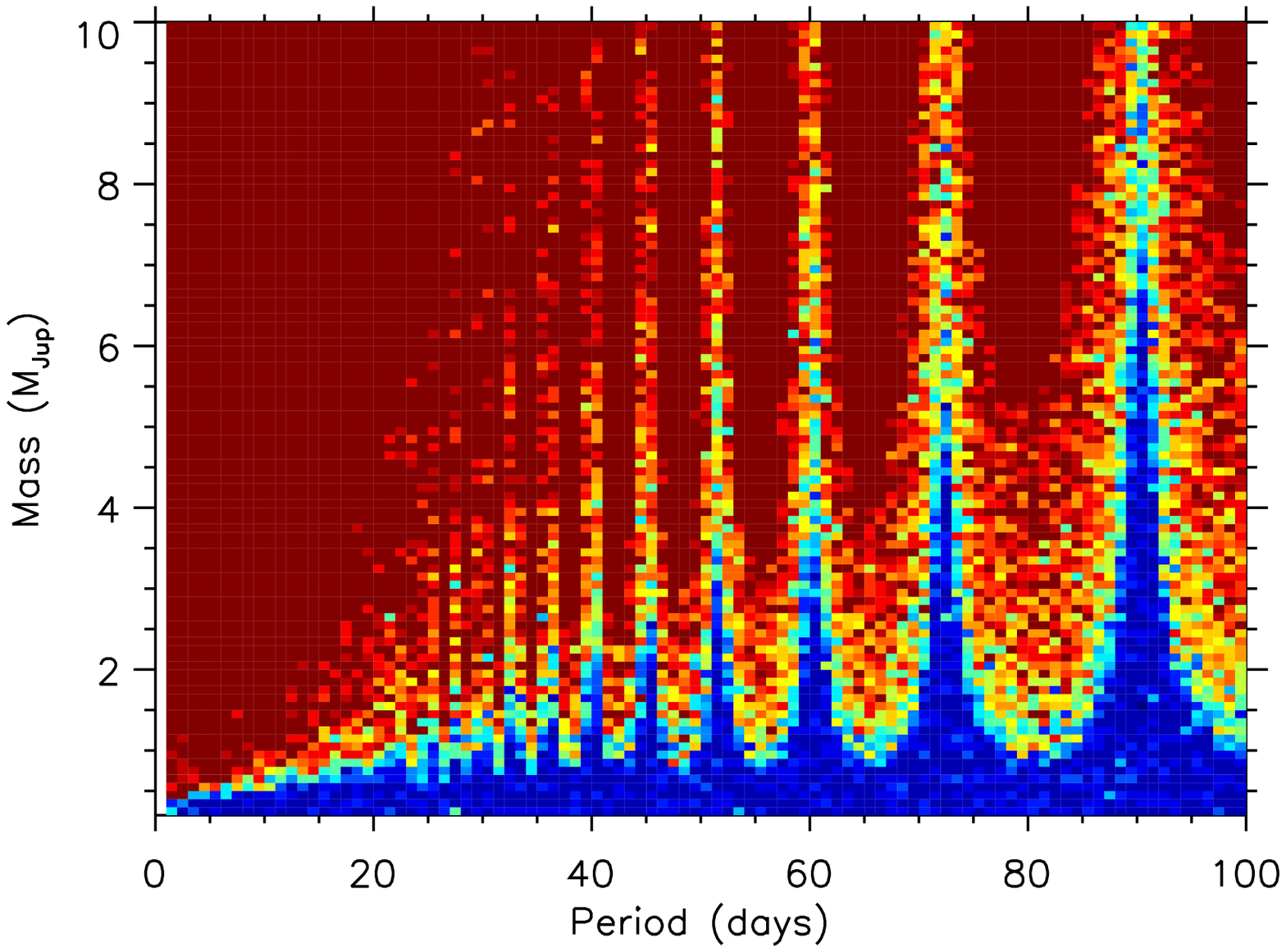}
                \includegraphics[scale=.44,angle=0]{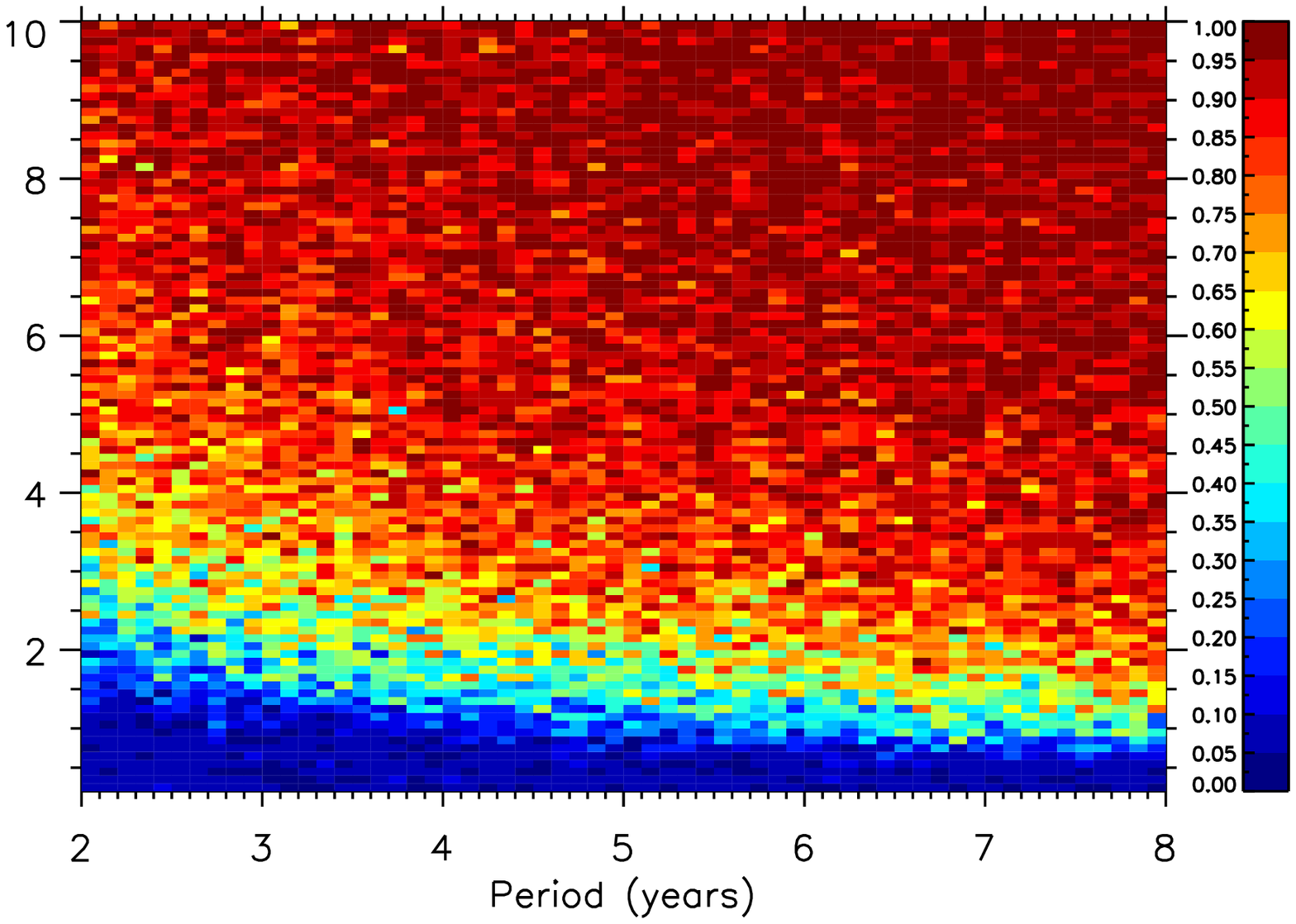}

                          \caption{Fraction of detected companions in a companion mass versus orbital period plot, based on Monte Carlo simulations using the RV data (left panel) and astrometric data (right panel) for GJ 1002. }

 \end{subfigure}

\end{center}
\end{figure}


\begin{figure}
\begin{center}
\begin{subfigure}
		              
		\includegraphics[scale=.44,angle=0]{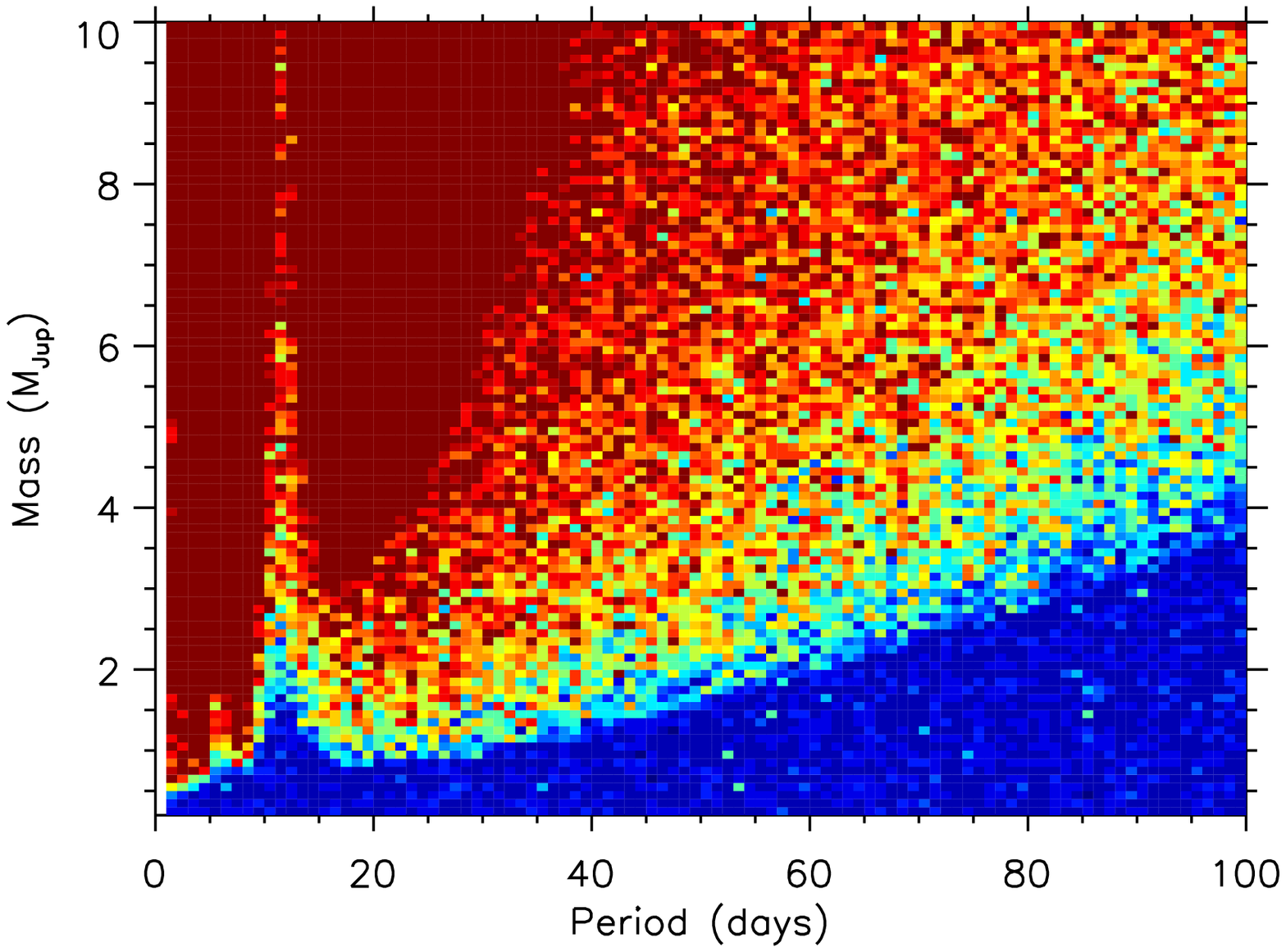}
                \includegraphics[scale=.44,angle=0]{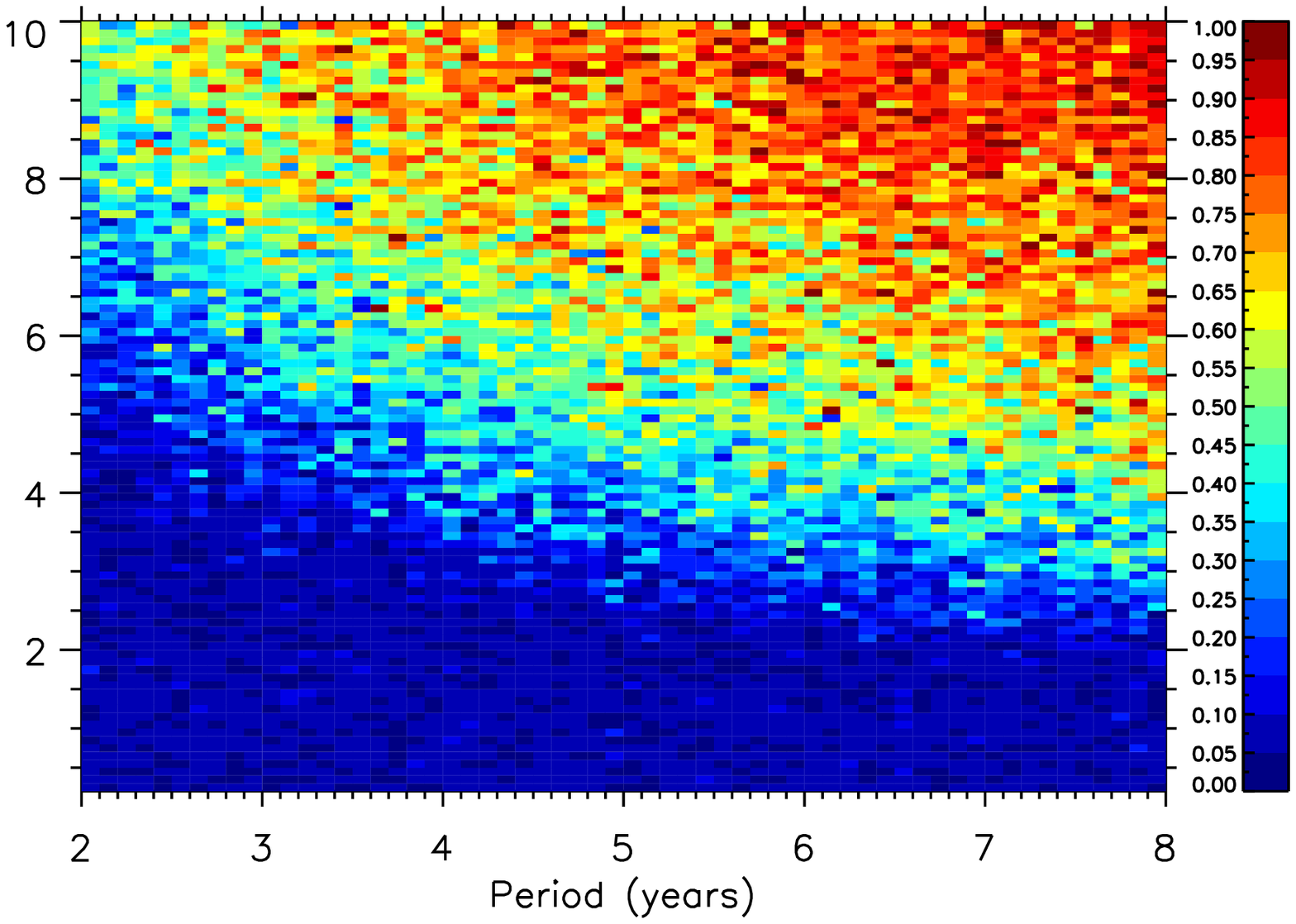}

                          \caption{Fraction of detected companions in a companion mass versus orbital period plot, based on Monte Carlo simulations using the RV data (left panel) and astrometric data (right panel) for GJ 1065. }

 \end{subfigure}

\end{center}
\end{figure}


\begin{figure}
\begin{center}
\begin{subfigure}
		              
		\includegraphics[scale=.44,angle=0]{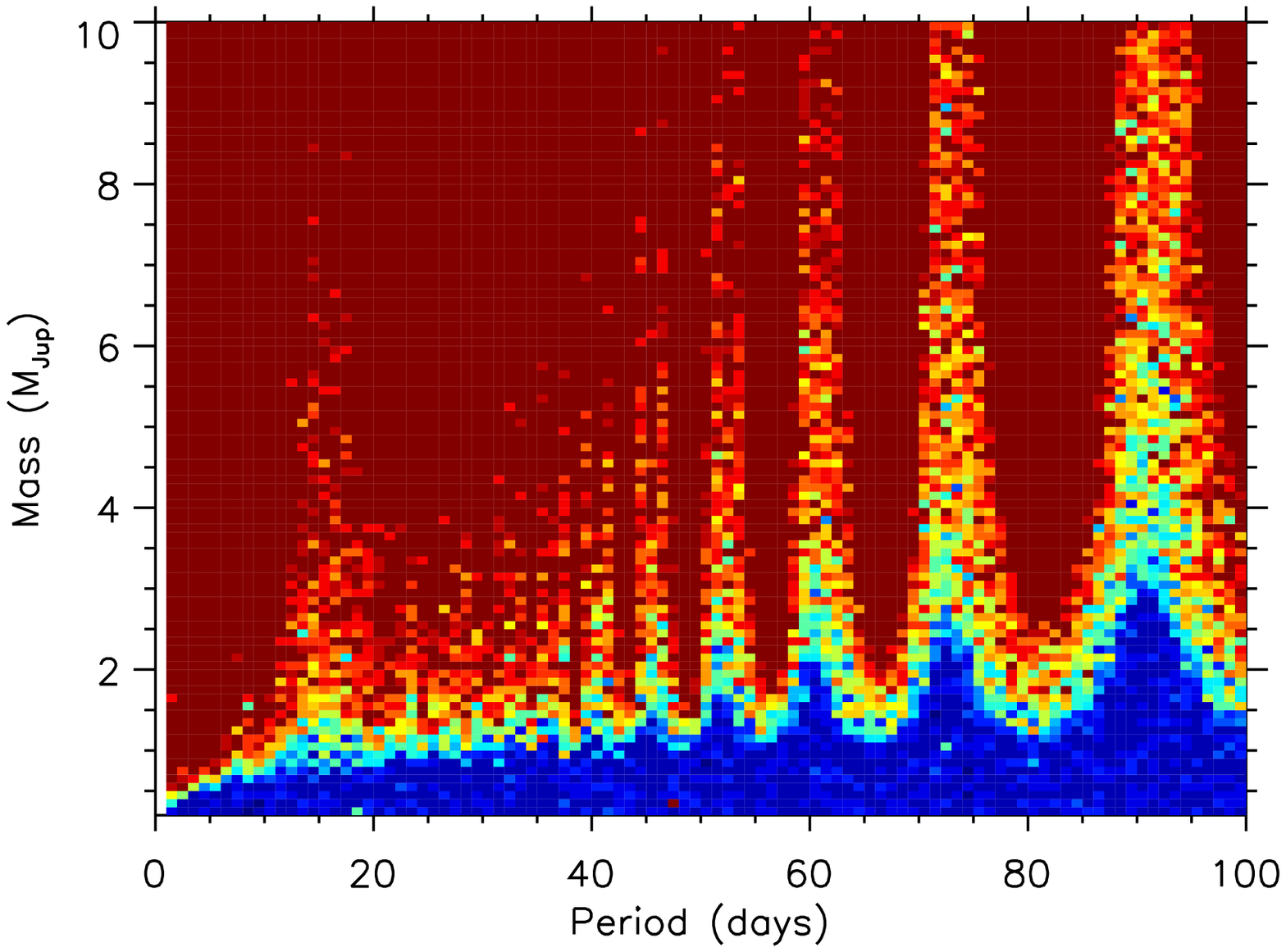}
                \includegraphics[scale=.44,angle=0]{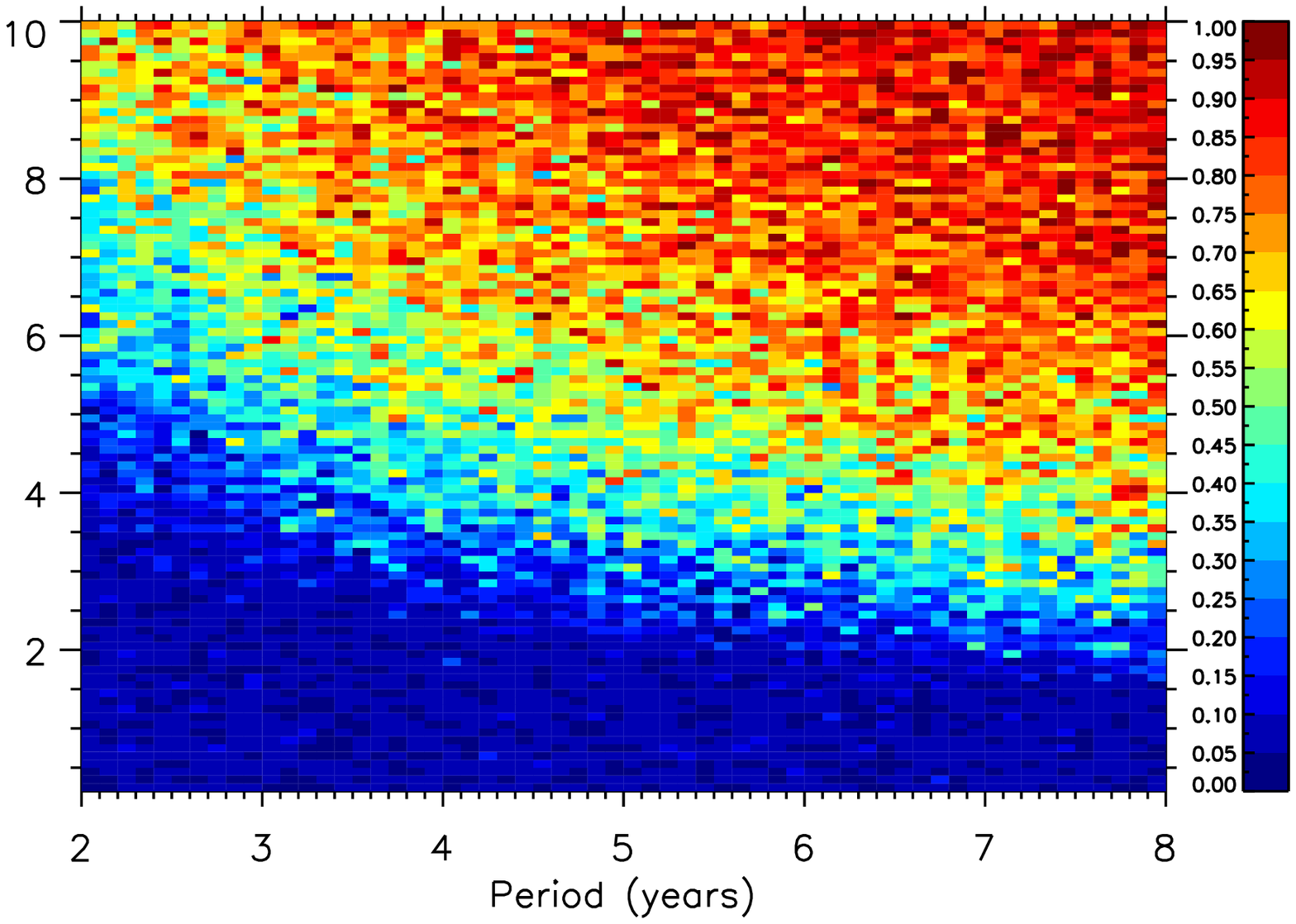}

                          \caption{Fraction of detected companions in a companion mass versus orbital period plot, based on Monte Carlo simulations using the RV data (left panel) and astrometric data (right panel) for GJ 1224.  }

 \end{subfigure}

\end{center}
\end{figure}


\begin{figure}
\begin{center}
\begin{subfigure}
		              
		\includegraphics[scale=.44,angle=0]{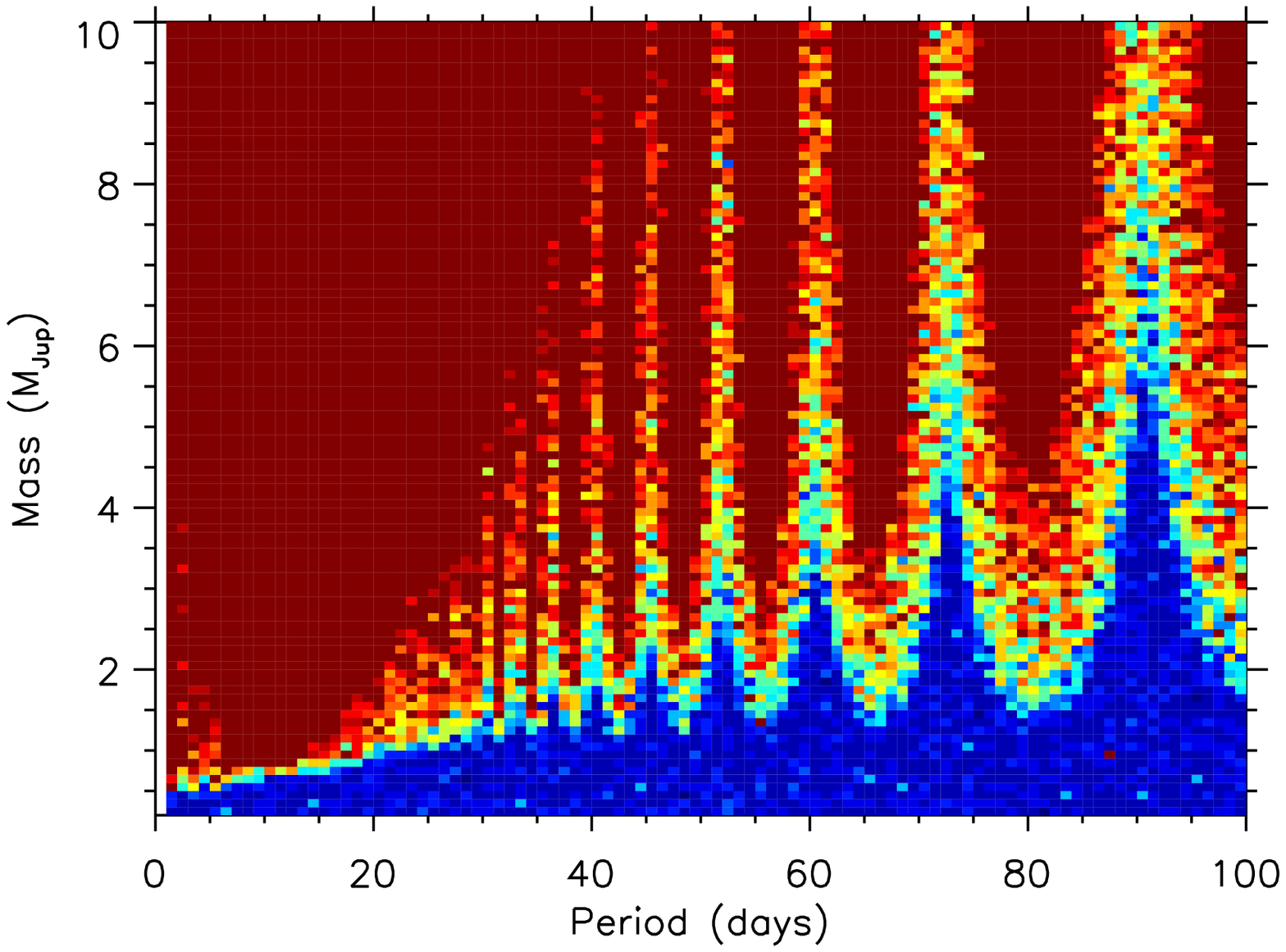}
                \includegraphics[scale=.44,angle=0]{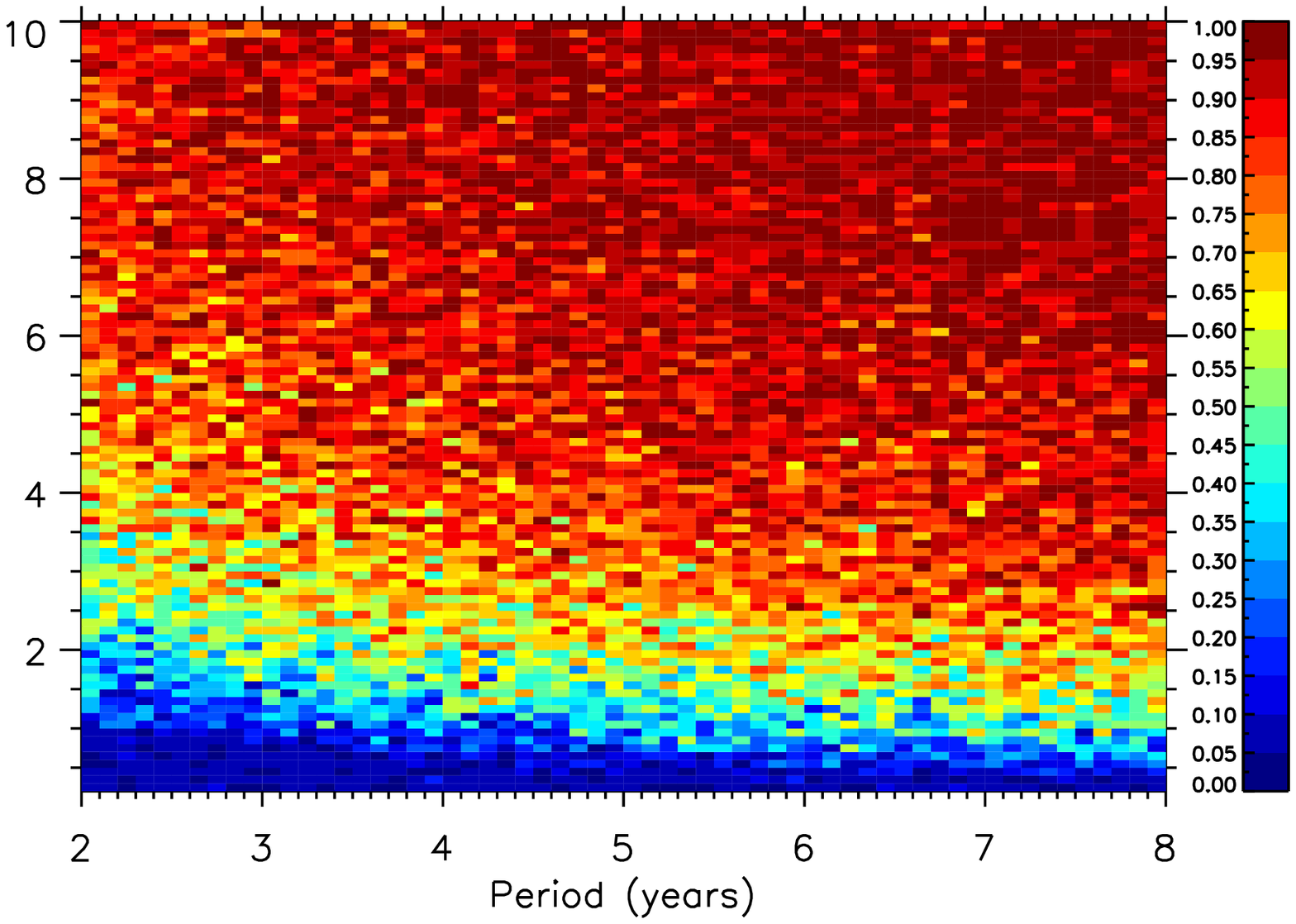}

                          \caption{Fraction of detected companions in a companion mass versus orbital period plot, based on Monte Carlo simulations using the RV data (left panel) and astrometric data (right panel) for GJ 1286.  }

 \end{subfigure}

\end{center}
\end{figure}


\begin{figure}
\begin{center}
\begin{subfigure}
		              
		\includegraphics[scale=.44,angle=0]{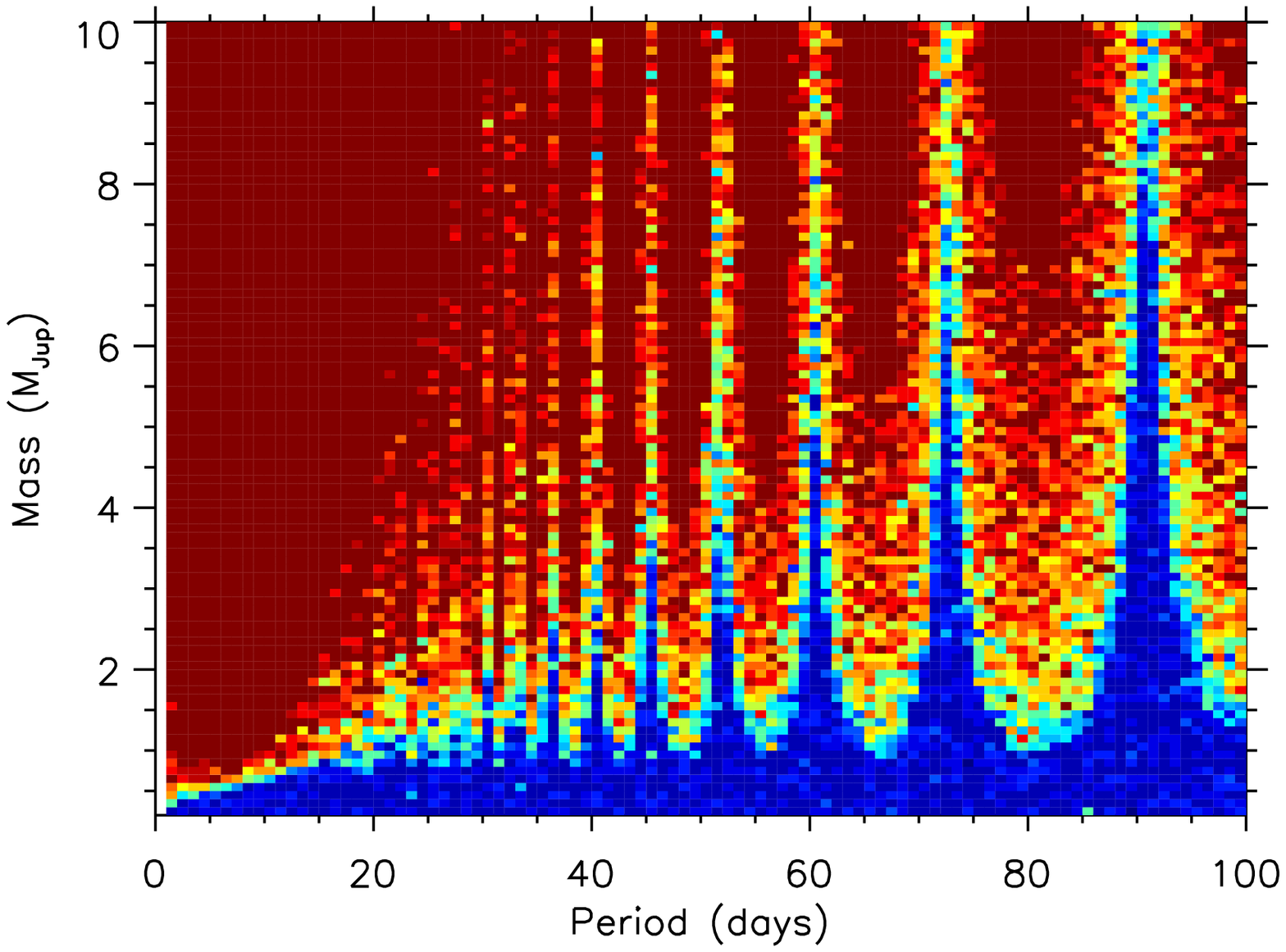}
                \includegraphics[scale=.44,angle=0]{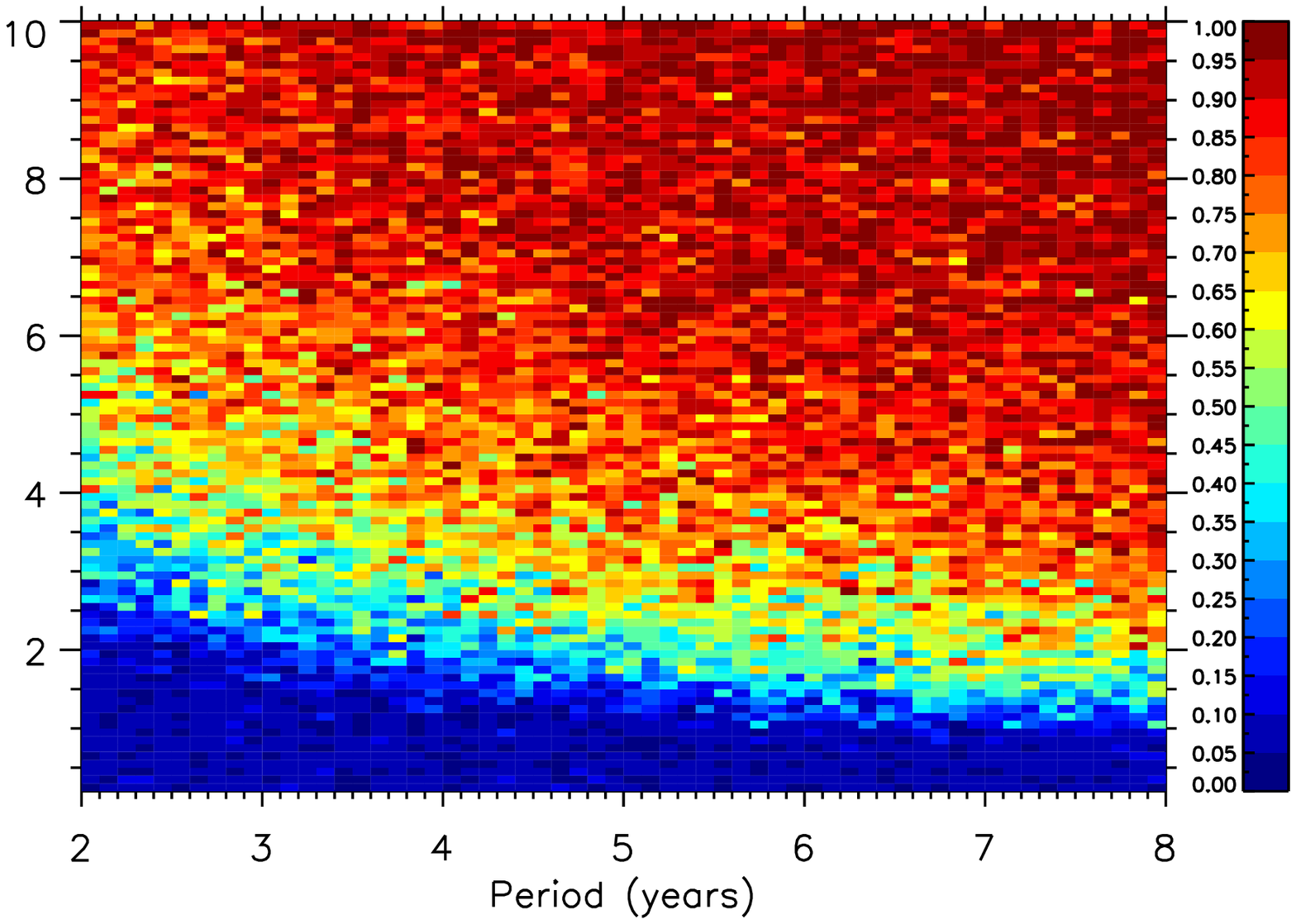}

                          \caption{Fraction of detected companions in a companion mass versus orbital period plot, based on Monte Carlo simulations using the RV data (left panel) and astrometric data (right panel) for LHS 1723.  }

 \end{subfigure}

\end{center}
\end{figure}


\begin{figure}
\begin{center}
\begin{subfigure}
		              
		\includegraphics[scale=.44,angle=0]{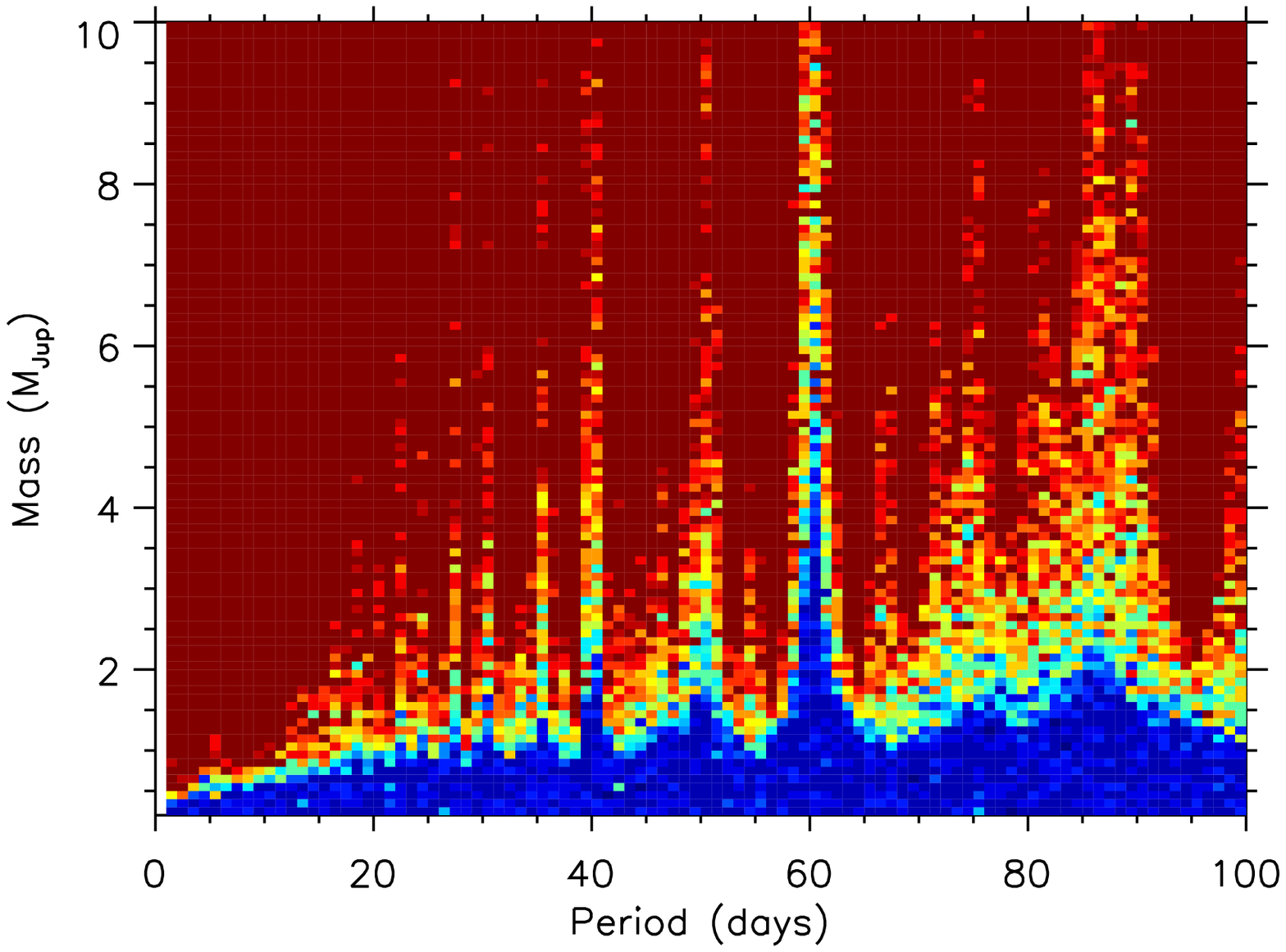}
                \includegraphics[scale=.44,angle=0]{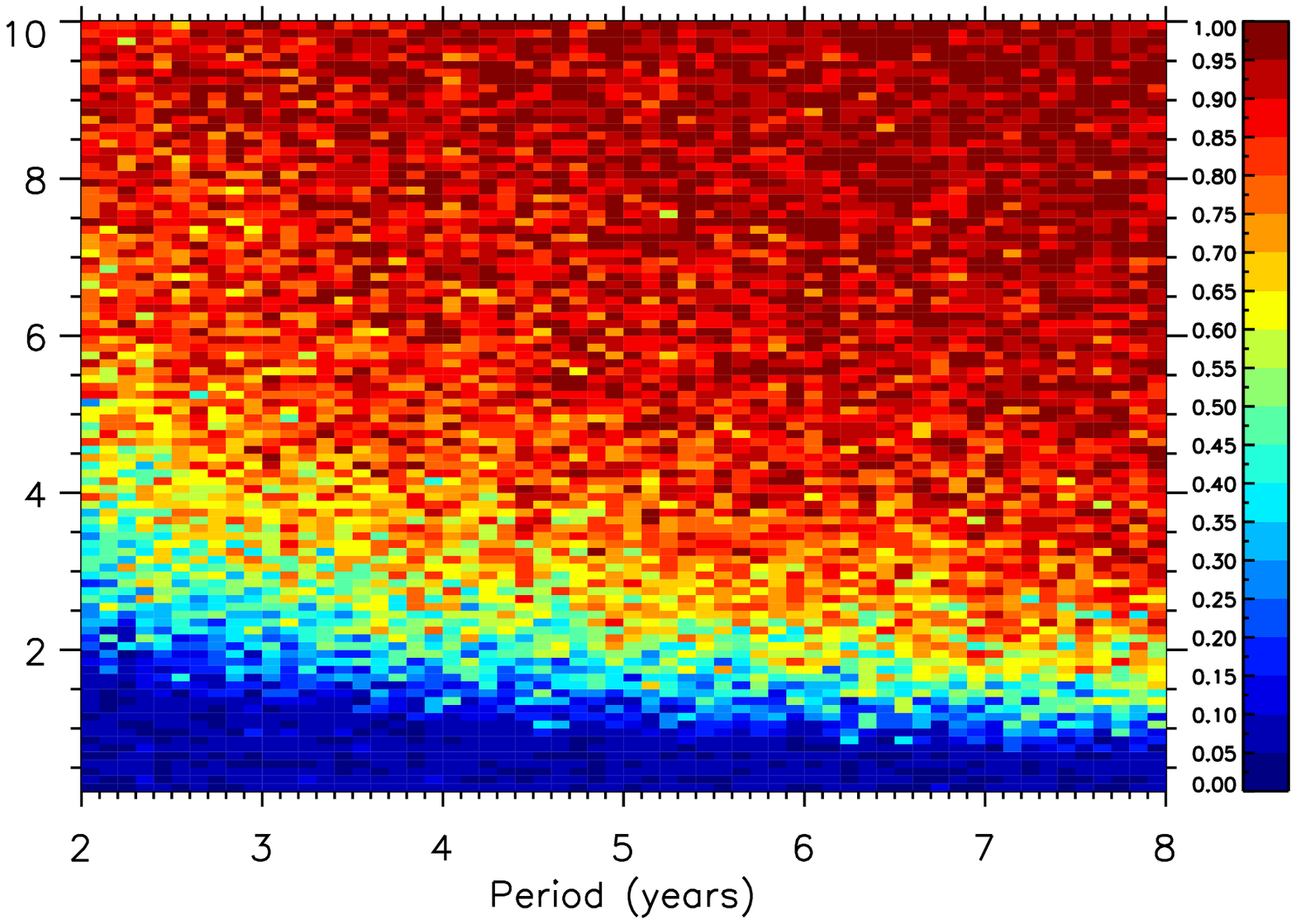}

                          \caption{Fraction of detected companions in a companion mass versus orbital period plot, based on Monte Carlo simulations using the RV data (left panel) and astrometric data (right panel) for LHS 3799.  }

 \end{subfigure}

\end{center}
\end{figure}


\end{document}